\documentclass[usenatbib]{mnras}

\usepackage[T1]{fontenc}

\DeclareRobustCommand{\VAN}[3]{#2}
\let\VANthebibliography\thebibliography
\def\thebibliography{\DeclareRobustCommand{\VAN}[3]{##3}\VANthebibliography}

\usepackage{graphicx}	
\usepackage{amsmath}	
\usepackage{txfonts}

\graphicspath{{Figures/}}
\usepackage{relsize}
\title[Magnetically-induced thermal mountains]{Gravitational radiation from thermal mountains on accreting neutron stars: sources of temperature non-axisymmetry}

\author[T. J. Hutchins \& D. I. Jones]{
T. J. Hutchins\thanks{E-mail: T.J.Hutchins@soton.ac.uk} \&  D. I. Jones
\\
Mathematical Sciences and STAG Research Centre, University of Southampton, Southampton SO17 1BJ, United Kingdom
}

\date{Accepted XXX. Received YYY; in original form ZZZ}

\pubyear{2022}

\begin{document}
\label{firstpage}
\pagerange{\pageref{firstpage}--\pageref{lastpage}}
\maketitle

\begin{abstract}
The spin-distribution of accreting neutron stars in low-mass X-ray binary (LMXB) systems shows a concentration of pulsars well below the Keplarian break-up limit. It has been suggested that their spin frequencies may be limited by the emission of gravitational waves, due to the presence of large-scale asymmetries in the internal temperature profile of the star. These temperature asymmetries have been demonstrated to lead to a non-axisymmetric mass-distribution, or `mountain', that generates gravitational waves at twice the spin frequency. The presence of a toroidal magnetic field in the interior of accreting neutron stars has been shown to introduce such anisotropies in the star's thermal conductivity, by restricting the flow of heat orthogonal to the magnetic field and establishing a non-axisymmetric temperature distribution within the star. We revisit this mechanism, extending the computational domain from (only) the crust to the entire star, incorporating more realistic microphysics, and exploring different choices of outer boundary condition. By allowing a magnetic field to permeate the core of the neutron star, we find that the likely level of temperature asymmetry in the inner crust ($\rho \sim 10^{13}$ g cm$^{-3}$) can be up to 3 orders of magnitude greater than the previous estimate, improving prospects for one day detecting continuous gravitational radiation. We also show that temperature asymmetries sufficiently large to be interesting for gravitational wave emission can be generated in strongly accreting neutron stars if \textit{crustal} magnetic fields can reach $\sim 10^{12}$ G. 
\end{abstract}




\section{Introduction}
\label{section:Introduction}

As the LIGO-Virgo-KAGRA collaboration approaches its 100$^{\text{th}}$ detection of a gravitational wave (GW) signal, the detection of a long-lived, (almost)-monochromatic \textit{continuous} gravitational waves remains elusive. Detections across three observing runs (O1, O2; \citet{Abbott_2019} and O3; \citet{theligoscientificcollaboration2021gwtc3}) have included binary black hole, binary neutron star, and most recently, binary neutron star - black hole coalescences \citep{Abbott_2021_Observation}. These merger events produce powerful, short-lived signals and belong to the class of \textit{transient} GW signals, that have facilitated advancements in the understanding of a range of astronomical phenomena. These include the physics of gamma ray bursts \citep{Abbott_2017}, as well as placing constraints on both the equation of state of dense matter \citep{Abbott_2018} and the Hubble Constant \citep{2017LIGO, theligoscientificcollaboration2021constraints}.  

In contrast to their transient counterparts, sources of continuous GWs produce signals that vary in both amplitude and frequency over much longer timescales. Single neutron stars (NS) are among the most promising source of continuous GW emission, with a number of different candidate mechanisms thought to be able to facilitate significant emission. These include: (i) fluid instabilities, particularly the excitation of the r-mode; (ii) non-axisymmetric quadrupolar deformations of the solid crust; (iii) free precession (see \citet{2018Glampedakis} for a recent review of these processes). The detection of continuous GW signals therefore presents a unique opportunity to explore the behavior of dense matter in a number of different astrophysical scenarios distinct from those relevant during binary coalescence.

In the second scenario, deformations on neutron stars are more often referred to as `mountains'. Such deformations can be divided into two classes.  There are so-called ‘\textit{magnetic} mountains’ formed via magnetic stresses built up in strongly magnetised neutron stars, and ‘\textit{elastic} mountains’, whereby deformations in solid crust of the NS are supported by elastic strains. 

Elastic mountains are a promising source of continuous GW radiation due to the ability of the solid crust of the neutron star to be strong enough to support significant stresses before cracking \citep{2002, 2006Haskell, 2013Owen, Gittins_2021, Morales:2022wxs}. Specific to accreting neutron stars, a promising sub-class of elastic mountains are the so-called `\textit{thermal} mountains', where deformations supported by elastic strains in the crust are thought to be formed due to the presence of large-scale temperature asymmetries that are misaligned from the star's rotation axis \citep{Bildsten_1998, 2002}. Assuming the crust could sustain the mountain, a description of how the neutron star might inherit such a non-axisymmetric temperature distribution in the first place has, until recently, been lacking, and motivates this paper.

\subsection{An Overview of the Problem}
\label{section:An Overview of the Problem}

Statistical analysis by \citet{Patruno_2017} of the spin-distribution of LMXBs suggests evidence for two distinct sub-populations. The majority of the neutron stars in LMXB exhibit, on average, a low spin frequency ($\Bar{\nu}_s \approx 300$ Hz). However, there are a small minority of stars which exhibit higher spin frequencies ($\Bar{\nu}_s \approx 575$ Hz). Millisecond pulsars are believed to be the descendants of LMXBs, with the most rapidly rotating pulsar, PSR J1748-2446ad, having a spin frequency of $\nu_s = 716$ Hz \citep{Hessels_2006}. These observations sit uneasily with the current understanding of accretion theory. Neutron stars from LMXBs are old \citep{1991PhR...203....1B}. Spin-up torques provided by long-term accretion implies that the neutron stars should have accreted enough angular momentum to possess rotation speeds approaching that of the Keplarian break-up frequency ($ \nu_k \approx 1 - 1.5$ kHz for most realistic equations of state; \citet{LATTIMER_2007}). Such considerations are therefore suggestive of the presence of some additional \textit{unknown} braking mechanism that is halting the spin up of these stars below that of the break-up limit.

The idea that GW torques from neutron star mountains may be dictating the spin equilibrium period in LMXBs was popularised by \citet{Bildsten_1998}. The idea echoed earlier discussions by \citet{1978MNRAS.182..423P} \& \citet{1984ApJ...278..345W}, that equilibrium might be achieved via gravitational wave emission from rotational instabilities. \citet{Bildsten_1998} posited a phenomenological argument that the rate of angular momentum lost via quadrupolar GW emission scales sharply with a high power of the spin frequency ($\nu_s^5$ for a deformed star) and may therefore naturally provide a ‘barrier’ to prevent further spin-up frequency past some equilibrium value. Encouragingly, recent timing observations of the pulsar PSR J1023+0038 has shown that the NS spins down $\sim 27\%$ faster during episodes of active accretion than in periods of quiescence \citep{2017}. This result can, at least qualitatively, be explained by the presence of gravitational wave torques from a transient ‘mountain’ inside the neutron star that forms during phases of accretion. 

The structure of isolated neutron stars and those found in LMXBs are markedly different. Accreted matter (mostly hydrogen, helium, and other light elements) settles on the neutron star surface and is subsequently buried under ever-increasing amounts of freshly accreted matter. The layers of hydrogen and helium within this envelope subsequently ignite. The resulting thermonuclear burning leads to the formation of a layer of nuclear ashes at the bottom of the envelope, consisting of mostly iron-peak nuclides as well as other heavier elements \citep{2000, Fantina}. Continual accretion from the companion star leads to further compression of these heavy nuclides that subsequently sink into the crust, replacing the primordial crust after approx. $\sim$ $5 \times 10^7$ years (assuming constant accretion at a rate, $\dot{M}$ / 10$^{-9}$ $M_{\odot}$ yr$^{-1}$; \citet{2002}). 

The accreted crust is defined by a series of density-dependant, non-equilibrium reactions (electron captures, pycnonuclear reactions and neutron emissions) that irreversibly changes the composition of the primordial crust. Interestingly, \citet{Cumming} showed that the electron-capture events also have weak temperature sensitivity. Consequently, in the regions of the crust that are hotter on average, electron-capture events can take place at lower density (closer to the star’s surface) than colder regions. Any temperature asymmetry in the accreted crust therefore results in `wavy' electron capture layers (\citet{Bildsten_1998}; and cf.Fig. 2 of \citet{2002}). If the temperature asymmetry happens to be misaligned with the rotation axis, then the mass distribution itself can inherit the non-axisymmetry and hence develop the required deformation, or `mountain'. 

There have been several studies of the level of temperature asymmetry one might realistically expect to provide significant GW emission. The earliest being \citet{2002} (hereafter UCB) that sought to explore in more detail the mechanism described by \citet{Bildsten_1998}. The authors considered two asymmetry-inducing scenarios, whereby either (i) non-axisymmetric nuclear burning during X-ray bursts could create local fluctuations in the amount of energy being released due to the nuclear reactions, or (ii) the burning of different atomic masses in different regions of the star could lead to variations in the charge-to-mass ratio ($Z^2/A$) and thus alter the star's thermal conductivity. 
UCB found that non-axisymmetries of the order 10\% in either the heating rate or the composition could provide the necessary asymmetry in order to build sufficiently large mountains. In this case, it was found that fractional temperature asymmetries at the order of the percent level ($\delta T / T \sim 1 \%$) were required to generate adequate mass quadrupoles. However,  UCB do not substantiate exactly \textit{how} a 10\% asymmetry in the heating rate or composition might arise, and instead simply assume this level of asymmetry \textit{a priori}. 

In fact, the initial description laid out by \citet{Bildsten_1998} also did not explore mechanisms to create the necessary temperature gradients. To reconcile this issue, \citet{Osborne_2020} (hereafter OJ) sought to provide substantiation by developing a mechanism for which the star might naturally inherit a non-axisymmetric temperature distribution. Like UCB, OJ also considered variations in the star's thermal conductivity. However, rather than assuming lateral variations in the star's composition, OJ exploited the likely (weak) internal magnetic fields of neutron stars in LMXBs to perturb the thermal conductivity tensor, thus rendering it anisotropic. By confining their calculation to the accreted crust only, OJ found that large \textit{internal} crustal toroidal magnetic fields ($B \sim 10^{13}$ G) - four orders of magnitude larger than inferred ($B \sim 10^9$ G) \textit{external} field strengths of LMXBs - were required in order to produce deformations large enough to facilitate GW emission at a significant level.  

In this paper, we seek to revisit and improve various aspects of the OJ mechanism for building a temperature asymmetry in the neutron star crust. Namely, we will also assume the neutron star to be threaded with an internal magnetic field, but extend the computational domain to include a self-consistent calculation of the thermal structure of the NS core (including descriptions of the relevant microphysics and the presence of baryon superfluidity - Section \ref{section:Thermal Structure of an accreting neutron star}). In doing so, we shall allow for the possibility of the magnetic field to penetrate the core, with the expectation that non-vanishing temperature perturbations at the crust-core transition will lead to greater asymmetries in the deep crust. 

Estimates of temperature asymmetry in magnetised neutron stars have also been provided by \citet{2020Singh}. The authors sought to model asymmetric accretion flow onto the polar cap of the neutron star, and exploit physical shifts in the location of electron capture layers induced by magnetic strains. It was found that, in general, the mass quadrupole generated from such a mechanism is too small to produce substantial GW emission, unless significant amounts of shallow heating sources are present in the outer regions of the crust.

To date, there have been a number of searches for continuous gravitational wave radiation within Advanced LIGO/Virgo data. These have included searches for continuous GW radiation from known neutron stars, e.g.\ accreting millisecond X-ray pulsars (AMXP) (e.g. \citet{2005LIGO, 2007LIGO, 2011LIGO, 2017LIGO, PhysRevD.105.022002}), and also `all-sky' searches for GW emission from unknown sources (e.g. \citet{2008LIGOAll, 2014LIGOAll, 2018LIGOAll, 2021LIGOAll}. The most recent and comprehensive of the targeted AMXP searches \citep{PhysRevD.105.022002} found no compelling evidence for continuous GW emission from any of the 20 targeted sources. 

However, \citet{PhysRevD.105.022002} were able to place new upper limits on the wave strain, and therefore on the neutron star ellipticity, defined by $\varepsilon \equiv (I_{xx} - I_{yy}) / I_{zz}$ (i.e. the fractional difference in neutron star's principal moments of inertia orthogonal to the spin axis, with the spin axis pointing along the z direction). The strictest constraint on the NS ellipticity among all of the sources targeted in the search was obtained from the accreting millisecond pulsar IGR J00291 + 5934, estimated to be no larger than $\varepsilon = 3.1 \times 10^{-7}$ at 95\% credibility.

It is therefore of great interest to be able to accurately calculate the size of a mountain that a temperature asymmetry in the neutron star crust might develop. This would not only provide estimates on the strength of gravitational wave emission one might reasonably expect from NS mountains, but also inform future gravitational wave searches towards particular accreting millisecond pulsars with parameters such as strong external magnetic fields or high accretion rates that might make them favourable for sustaining a thermal mountain.

\subsection{Outline of this paper}
\label{section:Outline of this paper}

We begin in Section \ref{section:Structure of an Accreted Crust} by computing the hydrostatic structure of an accreted neutron star for three different modern accreted crustal equations of state (as described by \citet{Fantina, Fantina_2022}). These models will serve as a basis for which a thermal background can be computed, presented in Section \ref{section:Thermal Structure of an accreting neutron star}. We extend the work of OJ in this regard by increasing the computational domain of the calculation to the centre of the star, rather than just the density range of the \citet{1990A&A...227..431H, 1990A&A...229..117H} crustal equation of state (EoS). The relevant microphysics for such a calculation is also described, including (i) discussions of crustal heating (Section \ref{subsection:Accretion_Heating}), (ii) contributions to neutrino cooling and the thermal conductivity in both the crust and core of the NS (Sections \ref{section:Neutrino Cooling} \& \ref{subsection:Thermal Conductivity} respectively), as well as (iii) effects on these two latter pieces of the microphysics due to both proton superconductivity and neutron superfluidity (Section \ref{subsection:The Superfluid Transition Temperatures}). 

In Section \ref{section: Temperature Perturbation due to an Internal Magnetic Field} we outline how large-scale temperature asymmetries are induced due to the presence of an internal toroidal magnetic field and compute the level of asymmetry. In particular, we critique the boundary conditions used by UCB (and by extension OJ). We find that allowing the magnetic field to penetrate into the core of the NS can raise the expected level of temperature asymmetry in the inner crust by $\sim 2 - 3$ orders of magnitude. 

In Section \ref{sec: The Resulting Deformations} we then place our results into the context of wider continuous GW searches, discussing how the expected level of temperature asymmetry in accreting neutron stars calculated here can be used to compute the NS ellipticity. Using the current upper limit on the ellipticity of IGR J00291 + 5934 as obtained by \citet{PhysRevD.105.022002}, we provide a proof-of-concept method to place upper limits on the strength of the magnetic field in the NS interior, which is largely unknown. Finally, in Section \ref{sec: Conclusions} we summarise our findings, paying special attention to some of the pieces still missing from our description of thermal asymmetries in accreting neutron stars.

\section{Structure of an Accreted Crust}
\label{section:Structure of an Accreted Crust}

The internal composition of an accreting neutron star is paramount for computing its thermal profile. Yet, whilst a plethora of different equations of state exist for \textit{non-accreted} neutron stars (see e.g. \citet{2016EOS}), there are few equations of state that describe the properties of \textit{accreting} neutron stars. \citet{1990A&A...227..431H, 2003HZ, 2008HZ} and \citet{2012PhRvC..85e5804S} are among those to have calculated the composition of accreting neutron stars, with \citet{1990A&A...229..117H} providing tabulated EoS data for the accreted crust. Indeed, the calculations of \citet{1990A&A...227..431H, 1990A&A...229..117H} (hereafter HZ) were used in both UCB and OJ for their calculations of temperature asymmetries in the accreted crust. To build our background neutron star model, we employ the set of equations of state derived from the set of Brussels–Montreal functionals BSk19, BSk20, and BSk21 (specific to that of accreting neutron stars) as obtained by \citet{Fantina, Fantina_2022}.

\begin{figure}
	\includegraphics[width=0.85\columnwidth]{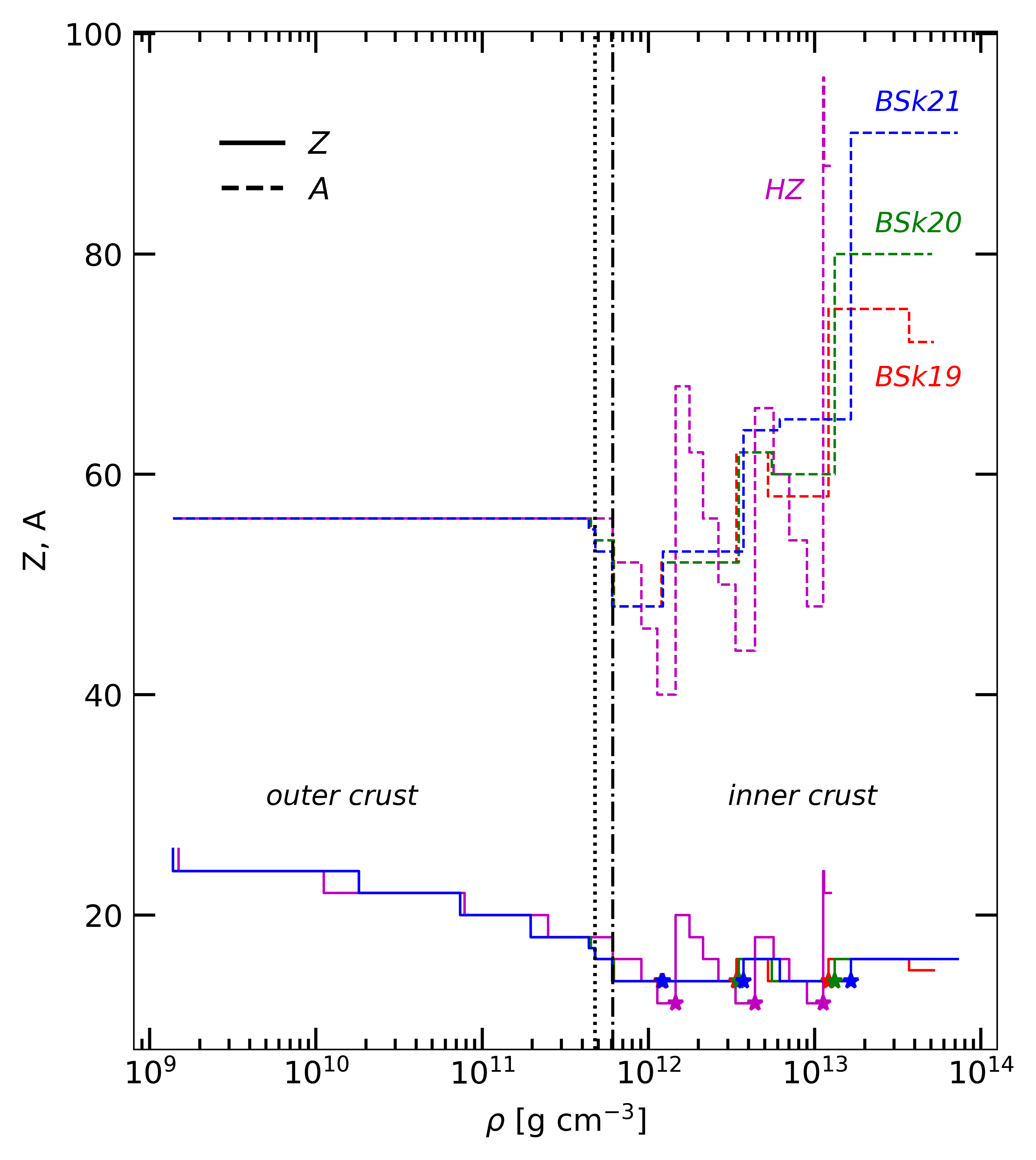}
	\centering
    \caption{Atomic number Z and mass number A of nuclear clusters in the crust of an accreting NS as a function of the density for the four EoSs indicated near the curves. The location of pycnonuclear reactions are marked by stars, and the vertical dotted and dash-dotted lines refer to the transition from the outer to inner crust (neutron drip) for the BSk19-21 models of \citet{Fantina, Fantina_2022} and \citet{1990A&A...227..431H,1990A&A...229..117H} (HZ) respectively.}
    \label{fig:Atomic Numbers}
\end{figure}

As a neutron star in an LMXB accretes matter, compression of the accreted material leads to the crust taking on a layered structure, with each layer defined by a particular density-dependant, non-equilibrium reaction. The variations of atomic number Z and mass number A with increasing density is shown in Fig. \ref{fig:Atomic Numbers}, using data taken from \citet{Fantina} and \citet{1990A&A...229..117H}. The different BSk19-21 and HZ models show identical compositions in the outer crust, albeit at different densities. In the inner crust however, the structure is notably different. The HZ model predicts a highly stratified crust with Z ranging from 10 - 18, whilst the BSk19-21 models predict a crustal structure with far fewer distinct capture layers. This is due largely to the inclusion of nuclear shell effects, which predict a freezing of the nuclear composition at $Z = 14$ \citep{Fantina}. Encouragingly, these findings coincide with the predictions of \citet{PhysRevC.69.052801} that $Z = 14$ is a \textit{magic proton number} in neutron star matter, making these nuclei more stable against both electron captures and pycnonuclear reactions in the inner crust.

Pycnonuclear reactions in the inner crust are marked as stars in Fig. \ref{fig:Atomic Numbers}, and (along with individual electron captures) release heat locally in the crust during periods of active accretion. These reactions are thought to source the X-rays observed from neutron stars in LMXBs during later periods of quiescence, as the star cools and heat is transported to the star's surface (see e.g. \citet{2017JApA...38...49W} for a review on the cooling of accreting neutron stars).  

In Fig. \ref{fig:Atomic Numbers}, we have assumed (as do both \citet{1990A&A...227..431H, 1990A&A...229..117H} and \citet{Fantina}) that the transition between successive capture layers is infinitely sharp. Strictly speaking however, the capture layers have a finite thickness. The thickness of the capture layers was calculated by UCB, who did so by integrating the capture rate equation (following \citet{1998ApJ...506..842B}) in order to track shifts in the position of the layers when a temperature perturbation was introduced. Here however, we adopt the approach of OJ, and assume instead that the transitions between capture layers are infinitely sharp, and smear the heat released at each transition over shells of constant (A, Z). Whilst less accurate than resolving the individual transition layers, this approximation should be sufficient for our needs since we are interested here in only the influence of a magnetic field on the star's thermal conductivity. 

In the thermal calculations that follow, we use four models: three low-mass (LM) neutron stars using each of the BSk19-21 EoS models respectively, and one high-mass neutron star using just the BSk21 EoS, which have the properties listed in Table \ref{tab:Model properties}. 

\begin{figure}
	\includegraphics[width=\columnwidth]{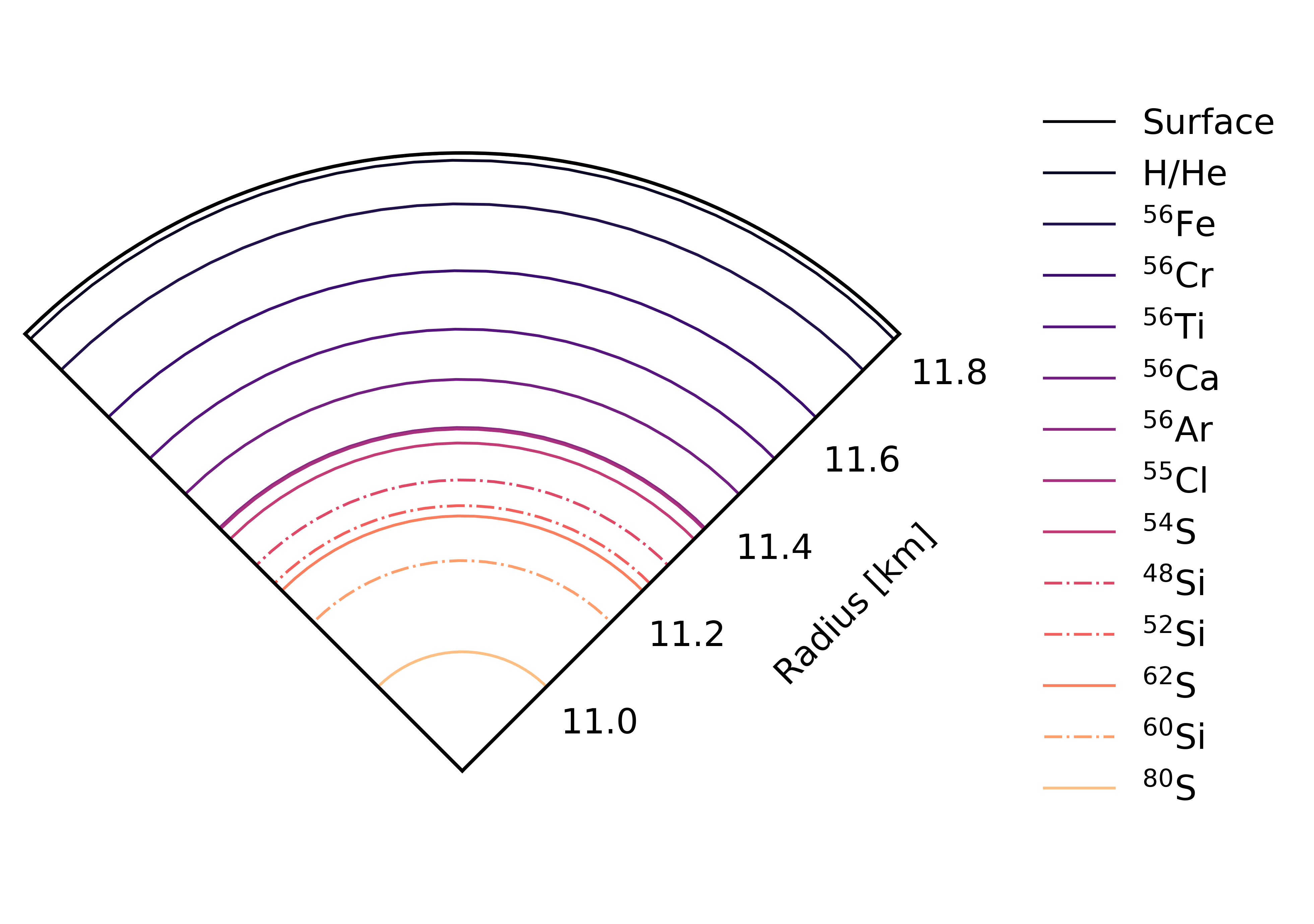}
	
    \caption{Hydrostatic structure of an accreting neutron star crust for the BSk20 (LM) model in Table \ref{tab:Model properties}. The coloured lines are the radial locations of the capture shells - of constant (A, Z) - obtained from solving the Tolman-Oppenheimer-Volkoff (TOV) equation. The dash-dotted lines indicate pycnonuclear reactions.}
    \label{fig:SchematicCrust}
\end{figure}

In order to fix the hydrostatic structure of our star, we pick a central density $\rho_c$ as indicated in Table \ref{tab:Model properties} and integrate the Tolman-Oppenheimer-Volkoff (TOV) equation from the center of the star outward to where the pressure drops to zero. We use the Python ODE solver \texttt{solve\_IVP} for the integration, with a fourth-order Runge-Kutta method with adaptive step-size and controlled accuracy. An example cross-sectional view of the location of the different capture layers in the accreted crust obtained from the TOV solution of the BSk20 (LM) model is shown in Fig. \ref{fig:SchematicCrust}. The 1.4$M_{\odot}$ star has a radius of 11.81 km, with the crust itself having a mass 0.018 $M_{\odot}$ and thickness ($\Delta R = R - R_{\text{core}}$) 1.00 km.

The maximum TOV masses of BSk19-21 are (to 3 s.f.) 1.86, 2.14 and 2.26 $M_{\odot}$ respectively \citep{2013A&A...560A..48P}. The maximum mass is effectively the same for both accreted and non-accreted NSs; cf.Fig. 3 of \citet{Fantina_2022}). Recent analysis of NICER XTI and XMM-Newton data on the millisecond pulsar PSR J0740+6620 indicates the NS has a gravitational mass $M^{68\%} = 2.08 \pm 0.07 M_{\odot}$ \citep{2021ApJ...918L..28M}, i.e.\ with a central value greater than the maximum mass of the BSk19 EoS, suggesting that BSk19 may be too soft to be viable. Note, however, that if the actual mass of PSR J0740+6620 lies $2\sigma$ (or more) below the central estimate (i.e.\ at or below approx 1.88 $M_{\odot}$; see  Table 7 of \citet{2021ApJ...918L..28M}, the BSk19 EoS remains viable.  For this reason, and given the scarcity of available accreted EoS models, we shall continue to include the accreted BSk19 EoS in our subsequent calculations.

The calculation of the star's hydrostatic structure is normally entirely dissociated from the thermal calculation (since the EoS is usually calculated at zero temperature, and therefore temperature independent).  The amount of heat released in the crust is proportional to the rate of accretion onto the NS, with higher accretion rates leading to hotter stars.  The resultant high temperatures, combined with the typically-low nuclear charge characteristic of the accreted crust (see Fig. \ref{fig:Atomic Numbers}) may cause it to melt \citep{2000, 2002}.   We monitor this possibility of melting  in each of our four NS models by tracking where the ratio of Coulomb energy to thermal energy,

\begin{equation} \label{eq:Coulomb to Thermal}
\Gamma_{\text{Coul}}  = \frac{Z^2e^2}{k_BT} \biggl( \frac{4\pi n_b}{3} \biggr)^{1/3} ,
\vspace*{3mm}
\end{equation}

\noindent
exceeds 175 and assume that the crust is solid wherever $\Gamma_{\text{Coul}} \geq \Gamma_{\rm m} = 175$ \citep{Haensel2007NeutronS1}. As was shown by UCB, it is the high density inner region of the NS crust that contributes most to the formation of the mass quadrupole. It is therefore the case that if the inner crust were to melt, it would render our (and indeed any) temperature asymmetry inducing mechanism, redundant. 

Note that $\Gamma_{\rm{m}} = 175$ is the canonical melting value for a one-component plasma (OCP) \citep{Haensel2007NeutronS1}. However, as we shall discuss in Section \ref{subsection:Thermal Conductivity}, the crustal layers of an accreting neutron star are expected to contain impurities, violating the OCP approximation. The Coulomb parameter at melting has recently been calculated for a multi-component plasma by \citet{Fantina_2020} for the outer crust of \textit{non-accreting} neutron stars, finding that $\Gamma_{\rm{m}}$ can be $\lesssim 20 \%$ less than the canonical value (cf. their Fig. 4), with a corresponding increase in the melting temperature. It is therefore worth bearing in mind that the accreted crust may possibly melt at greater temperatures than is indicated by the canonical $\Gamma_{\rm{m}} = 175$ result for a OCP. In this sense, our use of $\Gamma_{\rm{m}} = 175$ as a melting criterion is conservative.

\begin{table*}
	\centering
	\caption{Central density $\rho_c$, mass $M$, crustal mass $M_{\text{crust}}$, radius $R$, crust thickness $R_{\text{crust}}$ and total heat released per accreted nucleon $Q_{\text{tot}}$, for four different neutron star models: three low mass (LM) BSk19-21 models and a high mass (HM) BSk21 model.}
	\label{tab:Model properties}
    \begin{tabular}{l|cccccc}
        \hline
        Model & $\rho_c$ {[}10$^{15}$ g cm$^{-3}${]} & $M$ {[}$M_{\odot}${]} & $M_{\text{crust}}$ {[}$M_{\odot}${]} & $R$ {[}km{]} & $R_{\text{crust}}$ {[}km{]} &  $Q_{\text{tot}}$ {[}MeV{]} \\ \hline
        BSk19 (LM) & 1.321 & 1.40 & 0.014 & 10.80 & 0.83 & 1.535 \\
        BSk20 (LM) & 0.924 & 1.40 & 0.018 & 11.81 & 1.00 & 1.615 \\
        BSk21 (LM) & 0.732 & 1.40 & 0.018 & 12.64 & 1.11 & 1.651 \\
        BSk21 (HM) & 1.284 & 2.10 & 0.009 & 12.16 & 0.52 & 1.651 \\ \hline
    \end{tabular}
\end{table*}

\section{Thermal Structure of an accreting neutron star}
\label{section:Thermal Structure of an accreting neutron star}

Having specified the hydrostatic structure, we can now compute a steady-state thermal structure for our four NS models. In doing so, we shall employ a Newtonian formulation, rather than a general relativistic one (e.g. \citet{osti_4483768}). We make this choice since we aim to connect our temperature perturbations with the calculations of the neutron star ellipticity obtained by \citet{2002}, who worked in an entirely Newtonian framework. By employing the fully relativistic TOV calculation for the hydrostatic background, we were able to make use of realistic equations EoSs, which, in Newtonian theory, would have led to grossly unphysical density profiles.

We map our TOV hydrostatic solution to a Newtonian star by simply reinterpreting the TOV radial coordinate, pressure and energy density as their Newtonian equivalents (with the Newtonian mass density being the relativistic energy density divided by $c^2$). This choice respects the original equation of state (i.e. the relation $P = P(\rho)$ is preserved), but inevitably gives a star whose local acceleration due to gravity does not quite match the value that Newtonian gravity would require (i.e. does not quite satisfy Poisson's equation). This error is reasonably modest however, of the order $\sim 30\%$ in our calculations. We would seek to extend this calculation to a fully relativistic setting in subsequent work.

\noindent
From energy conservation, the heat flux $\rm{\boldsymbol{F}}$ is related to the net rate of heat energy generation per unit volume $Q$ as

\begin{equation} \label{eq:Luminoisty}
 \nabla \cdot \rm{\boldsymbol{F}} = Q = Q_{h} - Q_{\nu} \, ,
\vspace*{3mm}
\end{equation}
where $Q_{h}$ is the local energy deposited by nuclear reactions (and any shallow heating sources), and $Q_{\nu}$ is the local energy loss due to neutrino emission. The heat flux is related to the temperature $T$ via Fourier’s law:

\begin{equation} \label{eq:Fourier}
 \rm{\boldsymbol{F}} = - \kappa \, \nabla T \, ,
\vspace*{3mm}
\end{equation}

\noindent
where $\kappa$ is the thermal conductivity. Assuming the background NS is spherically symmetric, the luminosity $L$  is related to the heat flux via $L = 4 \pi r^2 \, F$, with $ F = |\boldsymbol{F}|$. From this, one can derive a system of two coupled ordinary differential equations (ODEs) for $L$ and $T$, with respect to the radial coordinate $r$ as

\begin{equation} \label{eq:dLdr}
 \frac{dL}{dr} = 4 \pi r^2 Q \, ,
\vspace*{3mm}
\end{equation}

\begin{equation} \label{eq:dTdr}
 \frac{dT}{dr} = -\frac{1}{\kappa} \frac{L}{4 \pi r^2} \, .
\vspace*{3mm}
\end{equation}

\noindent
Solving the above ODEs requires an accurate description of the heating term $Q_{h}$, neutrino cooling $Q_{\nu}$, and thermal conductivity $\kappa$, as well as a set of inner and outer boundary conditions at each end of the integration. These are discussed in turn in the following sections.

\subsection{Accretion Heating}
\label{subsection:Accretion_Heating}

\begin{figure}
	\includegraphics[width=0.85\columnwidth]{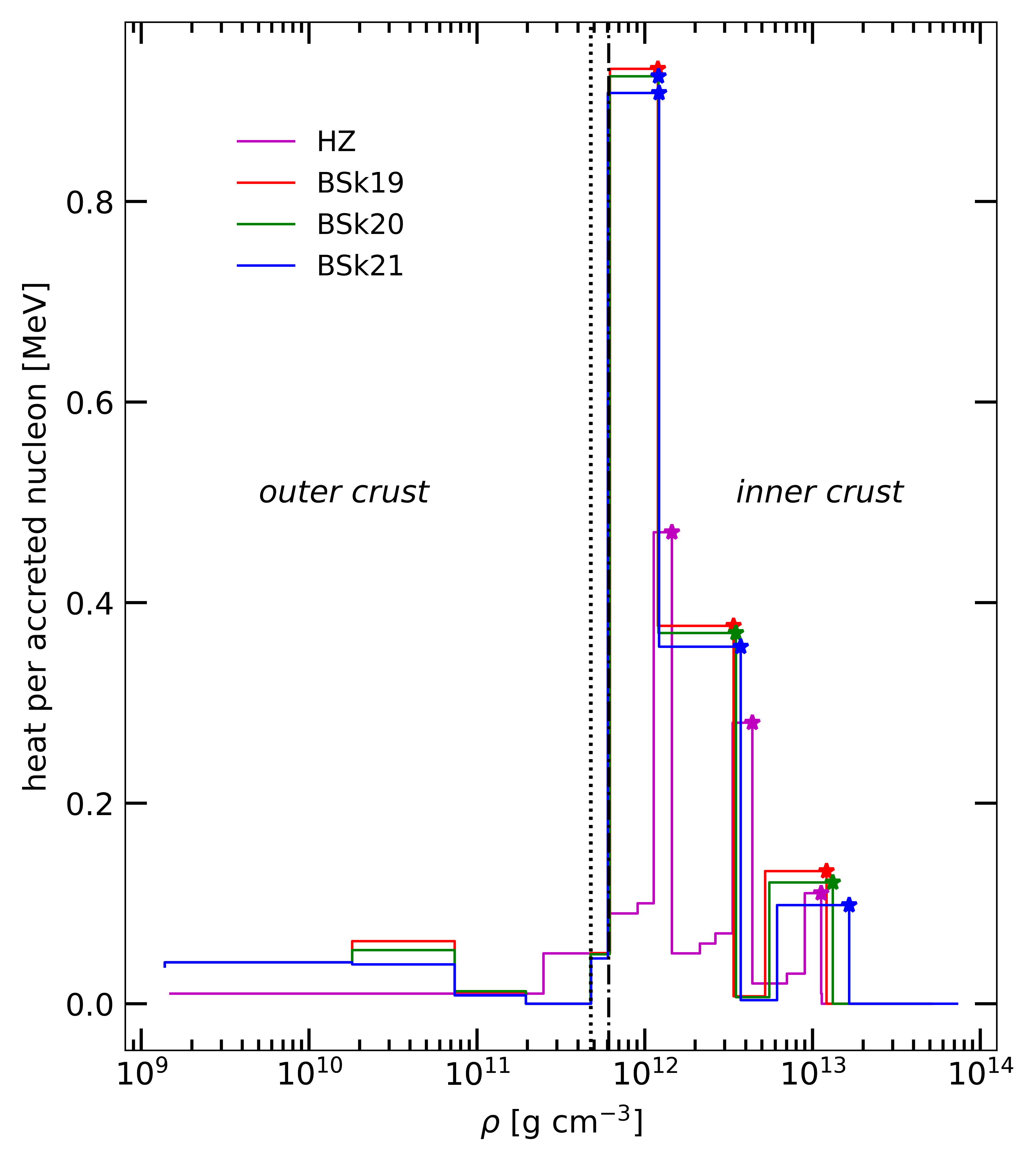}
    \caption{Heat deposited (per accreted nucleon) in the crust of an accreting NS from deep crustal heating reactions as a function of the density for the BSk19-21 (Tables A.3 - A.1 in \citet{Fantina}) and HZ (Tables 1 and 2 of \citet{1990A&A...227..431H}) EoSs. The step-like features represent our choice to smear the heat released at each transition over shells of constant (A, Z). Locations of pycnonuclear reactions are marked by stars, and the vertical dotted and dash-dotted lines refer to the outer to inner crust transition (neutron drip) for the BSk19-21 and HZ EoSs respectively.}
    \label{fig:Heat}
\end{figure}

The total amount of heat released per accreted nucleon in each of the BSk19-21 EoS models is shown in the final column of Table \ref{tab:Model properties}. How much of this heat is deposited in each of the capture layers of the different models with increasing density is shown in Fig. \ref{fig:Heat} (see also Tables A.3 - A.1 in \citet{Fantina}). 

Almost the entirety of the heat produced in the BSk19-21 models is released in the inner crust ($ \rho \sim 10^{12} - 10^{13}$ g cm$^{-3}$) via pycnonuclear reactions. For comparison, in Fig. \ref{fig:Heat} we also include the heat release per capture layer as predicted by \citet{1990A&A...227..431H}, labelled `HZ'. Despite the number of pycnonuclear reactions being the same across all models, almost twice the amount of heat is produced by these reactions in BSk19-21 ($\sim 1.4$ MeV) than the HZ model ($0.86$ MeV). \citet{Fantina} attribute this additional heating in the inner crust to the inclusion of the nuclear shell effects that the HZ model neglects. 

However, in addition to this `deep crustal heating' (DCH), analysis of light curves from transiently accreting LMXBs indicates the presence of some as-yet unknown heat sources in the outer layers ($\rho \leq 10^{10}$ g cm$^{-3}$) of the star (e.g. \citet{2020Chamel}). The \textit{average} amount of additional heating required by cooling simulations to fit the observational data is around 1-2 MeV per accreted nucleon, roughly the same amount of heat released from the DCH. Proposed explanations of this `shallow crustal heating' (SCH) vary, including uncertainties in both the accretion rate \citep{Ootes} and envelope constitution \citep{10.1093/mnras/sty825}, convection in the liquid ocean \citep{2015ComposDrive}, and differential rotation between the ocean and solid crust \citep{Rotation}.

We choose to follow the prescription of OJ and smear the heat deposited in each capture layer over shells of constant (A, Z), with the width of the shell being defined by the density region between two consecutive capture layers (see Fig. \ref{fig:Heat}). The amount of heat deposited in each capture layer (per unit volume per unit time) is a function of the accretion rate $\dot{M}$. By smearing the deposited heat over whole shells, the local energy deposited in each shell is then

\begin{equation} \label{eq:Heat}
     Q_{h} = Q_{\rm{nuc}} + Q_{\rm{S}} = \frac{\dot{M}\epsilon_{\rm{nuc}}}{\frac{4}{3}\pi(r^3_{i}-r^3_{i+1})} + Q_{\rm{S}} \,,
\vspace*{3mm}
\end{equation}

\noindent
where $\epsilon_{nuc}$ is the heat deposited from DCH per nucleon in a given capture shell (given in the seventh column of Tables A.3 - A.1 in \citet{Fantina}) and $r_{i}$ \& $r_{i+1}$ are the radii at the $i^{\rm{th}}$ capture layer as depicted in Fig. \ref{fig:SchematicCrust}. In Eq. (\ref{eq:Heat}), $Q_S$ is the shallow heating term which affects only the lowest-density regions of the crust ($\rho_S \leq 10^{10}$ g cm$^{-3}$). At densities $\rho < \rho_S$, we add an extra 0.5 MeV per nucleon in each of three compositional shells defined by the densities $\rho = 1.00 \times 10^7$ g cm$^{-3}$ (the approximate location of the base of the H/He layer), and $\rho = 1.38 \times 10^9$ g cm$^{-3}$,  $\rho = 1.81 \times 10^{10}$ g cm$^{-3}$,  $\rho = 7.37 \times 10^{10}$ g cm$^{-3}$ (corresponding to the lowest-density compositional shells, which are the same across each of the BSk19-21 EoS models) to give a total of 1.5 MeV per accreted nucleon of SCH. Accordingly, at densities whereby $\rho > \rho_S$, we set $Q_S = 0$, and the heat deposited into the crust is then supplied solely by the relevant DCH reactions.

\subsection{Neutrino Cooling}
\label{section:Neutrino Cooling}

We now describe the neutrino cooling mechanisms that we include in our model. A summary of these is given in Table \ref{NeutrinoCoreTable}. 

Electron-ion bremsstrahlung is responsible for most of the neutrino luminosity in the crust of accreting neutron stars \citep{2000}. The level of neutrino emission however is largely dictated by the state of the ions in the crust during the accretion. In the regions of the crust where the ions are crystallised ($\Gamma_{\rm{Coul}} \geq 175$ in Eq. \ref{eq:Coulomb to Thermal}) the bremsstrahlung rates are significantly suppressed due to the separation of electron energy bands \citep{PhysRevLett.72.1964, 1996AstL...22..491Y}. However, if in some regions of the crust the ions are liquefied ($\Gamma_{\rm{Coul}} < 175$), then collisions of relativistic degenerate electrons with atomic nuclei in a Coulomb liquid allow for much more efficient neutrino emission \citep{1996A&A...314..328H}.

If the crust is sufficiently hot ($T \gtrsim 10^9$ K), then plasmon decay into a neutrino-antineutrino pair (facilitated by interactions of the electrons with plasma microfields) can also become a significant source of neutrinos \citep{2001Yakovlev, 2007Kantor}. It is usually the case however, that regardless of the state of the crust, the neutrino luminosity is mostly dominated by the conditions inside the core of the neutron star. 

Neutrino emission in the core is determined by numerous different nuclear reactions that often scale with density, as well as (to a steep power) temperature  (see e.g. the review by \citet{2001Yakovlev}). These mechanisms can be subdivided into two categories: \textit{slow} and \textit{fast} processes. The fast processes are the most efficient neutrino emission mechanisms, but have density thresholds that restricts their availability to the innermost portion of the core of only the densest neutron stars. The most powerful of these mechanisms are the direct Urca processes. Direct Urca (DUrca) can only proceed in regions where the proton fraction is sufficiently large. If the Fermi momenta of the protons and electrons (and muons in npe$\mu$ matter) are small relative to the neutron Fermi momenta, then the process is forbidden via momentum conservation \citep{2004Yakovlev}. Electron DUrca only operates above a critical proton fraction $Y_{p, \, e} ^c \approx 0.11$, with muon DUrca opening up at a slightly higher still proton fraction $Y_{p, \, \mu} ^c \approx 0.14$ \citep{2008Aguilera}.

In the majority of neutron stars where DUrca is forbidden, the neutrino emissivity in the core is dominated by the slow processes, namely modified Urca (MUrca) and nucleon-nucleon (NN) bremsstrahlung. The MUrca process features an additional `spectator' nucleon that ensures that momentum (and energy) is always conserved. The spectator nucleon allows two separate sequences of reactions - a neutron and proton branch - depending on the spectator nucleon involved \citep{1995A&A...297..717Y}. Similarly, in the case of the bremsstrahlung interactions, there are three different (nn, np, and pp) processes in npe$\mu$ matter.

These reactions are significantly influenced by the presence of baryon superfluidity. If the temperature inside the star is less than that of some critical temperature $T_c$, an energy gap in the baryon energy spectrum renders the protons (and/or) neutrons `inactive' and suppresses the neutrino emissivity involving these particles by (approximately) exponential Boltzmann factors \citep{2004Yakovlev, 2008Aguilera}. Moreover, in addition to suppressing neutrino emission of the fast and slow processes, baryon superfludity also facilitates an extra neutrino emission mechanism unique to the formation and breaking of Cooper pairs (Cooper Pair Breaking and Formation emissivity - CPBF). Unlike the suppression behavior of the other neutrino interactions, the strength of the Cooper pairing emissivity sharply increases as the temperature falls below that of $T_c$ - reaching some maximum at $T \approx T_c / 5$ before decreasing again \citep{2001Yakovlev}.

In the subsequent calculations, we include all relevant neutrino emission mechanisms in both the crust and core listed in Table \ref{NeutrinoCoreTable}. The general form of each emission mechanism including their respective temperature and density dependence are given. In the presence of baryon superfluidity, the suppression of a given neutrino processes involving a particular baryon species is accounted for by the exponential reduction functions $\mathcal{R}$. The exact form of each emission mechanism and any associated reduction factors (which are usually complicated functions of the critical temperature $T_c$) can be found in the references given in Table \ref{NeutrinoCoreTable}. 

\begin{table}
\caption{Neutrino emission processes in the crust and core of a weakly-magnetised accreting neutron star used in this work. A precise, detailed description of each of the neutrino emissivities $Q$ and any associated microphysical quantities can be found in the given references.}
\resizebox{\columnwidth}{!}{
\renewcommand{\arraystretch}{0.15}
\begin{tabular}{l|l}
\hline
\hline

Process &$Q\,[\rm{erg} \, \rm{cm}^{-3} \, \rm{s}^{-1}]$ 
                    \\ 
                    
\hline
\multicolumn{1}{l}{\textbf{Interactions in the crust}} &  \\ 
\noalign{\smallskip}
\textit{Plasmon decay}$\, {}^{[\textbf{1}]}$ &\\
\rule{-1pt}{3ex} 
$\tilde e\rightarrow \tilde e\nu\bar{\nu}$ &
$Q_{pl} \approx 1.03\times 10^{23} \, I_{\rm pl} \mathlarger{\sum}\limits_{\nu} C^2_V$
 \\
\hline
\textit{$e-A$ Bremsstrahlung  (crystallised)}$\,{}^{[\textbf{2}]}$ &\\
$e (A,Z)  \rightarrow e (A,Z) \nu \bar{\nu}$&
$ Q_{br} \approx 3.229 \times 10^{11}\, \rho_{12}\, (Z^2 / A) \,(1 - X_n)\,L\,T^6_8$
 \\

\rule{-2pt}{4ex} 
\textit{$e$-$A$ Bremsstrahlung (liquefied)}$\,{}^{[\textbf{3}]}$ &\\
$e (A,Z)  \rightarrow e (A,Z) \nu \bar{\nu}$&
$ Q_{br} \approx 3.229 \times 10^{17}\, \rho_{12}\, (Z^2 / A) \,(1 - X_n)\,L\,T^6_8$
 \\
\hline\noalign{\smallskip}
\multicolumn{1}{l}{\textbf{Interactions in the core}}\\
\noalign{\smallskip}

\textit{Direct Urca}$\,{}^{[\textbf{4, 1}]}$ &\\
$n\rightarrow pe\bar{\nu}_e, pe \rightarrow n\nu_e$ & 
$Q^{(D)}_e \approx 4\times 10^{27}\, \biggl(\frac{n_e}{n_0}\biggr)^{1/3} \, \mathcal{R}^{DU} \, T^6_9\,\Theta_{npe}$
\\ 
$n\rightarrow p\mu\bar{\nu}_{\mu},p \mu \rightarrow n\nu_{\mu} $ 
& $Q^{(D)}_{\mu} \approx Q^{(D)}_e \, \Theta_{np\mu} $
\\

\hline
\textit{Modified Urca ($n$-branch)}$\,{}^{[\textbf{5, 1}]}$ &\\ 
\rule{-2pt}{3ex}  

$nn\rightarrow pne\bar{\nu}_e$ &\\
$pne\rightarrow nn\nu_e$ 
 & 
$Q^{Mn}_e \approx 8.1 \times 10^{21} \, \biggl(\frac{n_p}{n_0}\biggr)^{1/3} \,T^8_9 \, \mathcal{R}^{MU}_n $   \\ 
\rule{-2pt}{3ex} 
$nn\rightarrow pn\mu\bar{\nu}_{\mu}$ &\\
$pn\mu\rightarrow nn\nu_{\mu}$ 
 & 
$Q^{Mn}_{\mu} \approx Q^{Mn}_e \, \biggl(\frac{n_{\mu}}{n_e}\biggr)^{1/3} \,  \mathcal{R}^{MU}_p / \mathcal{R}^{MU}_n$   \\ 

\rule{-2pt}{4ex} 

\textit{Modified Urca ($p$-branch)}$\,{}^{[\textbf{5, 1}]}$ &  \\ 
\rule{-1pt}{3ex} 
$np\rightarrow ppe\bar{\nu}_e$ &\\
$ppe\rightarrow np\nu_e$  
 & 
$Q^{Mp}_e \approx Q^{Mn}_e \, \biggl [ \frac{(p_{Fe} + 3p_{Fp} - p_{Fn})^2}{8p_{Fe}p_{Fp}} \biggr] \, \mathcal{R}^{MU}_p / \mathcal{R}^{MU}_n  \, \Theta_{Mpe} $  \\ 
\rule{-1pt}{3ex} 
$np\rightarrow pp\mu\bar{\nu}_{\mu}$ &\\
$pp\mu\rightarrow np\nu_{\mu}$  
 & 
$Q^{Mp}_{\mu} \approx Q^{Mn}_{\mu} \, \biggl [ \frac{(p_{F\mu} + 3p_{Fp} - p_{Fn})^2}{8p_{F\mu}p_{Fp}} \biggr] \, \mathcal{R}^{MU}_p / \mathcal{R}^{MU}_n  \, \Theta_{Mp\mu}$   \\ 
\hline
\textit{NN-Bremsstrahlung}$\,{}^{[\textbf{5}]}$  & \\ 
$nn\rightarrow nn \nu \bar{\nu}$ & 
$Q^{(nn)} \approx 7.5\times 10^{19} \, \biggl(\frac{n_n}{n_0}\biggr)^{1/3} \, T^8_9 \, \mathcal{R}^{nn}$ \,
 \\ 

$np\rightarrow np \nu \bar{\nu}$ &
$Q^{(np)} \approx 1.5\times 10^{20} \, \biggl(\frac{n_p}{n_0}\biggr)^{1/3} \, T^8_9 \, \mathcal{R}^{np}$ \,    \\ 

$pp\rightarrow pp \nu \bar{\nu} $ &
$Q^{(pp)} \approx 7.5\times 10^{19} \, \biggl(\frac{n_p}{n_0}\biggr)^{1/3} \, T^8_9 \, \mathcal{R}^{pp}$     \\ 

\hline\noalign{\smallskip}
\multicolumn{1}{l}{\textbf{Interactions in the crust \& core}}\\ 
\noalign{\smallskip}

\textit{CPBF}$\,{}^{[\textbf{6}]}$ &\\
$\tilde B + \tilde B \rightarrow \nu \bar{\nu}$
& $ Q^{(CP)} \approx 1.17\times 10^{21} \, \biggl(\frac{n_N}{n_0}\biggr)^{1/3} \, a F({\nu})\, T^7_9 $ \\
\hline 
\hline
\end{tabular} 
}
Ref. (1)~\citet{2001Yakovlev}; (2)~\citet{1996AstL...22..491Y}; (3)~\citet{1996A&A...314..328H}; 
(4)~\citet{PhysRevLett.66.2701}; (5)~\citet{1995A&A...297..717Y}; (6)~\citet{yakovlev1998neutrino, 1999A&A...343..650Y}
 
 \label{NeutrinoCoreTable} 

\end{table} 

\subsection{The Superfluid Transition Temperatures}
\label{subsection:The Superfluid Transition Temperatures}

LMXBs are almost certainly superfluid and superconducting over some range of densities. Pairing within nuclear matter is expected to have an important role in dictating the NS thermal structure, potentially strongly modifying both the star's neutrino emissivity as well as the thermal conductivity. Neutrons in the inner crust and protons in the core are thought to pair in the singlet $(^1S_0)$ state \citep{1966ApJ...145..834W}. Neutrons in the core, on the other hand, are expected to pair in a triplet $(^3P_2)$ state \citep{Hoffberg:1970vqj, 10.1143/PTP.44.905}. There is, however, uncertainty in both the values of the critical temperatures of the protons and neutrons, and exactly what density ranges the superconducting/superfluid regions extend in the NS interior (see e.g. Fig. 5 of \citet{2015SSRv..191..239P}). 

Additionally, many of the calculations for the critical temperature lack convenient fitting formulae in terms of the density. Therefore, in order to be able to introduce effects of superfluidity into our own work, we follow the prescription of \citet{2000}. It can be shown that the critical temperatures are (approximately) quadratic functions of the Fermi wavevector $k_{n,p} = (3\pi^2 n_{n,\, p})^{1/3}$, where $n_{n, \, p}$ is the number density of neutrons and protons respectively. Eq. (12) of \citet{2000} gives the functional form of $T_c$ for each of the singlet ($^1S_0$) proton and neutron states, as well as the triplet ($^3P_2$) neutron state as

\begin{equation}\label{eq:critical temps}
    T_c(k_{n, \, p}) = T_{c0} \biggl[ 1 - \frac{(k_{n, \, p} - k_0)^2}{(\Delta_{k_{n, \, p}} / 2)} \biggr],
    \vspace*{3mm}
\end{equation}

\noindent
where $T_{c \, 0}$, $k_0$, and $\Delta_{k_{n, \, p}}$ are parameters chosen by \citet{2000} to reproduce the critical transition temperatures calculated by \citet{1985NuPhA.437..487A, 1985NuPhA.442..163A} for the $^1$S$_0$ singlet states and $^3$P$_2$ triplet state respectively. The parameters $T_{c \, 0}$, $k_0$, and $\Delta_{k_{n, \, p}}$ are listed in Table \ref{tab:TcParams}, and are valid for the regime whereby $k_{n, \, p} < \Delta_{k_{n, \, p}}$, with $T_c$ vanishing outside this range. 

\begin{figure*}
\centering
	\includegraphics[width=\textwidth]{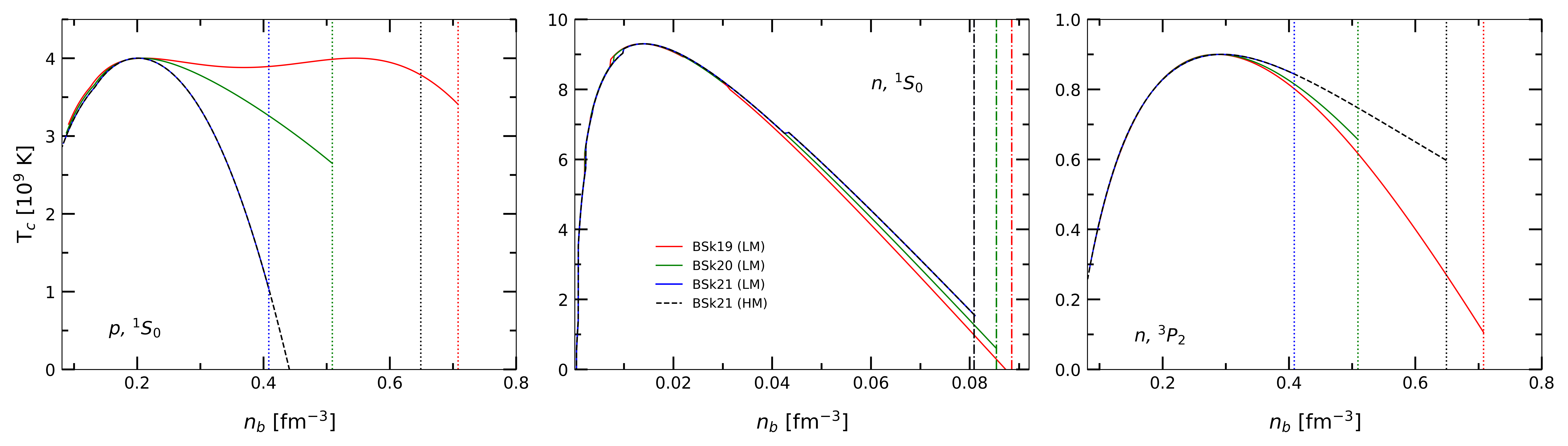}
    \caption{Superfluid transition temperatures $T_\text{c}$ as a function of baryon density for the four NS models in Table \ref{tab:Model properties}. The left panel shows the proton $^1$S$_0$ pairing in the core, the middle panel the neutron $^1$S$_0$ pairing in the crust, and the right panel shows the neutron $^3$P$_2$ pairing in the core. Dotted lines in the left and right panels indicate the central number densities of each model. The dash-dotted lines in the middle panel indicate the crust-core transitions of each model.}
    \label{fig:p and n transition Temps}
\end{figure*}

Fig. \ref{fig:p and n transition Temps} shows $T_\text{c}$ for the proton $^1$S$_0$ and neutron $^3$P$_2$ states in the core, as well as the neutron $^1$S$_0$ state in the inner crust of each of the NS models in Table \ref{tab:Model properties}. The neutron critical temperatures are almost identical in the inner crust and outer core, but begin to diverge towards their respective central densities (denoted by dotted lines). The proton critical temperatures on the other hand are more disparate, and we find that $T_\text{c}$ actually vanishes at higher n$_\text{b}$ for the BSk21 EoS. This cutoff is because the BSk21 model has a higher proton fraction than either BSk19 or BSk20 at a given density (see Fig. \ref{fig:Core_Fractions}) and can therefore exceed the condition whereby $k_{n, \, p}$ must be less than $\Delta_{k_{n, \, p}}$. Clearly, each of our 4 NS models are an admixture of both normal/superfluid neutrons and normal/superconducting protons in differing proportions. Consequently, the different microphysical processes will affect the thermal structure to a varying degree, dependant upon the ratio $T/T_\text{c}$.

The dependence of $T_c$ on the Fermi wavevector $k_{n, \, p}$ requires knowledge of the nucleon distributions in each of our NS models. The accreting BSk19-21 EoS models predict a $npe\mu$ matter composition in the core of the neutron star, following the composition and EoS of non-accreted NSs calculated by \citet{PhysRevC.82.035804, 2011PhRvC..83f5810P, 2012PhRvC..85f5803P} - for which the cores in both the accreting and non-accreting case are assumed to be the same. The number fractions of these particles (relative to the total number of nucleons) are $Y_n$, $Y_p$, $Y_e$ and $Y_{\mu}$ respectively, where $Y_n = 1 - Y_p$, with $Y_p = Y_e + Y_{\mu}$ (since the core is assumed to be electrically neutral). Analytical representations of these EoS models in the core were obtained by \citet{2013A&A...560A..48P}, who provide convenient fitting formulae for calculating the lepton fractions (cf. their Eq. (9)) $Y_e$ and $Y_{\mu}$ as

\begin{table}
	\centering
	\caption{Parameters to calculate proton superconducting and neutron superfluid transition temperatures (Table 2 of \citet{2000}).}
	\label{tab:TcParams}
    \begin{tabular}{l|cccc}
        \hline
    \textbf{State}  & $T_c$ {[}MeV{]} & $k_0$ {[}fm$^{-3}${]} & $\Delta_k$ {[}fm$^{-3}${]} \\ \hline
    neutron $^1S_0$ & 0.802           & 0.7                   & 1.2                        \\
    proton $^1S_0$  & 0.345           & 0.7                   & 1.0                        \\
    neutron $^3P_2$ & 0.0076          & 2.0                   & 1.6                        \\ \hline
    \end{tabular}
\end{table}

\begin{equation} \label{eq:Particle Fractions}
    Y_{e, \, \mu} = \frac{q_1^{(e, \, \mu)}+q_2^{(e, \, \mu)}n_b + q_3^{(e, \, \mu)}n_b^4}{1 + q_4^{(e, \, \mu)}n_b^{3/2}+q_5^{(e, \, \mu)}n_b^4} \, \text{exp}(-q_6^{(e, \, \mu)} n_b^5),
    \vspace*{3mm}
\end{equation}

\noindent
where the parameters $q^{(e,\, \mu)}_i$ are given in Table \ref{tab:ParticleFractions} (reproduced from Table 6 of \citet{2013A&A...560A..48P}). As pointed out by the authors, it is possible for Eq. (\ref{eq:Particle Fractions}) to return a negative value, in which case it should be replaced by zero. Such a scenario is possible for the muons, in regions of the core where they are forbidden (i.e where the muon chemical potential exceeds than of the electron chemical potential; $\mu_{\mu} > \mu_{e}$).  

\begin{figure*}
	\includegraphics[width=0.9\textwidth]{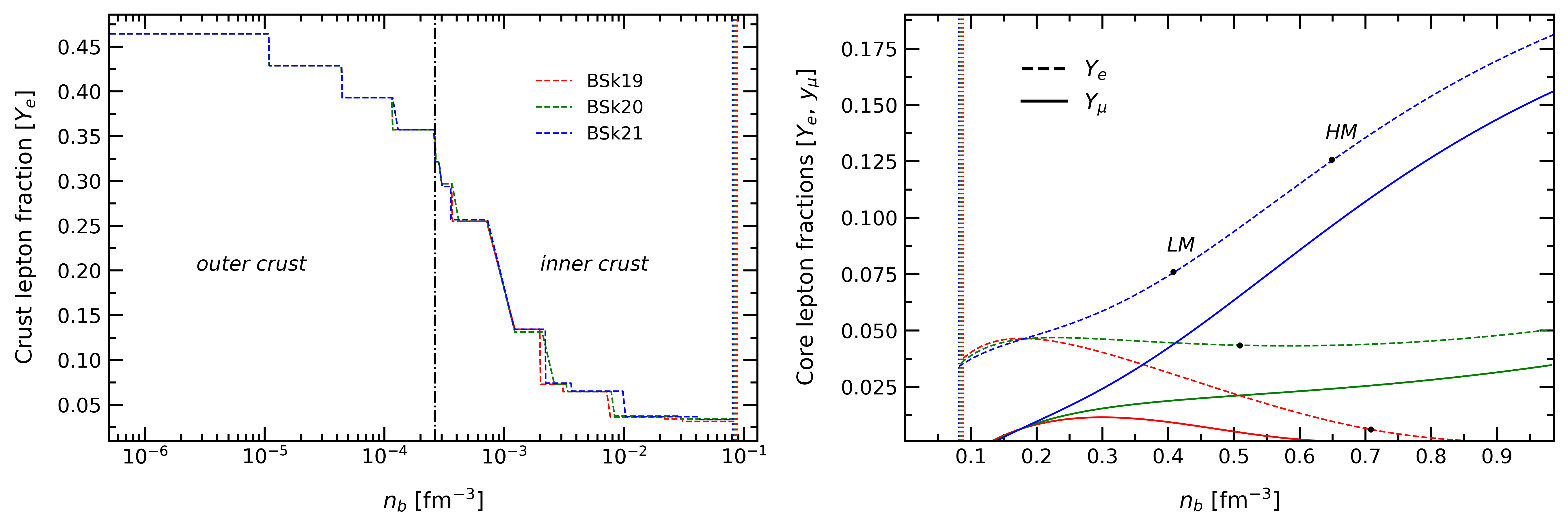}
	
    \caption{Left: Number fraction of electrons $Y_e$ in the crust of an accreting neutron star (relative to the number of nucleons) as a function of the baryon number density $n_b$ for the three \textit{accreting} BSk19-21 EoSs. The vertical dash-dotted line indicates the average location of the neutron drip transition across the three EoSs. Right: Number fractions of electrons $Y_e$ (solid lines) and muons $Y_{\mu}$ (dashed lines) in the core of an accreting neutron star as functions of the baryon number density $n_b$. Dotted vertical lines indicate the location of the crust-core transition for each EoS respectively. The black dots indicate the location of the central density used in each model in Table \ref{tab:Model properties}. For the BSk21 EoS model, the low-mass (LM) and high-mass (HM) models are labelled separately. This data was obtained using Eq. (9) and Table 6 from \citet{2013A&A...560A..48P} (cf. their Fig. 4), reproduced here in Eq. (\ref{eq:Particle Fractions}) and Table \ref{tab:ParticleFractions}.}
    \label{fig:Crust_Fractions}
\end{figure*}

\begin{figure*}
	\includegraphics[width=0.9\textwidth]{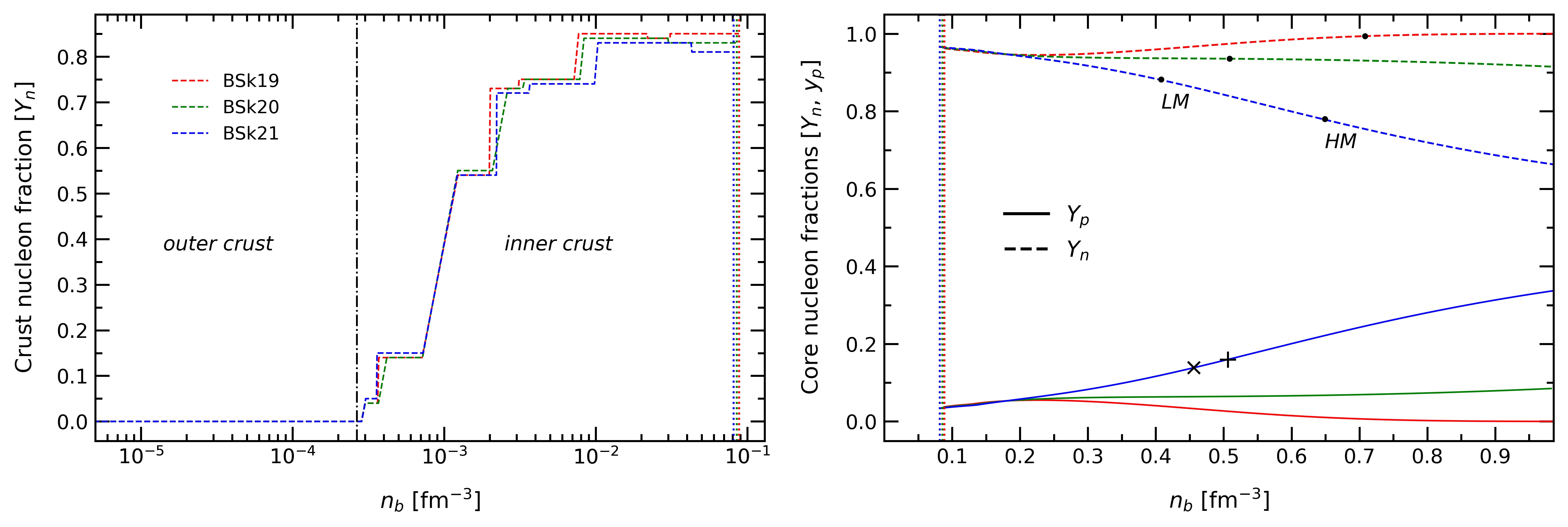}
	
    \caption{Left: Number fraction of free neutrons $Y_n$ in the crust of an accreting neutron star as a function of the baryon number density $n_b$ for the three \textit{accreting} BSk19-21 EoSs. Right: Number fractions of neutrons $Y_n$ (solid lines) and protons $Y_{p}$ (dashed lines) in the core of an accreting neutron star as functions of the baryon number density $n_b$. The cross and plus refer to the minimum threshold densities required for electron direct Urca and muon direct Urca respectively for the BSk21 EoS.}
    \label{fig:Core_Fractions}
\end{figure*}

Figs. \ref{fig:Crust_Fractions} and \ref{fig:Core_Fractions} give the particle fractions per baryon $Y_{n, \, p, \, e, \, \mu}$ in both the crust and core respectively, as a function of the baryon density. Momentum conservation requires that the triangle inequality $|p_{\rm{Fp}} - p_{\rm{Fe}}| \leq p_{\rm{Fn}} \leq p_{\rm{Fp}} + p_{\rm{Fe}}$ be satisfied for DUrca to be permitted. We find that the triangle inequality is never satisfied in the accreting BSk19 or BSk20 models, and calculate the electron $n_{\rm{DUrca} \, , \rm{e}}$ and muon $n_{\rm{DUrca}, \, \mu}$ DUrca thresholds to be 0.46 fm$^{-3}$ and 0.51 fm$^{-3}$ respectively for the accreting BSk21 model (indicated by the $\times$ and $+$ in the right panel of Fig. \ref{fig:Core_Fractions}). This equates to electron and muon DUrca becoming a contributory (or even dominant) cooling mechanism in stars assuming the BSk21 EoS with masses greater than 1.60 and 1.78 $M_{\odot}$ respectively. 

\begin{table}
	\centering
	\caption{Values of $q^{(e,\mu)}_i$ for particle fractions (Eq. (\ref{eq:Particle Fractions})) of the BSk19-21 EoSs (Table 6 of \citet{2013A&A...560A..48P}). }
	\label{tab:ParticleFractions}
    \begin{tabular}{l|cccc}
        \hline
        \multicolumn{1}{c|}{i} & BSk19    & BSk20         & BSk21    \\ \hline
                               &          & $q^{(\text{e})}_i$   &          \\
        1                      & - 0.0157 & - 0.0078      & 0.00575  \\
        2                      & 0.9063   & 0.075         & 0.4983   \\
        3                      & 0.0      & 0.508         & 9.673    \\
        4                      & 26.97    & 22.888        & 16.31    \\
        5                      & 106.5    & 0.449         & 38.383   \\
        6                      & 5.82     & 0.00323       & 0.0        \\ \hline
                               &          & $q^{(\mu)}_i$ &          \\
        1                      & - 0.0315 & - 0.0364      & - 0.0365 \\
        2                      & 0.25     & 0.2748        & 0.247    \\
        3                      & 0.0      & 0.2603        & 11.49    \\
        4                      & 12.42    & 12.99         & 24.55    \\
        5                      & 72.4     & 0.0767        & 48.544   \\
        6                      & 19.5     & 0.00413       & 0.0      \\ \hline    
    \end{tabular}
\end{table}

\subsection{Thermal Conductivity}
\label{subsection:Thermal Conductivity}

In this section we describe the heat conduction mechanisms that we include in our model. A summary of these is given in Table \ref{ThermCondTable}. 

The primary carriers of heat inside neutron stars crusts are relativistic electrons. In the core, the heat can be transported by electrons, muons (if available), as well as neutrons if they are normal. If the neutrons are superfluid, then their contribution to the conductivity is suppressed due to pairing effects \citep{2001Baiko}. Under the relaxation-time approximation, the thermal conductivity is \citep{ziman_1972},

\begin{equation} \label{eq:ThermalConductivity}
    \kappa_X = \frac{\pi^2 k_B \, n_X}{3m^*_X} \, T \tau_X
    \vspace*{2mm}
\end{equation}

\noindent
where n$_X$ is the number density of the heat carrier (denoted by the subscript X, i.e. X = e, $\mu$, n), m$^\ast _X$ is its effective mass, and $\tau_X$ is the collision relaxation time. The total thermal conductivity is a linear sum of the individual contributions from each carrier, i.e.

\begin{equation} \label{eq:ThermalConductivitySum}
    \kappa = \kappa_e + \kappa_{\mu} + \kappa_n \, .
    \vspace*{2mm}
\end{equation}

\noindent
In the crust, the relaxation time is simply the inverse of the sum of the frequencies of the different scattering mechanisms. In regions where the crust is solid ($\Gamma_{\rm{Coul}} \geq 175$), 

\begin{equation} \label{eq:SolidRelaxationTime}
    \tau = \frac{1}{\nu} = \frac{1}{\nu_{ep} + \nu_{eQ}},
    \vspace*{3mm}
\end{equation}

\noindent
and in regions where the crust is liquid ($\Gamma_{\rm{Coul}} < 175$),

\begin{equation} \label{eq:LiquidRelaxationTime}
    \tau = \frac{1}{\nu} = \frac{1}{\nu_{eQ} } \, ,
    \vspace*{2mm}
\end{equation}

\noindent
where $\nu_{ep}$ and $\nu_{eQ}$ are the scattering frequencies from electron-phonon and electron-impurity collisions respectively (see the appendix of \citet{2009Bildsten}). Strictly speaking, there is also a contribution from electron-electron ($\nu_{ee}$) collisions. However, strong degeneracy of the relativistic electrons in the NS crust restricts the available phase space, and therefore suppresses this form of scattering over most of the crust \citep{2000}. 

In our description of the accreted crust in Section \ref{section:Structure of an Accreted Crust}, it was assumed that each layer of the crust is a shell of constant (A, Z), as depicted in Fig. \ref{fig:Atomic Numbers}. This is because both the BSk19-21 and HZ models assume the ashes of X-ray bursts at the bottom of the envelope consist of pure $^{56}$Fe only, and also go on to make the one-component plasma approximation.  In reality however, nuclei far beyond the iron group with masses A $\sim$ 60 – 100 are expected to be formed during hydrogen/helium burning as a result of rapid proton captures via the rp-process \citep{1999Schatz}. It is therefore highly likely that the layers of the crust will be to some degree an admixture of different nuclei as the heavier elements sink into the crust \citep{2020Chamel}. The exact distribution of these nuclei in the different crustal layers is largely unknown. Instead, the likely presence of other nuclear species is usually incorporated into simulations via the so called `impurity factor' Q$_{\text{imp}}$, given as

\begin{equation}\label{eq:Impurity}
    Q_{\text{imp}} \equiv \frac{1}{n_{\text{ion}}} \displaystyle\sum_{i} n_i(Z_i - \langle Z \rangle)^2 \, ,
    \vspace*{3mm}
\end{equation}

\noindent
where the sum $i$ is over all the different species of ions, with atomic number $Z_i$ and mean atomic number $\langle Z \rangle$. Modelling of transiently accreting LMXBs suggests typical values for Q$_{\text{imp}}$ are of the order $\sim$ 1 - 10, but can reach as high as Q$_{\text{imp}}$ $\sim$ 100 in particular circumstances (\citet{2019Ootes} and references therein). 

In the core, we include contributions to the thermal conductivity from the electrons, muons and neutrons. The conduction of heat via the electrons and muons is limited by Coulomb collisions between themselves and other charged particles. The relationship between relaxation times and scattering frequencies is more complex in the core than in the crust.  For the leptons, the total effective electron and muon scattering frequencies are \citep{1995NuPhA.582..697G},

\begin{align} \label{eq:CoreRelaxationTime}
    \centering
        \tau_e = \frac{\nu_{\mu} - \nu'_{e\mu}}{\nu_{e}\nu_{\mu} - \nu'_{e\mu}\nu'_{\mu e}}, \qquad \tau_{\mu} = \frac{\nu_{e} - \nu'_{\mu e}}{\nu_{e}\nu_{\mu} - \nu'_{e\mu}\nu'_{\mu e}} \, ,
\end{align}

\vspace{3mm}

\noindent
where the collision frequencies $\nu_{e}$ and $\nu_{\mu}$ are the sum of the partial collision frequencies $\nu_{ei}$ and $\nu_{\mu i}$,

\begin{equation} \label{eq:PartialFreqs}
\begin{gathered} 
    \nu_{\text{e}} = \sum_{i} \nu_{ei} = \nu_{\text{ep}} + \nu_{\text{ee}} + \nu_{\text{e$\mu$}} \, , \\
    \nu_{\mu} = \sum_{i} \nu_{\mu i} = \nu_{\text{$\mu$p}} + \nu_\text{$\mu \mu$} + \nu_\text{$\mu$e} \, .
\end{gathered}
\vspace*{3mm}
\end{equation}

\noindent
The additional terms $\nu'_{e\mu}$ and $\nu'_{\mu e}$ in Eq. (\ref{eq:CoreRelaxationTime}) represent two additional effective `cross-collision' frequencies which couples the heat transport between electrons and the muons. 

In contrast to the leptons, the heat conduction via neutrons is limited by their collisions with other nucleons due to the strong interaction. The heat transport via neutrons is therefore effectively independent from the Coulomb-scattering leptons, and was studied in detail by \citet{2001Baiko}. For the neutrons, the relationship between the relaxation time and the  effective nucleon-nucleon scattering frequencies is

\begin{equation}\label{eq:NeutronTau}
    \tau_n = \frac{1}{\nu_{nn}+ \nu_{np}},
    \vspace*{3mm}
\end{equation}

\noindent
where $\nu_{nn}$ and $\nu_{np}$ correspond to the frequency of neutron-neutron and neutron-proton interactions respectively. 

More recently, \citet{Shternin:2013pya} reconsidered nucleon-nucleon interactions using the Brueckner-Hartree-Fock (BHF) method, showing that three-body forces suppress the thermal conductivity from these interactions. However, \citet{2015SSRv..191..239P} point out that it is sufficient to simply multiply the conductivities obtained by \citet{2001Baiko} by a factor of 0.6 in order to reproduce the results obtained by \citet{Shternin:2013pya}, within an accuracy of a few percent. For simplicity, we therefore adopt this procedure when implementing the equations of \citet{2001Baiko} in our subsequent calculations.

The thermal conductivity in the core is a combination of all the different scattering interactions, that scale with temperature as well as the respective number densities of the particles involved. The general form of each of the scattering mechanisms in both the crust and core are listed in Table \ref{ThermCondTable}. Much like the neutrino emissivity, the presence of baryon superfluidity will influence the heat conduction within the core of the NS as well. For the leptons, the effect is two-fold: reducing the lepton-proton collision frequencies, as well as influencing the screening momentum of the interacting particles \citep{1995NuPhA.582..697G}. 

The nucleon-nucleon interactions are also impacted by baryon superfluidity, either enhancing or suppressing the neutron thermal conductivity depending on the relative strength of the superfluidity of the proton and neutrons respectively in the core (cf. Fig. 4 and the relevant discussion in \citet{2001Baiko}). In general, the heat transport is very sensitive to the superfluid energy gap, being suppressed significantly if the energy gap is large enough. The exact form of the superfluid reduction factors for the lepton-proton and nucleon-nucleon interactions can be found in \citet{1995NuPhA.582..697G} and \citet{2001Baiko} respectively (and references therein).

\begin{table}

\caption{Scattering frequencies for processes that contribution to the thermal conductivity in the crust and core of a weakly-magnetised accreting neutron star used in this work. A precise, detailed description of each can be found in the given references.}
\resizebox{\columnwidth}{!}{%
\renewcommand{\arraystretch}{0.05}
\begin{tabular}{l|l}
\hline
\hline
Conduction Mechanism &$\nu$ [s$^{-1}]$
                    \\ 
\hline
\multicolumn{1}{l}{\textbf{Mechanisms in the crust}$\, {}^{[\textbf{1}]}$}   \\ 
\noalign{\smallskip}
electron-phonon scattering 
 &
$\nu_{ep} \approx 1.25 \times 10^{18} \, T_8$   \\ 
electron-impurity scattering  
 & 
$\nu_{eQ} \approx 1.77 \times 10^{18} \, \frac{Q_{\text{imp}}}{\text{Z}} \, \biggl( \frac{\rho_{12}}{\mu_e} \biggr)^{1/3}$    \\ 
\hline
\multicolumn{1}{l}{\textbf{Mechanisms in the core}}\\
\noalign{\smallskip}
\textit{Lepton conduction}$\, {}^{[\textbf{2}]}$  \\ 
electron-proton scattering &
$\nu_{ep} \approx 1.15 \times 10^{12} \, \biggl( \frac{p_{Fe}}{q_0} \biggr)^3 \, \biggl( \frac{n_0}{n_e} \biggr) \, T^2_8 \, \mathcal{R}_p $  \\
muon-proton scattering &
$\nu_{\mu p} = \nu_{ep} \, \biggl( \frac{n_e}{n_{\mu}} \biggr)^2$  \\
\hline
electron-muon scattering & 
$\nu_{e\mu} \approx 1.43 \times 10^{11} \, \biggl(\frac{p_{F\mu}}{q_0}\biggr)^3 \, \biggl(\frac{n_0}{n_e}\biggr)^{1/3} \, \biggl[1+\frac{1}{2}\biggl(\frac{n_{\mu}}{n_e}\biggr)^{2/3}\biggr]^{1/3} \, T^2_8$  \\
muon-electron-scattering &
$\nu_{\mu e} = \nu_{e\mu} \, \biggl(\frac{n_e}{n_{\mu}}\biggr)^{1/3}$  \\
\hline
electron-electron scattering &
$\nu_{ee} \approx 3.58 \times 10^{11} \, \biggl( \frac{p_{Fe}}{q_0} \biggr)^3 \, \biggl( \frac{n_0}{n_e} \biggr)^{1/3} \, T^2_8$  \\
muon-muon scattering &
$\nu_{\mu \mu} = \nu_{ee} \, \biggl( \frac{n_{\mu}}{n_e} \biggr) \, \biggl[ 1 + \frac{6}{5} \biggl(\frac{n_{\mu r}}{n_{\mu}} \biggr)^{2/3} + \frac{2}{5} \biggl(\frac{n_{\mu r}}{n_{\mu}} \biggr)^{4/3} \biggr]$  \\
\hline
Cross-scattering terms &
$\nu'_{e \mu} \approx 1.43 \times 10^{11} \, \biggl( \frac{p_{Fe}}{q_0} \biggr)^3 \, \biggl( \frac{n_0}{n_e} \biggr)^{1/3} \, \biggl( \frac{n_{\mu}}{n_e} \, \biggr)^{2/3} \, T^2_8$ \\
& 
$\nu_{\mu e} = \nu'_{e \mu} \, \biggl( \frac{n_e}{n_{\mu}} \biggr)$   \\ 
\hline
\textit{Baryon conduction}$\, {}^{[\textbf{3}]}$ \\ 
neutron-neutron scattering & 
$\nu_{nn} \approx 3.48 \times 10^{15} \, T^2_8 \, \biggl \{ \mathcal{S}_{n2} \mathcal{R}_{n2} + 3\mathcal{S}_{n1} [\mathcal{R}_{n1} - \mathcal{R}_{n2}] \biggr \}$ \\
neutron-proton scattering &
$\nu'_{np} \approx 3.48 \times 10^{15} \, T^2_8 \, \biggl \{ \mathcal{S}_{p2} \mathcal{R}_{p2} + 0.5 \mathcal{S}_{p1}  [3\mathcal{R}_{p1} - \mathcal{R}_{p2}] \biggr \}$  \\
\hline
\hline
\end{tabular}
}%
\\
Ref. (1)~\citet{2009Bildsten}; (2)~\citet{1995NuPhA.582..697G}; (3)~\citet{2001Baiko}
 
\label{ThermCondTable} 

\end{table} 

\subsection{Boundary Conditions and Method of Solution}
\label{Boundary Conditions and Method of Solution}
 
The system of ODEs Eqs. (\ref{eq:dLdr}) \& (\ref{eq:dTdr}) for the thermal background is a boundary value problem with a set of outer and inner boundary conditions. The outer boundary is set by the interface whereby the crust meets the accreted material falling onto the star. For simplicity, we set the interface at the approximate location of the bottom of the hydrogen/helium burning layer at $\rho = 10^7$ g cm$^{-3}$ \citep{2008}. When thermonuclear burning in the envelope is stable, we follow the prescription of UCB and assume that the outer boundary condition is fixed by the temperature at the base of the hydrogen/helium burning layer (as computed by \citet{1999Schatz}), 

\begin{equation} \label{eq:T_OB}
    T_{\text{OB}} = 5.3 \times 10^8 \text{K} \, \biggl(\frac{\dot{m}}{\dot{m}_{\text{Edd}}}\biggr)^{2/7},
    \vspace*{3mm}
\end{equation}

\noindent
where $\dot{m}$ is the \textit{local} accretion rate and $\dot{m}_{\text{Edd}}$ is the local Eddington limit. In the calculations that follow, we will henceforth assume uniform accretion over the surface of the star, and therefore parameterise our results in terms of $\dot{M}$, the \textit{global} rate of mass accretion.

Burning is assumed to stable when the NS is accreting at a level $\dot{M} = 0.1 - 1 \dot{M}_{\text{Edd}}$ ($\dot{M}_{\text{Edd}} = 2\times 10^{8} \, M_{\odot} \, {\rm yr}^{-1}$), since the rate of nuclear energy generation in this regime is less temperature-sensitive than the radiative cooling \citep{1999Schatz}. If the accretion rate is much lower however, then burning in the upper atmosphere can lead to type I X-ray bursts. These X-ray bursts are short-lived and occur periodically over the accretion episode. The bursts produce a rapid increase in the observed luminosity as burning of the accreted material quickly breaks out into thermonuclear runaway. 

Since we seek only a steady-state solution to our background thermal equations, we will interpret $\dot{M}$ as a \textit{time-averaged} accretion rate. On a timescale much larger than that of an individual burst, we can average over all of the bursts in a given accretion episode. Such a choice allows us to apply Eq. (\ref{eq:T_OB}) even at low accretion rates, as we can effectively assume a constant $T_{\rm OB}$ even if burning in the upper atmosphere is unstable. 

For the inner boundary condition, initially one might naively set $\left. L \right|_{r = 0} = 0$ for the centre of the star. However, the ODEs Eqs. (\ref{eq:dLdr}) \& (\ref{eq:dTdr}) are singular at the origin due to our choice of spherical coordinates. We therefore expand all variables via Taylor series about the center in order to obtain an approximate solution of the coupled ODEs at small radii. To leading order, we approximate the solutions to Eqs. (\ref{eq:dLdr}) \& (\ref{eq:dTdr}) near the origin as

\begin{equation} \label{eq:BavkgroundTaylor L}
    L_{IB} \approx \frac{4}{3} \, \pi \, Q(\rho_c, T_{\rm{cent}}) \, R_{IB}^3 \, ,
\end{equation}

\begin{equation} \label{eq:BavkgroundTaylor T}
    T_{IB} \approx T_{\rm{cent}} - \frac{1}{6}\frac{Q(\rho_c, T_{\rm{cent}})}{\kappa(\rho_c, T_{\rm{cent}})} \, R_{IB}^2 \, ,
\vspace*{3mm}
\end{equation}

\noindent
where $\rho_c$ is the central density of the star (given in Table \ref{tab:Model properties}), $T_{\rm{cent}}$ (not to be confused with the critical temperature $T_c$) is the (initially unknown) value of the temperature at the center of the star and $R_{\rm{IB}}$ is the radius at the inner boundary of the computational domain. We choose the radius at the inner boundary of the computational domain such that $0 < R_{\rm{IB}} << R_{\rm{OB}}$, where $R_{\rm{OB}}$ is the radial location of the base of the H/He burning layer which we obtain from our TOV solution. 

To obtain the unknown value of $T_{\rm{cent}}$, we use the Python ODE solver \texttt{solve\_BVP} (which implements a 4$^{\text{th}}$ order collocation algorithm as outlined by \citet{10.1145/502800.502801}) to obtain the value of $T_{\rm{cent}}$ that satisfies the outer boundary condition Eq. (\ref{eq:T_OB}). An initial guess for the integration is constructed by fixing the temperature throughout the entire star to that of $T_{\text{OB}}$, which is a function of the predetermined accretion rate only. 

\subsection{Background Thermal Structure}\label{Results: Background Thermal Structure}

\begin{figure*}
    \centering
	\includegraphics[width=0.90\textwidth]{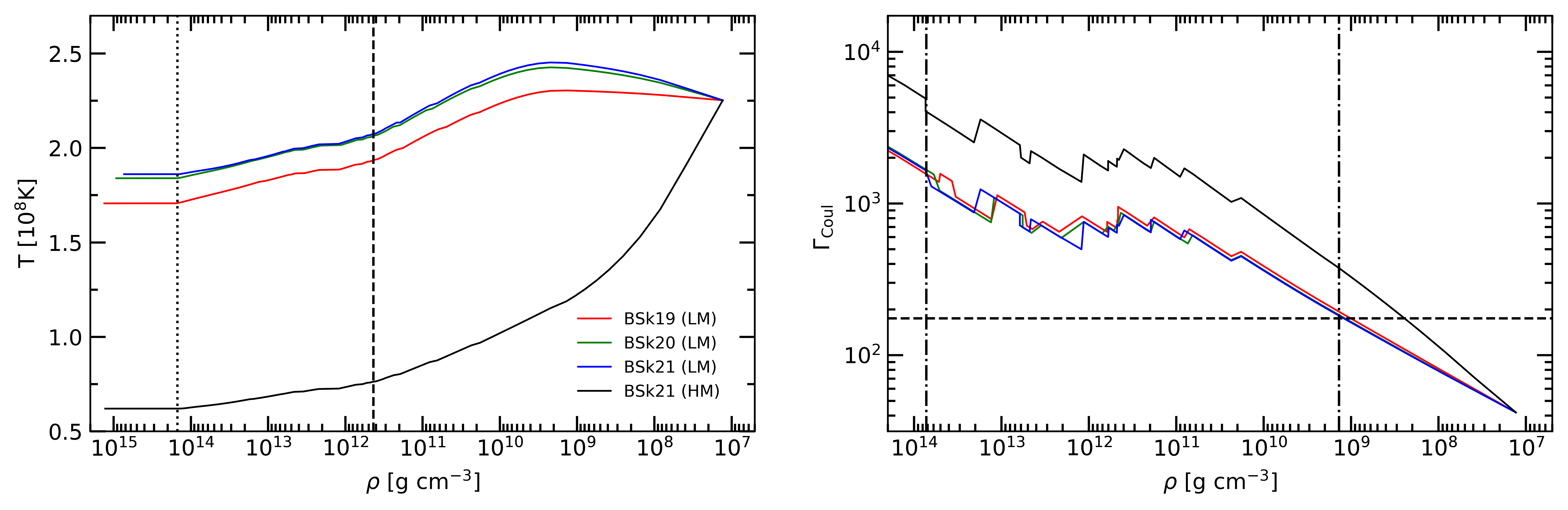}
    \caption{\textit{Left}: Temperature profiles (in units of 10$^8$ K) as a function of density for the different NS models listed in Table \ref{tab:Model properties}. The vertical dotted line indicates the average location of the crust-core transition for the different models, whilst the dashed line indicates the average location of neutron drip. \textit{Right}: The Coulomb parameter indicating the physical state of ions in the accreted crust as a function of the density. The horizontal dashed line is the value $\Gamma_{\rm{Coul}} = 175 $; a one component plasma forms a  solid Coulomb lattice when $\Gamma_{\rm{Coul}} \geq 175 $. The vertical dash-dotted lines bound the heat-producing region of the BSk19-21 EoSs (Tables A.3 - A.1 in \citet{Fantina}). Here $Q_{\rm{imp}} = 1$, $Q_{\rm{S}} = 1.5$ MeV and $\dot{M} = 0.05 \dot{M}_{\rm{Edd}}$ }
    \label{fig:BSk Background Parameters}
\end{figure*}

We have calculated the background thermal structure of an accreting neutron star for the four different BSk19-21 EoS models listed in Table \ref{tab:Model properties}. The left panel of Fig. \ref{fig:BSk Background Parameters} shows the temperature profiles of each model as a function of the density, assuming an impurity parameter $Q_{\text{imp}} = 1$, shallow heating term $Q_{\rm{S}} = 1.5$ MeV and accretion rate $\dot{M} = 0.05 \dot{M}_{\text{Edd}}$.

For the low-mass stars, the radial temperature gradient ($dT/dr$) changes from positive in the inner crust to negative in the outer crust. The temperature gradient is negative when heat flows to the surface, and positive when the heat is conducted down into the core. Over the majority of the crust, heat is mostly conducted into the core, where it is then radiated away as neutrinos. For the high-mass star however, the temperature gradient is positive over the entire crust. In this scenario the heat is overwhelmingly conducted down through the crust into the core. Clearly, the high-mass NS model produces a much cooler star than any of the low-mass models. This is due to the presence of DUrca emission in the high-mass case that effectively radiates away heat produced from both the shallow and deep crustal heating mechanisms. 

As was the topic of discussion in \citet{2000}, the accretion of matter onto the neutron star may cause parts of the crust to melt. Any such molten regions will not be able to contribute to building an elastic mountain (thermal or otherwise) due to the vanishing of any existent shear stresses. The right panel of Fig. \ref{fig:BSk Background Parameters} shows the Coulomb parameter (Eq. \ref{eq:Coulomb to Thermal}) as a function of the density for each of the temperature profiles in the left panel of Fig. \ref{fig:BSk Background Parameters}. The cooler high-mass NS model results in a crust that is entirely solid ($\Gamma_{\rm{Coul}} \geq 175$) over the heat producing region (10$^{9} \lesssim \rho \lesssim $ 10$^{14}$ g cm$^{-3}$). In the hotter low-mass stars, the inner crust is also solid for each EoSs model, but does turn liquid at densities $\rho \lesssim 2 \times 10^{9}$ g cm$^{-3}$. In both cases, the very outer layers of the crust just below the envelope are liquid.

In the analysis of their own deformation producing mechanism, UCB found that it is the deeper capture layers that contribute the most to the formation of the mass quadrupole. The increased density, together with a greater shear modulus in the inner crust allow this portion of the crust to support greater stresses than the outer layers. Therefore, as long as the heat producing region of our crust is solid, then we are likely in a regime whereby the crust can sustain a mountain developed from a non-axisymmetric temperature distribution.

Before discussing how exactly the temperature asymmetry may be introduced, we first consider the implications of some of the background parameters, namely the impurity parameter Q$_{\rm{imp}}$ and shallow heating term Q$_{\rm{S}}$, on the star's temperature profile. The accretion rate $\dot{M}$ will also be considered in Section \ref{sec: Perturbed Thermal Structure Results}. 

\begin{figure*}
    \centering
	\includegraphics[width=0.90\textwidth]{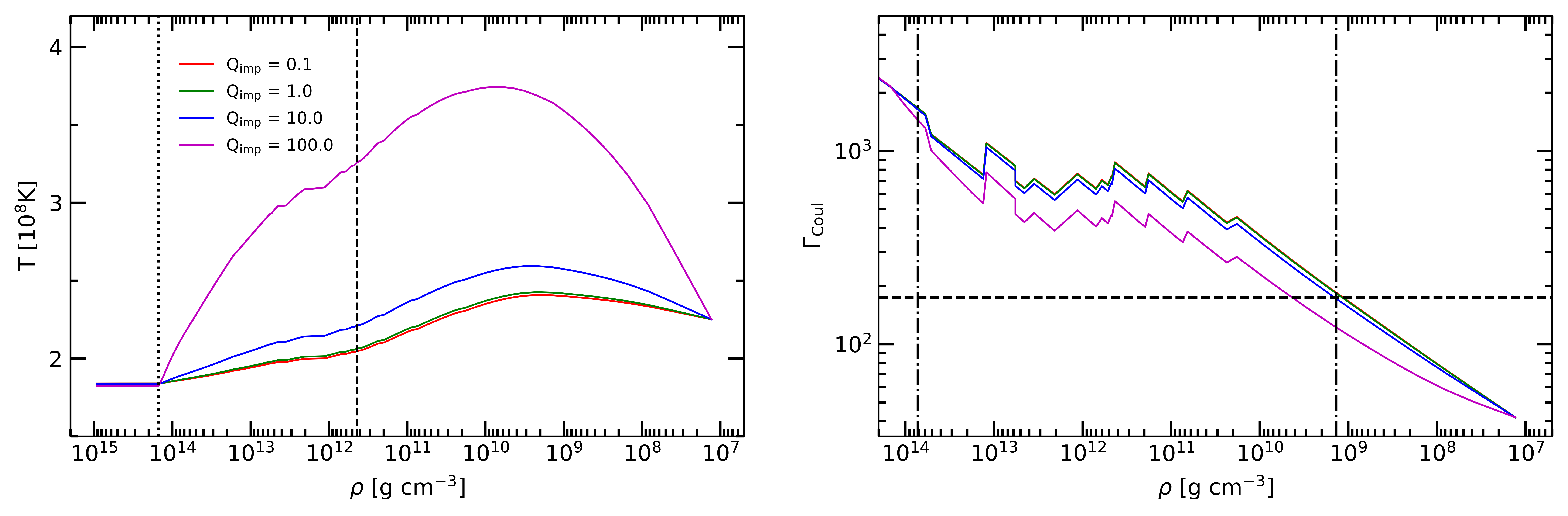}
    \caption{Temperature profiles (\textit{left}) and corresponding Coulomb parameter values (\textit{right}) for the low-mass BSk20 EoS model in Table \ref{tab:Model properties} for different values of the impurity parameter $Q_{\rm{imp}}$ indicated in the legend. Here $Q_{\rm{S}} = 1.5$ MeV and $\dot{M} = 0.05 \dot{M}_{\rm{Edd}}$. Vertical and horizontal lines have same meaning as in Fig. \ref{fig:BSk Background Parameters}.}
    \label{fig:BSk Impurity Background Parameters}
\end{figure*}

\begin{figure*}
    \centering
	\includegraphics[width=0.90\textwidth]{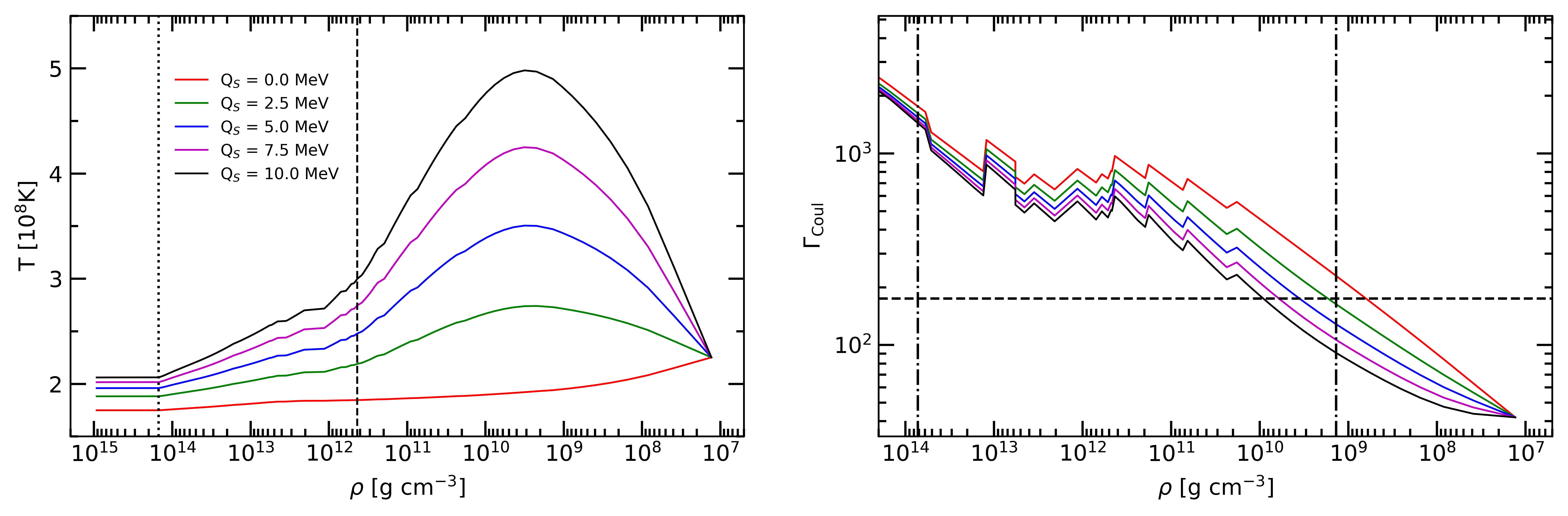}
    \caption{Temperature profiles (\textit{left}) and corresponding Coulomb Parameter values (\textit{right}) for the low-mass BSk20 EoS model in Table \ref{tab:Model properties} for different values of the shallow heating term $Q_{\rm{S}}$ indicated in the legend. Here $Q_{\rm{imp}} = 1.0$ and $\dot{M} = 0.05 \dot{M}_{\rm{Edd}}$. Vertical and horizontal lines have same meaning as in Fig. \ref{fig:BSk Background Parameters}.}
    \label{fig:BSk Shallow Background Parameters}
\end{figure*}

In producing Fig. \ref{fig:BSk Background Parameters}, Q$_{\rm{imp}}$ and Q$_{\rm{S}}$ were assumed to be 1.0, and 1.5 MeV respectively, taken as averages from observational constraints. However, these quantities can, in principle, be much larger. For example, Q$_{\rm{imp}}$ may exceed 100 at the end of stable burning \citep{1999Schatz}, and Q$_{\rm{S}}$ may be required to be as much as 17 MeV per accreted nucleon to fit observational data from some soft X-ray transients (see Table 1 of \citet{2020Chamel}). In the left panels of Figs. \ref{fig:BSk Impurity Background Parameters} and \ref{fig:BSk Shallow Background Parameters}, we show how the background temperature profile depends upon the value of Q$_{\rm{imp}}$ and Q$_{\rm{S}}$ respectively, for the low-mass BSk20 EoS model with a fixed accretion rate of $\dot{M} = 0.05 \dot{M}_{\rm{Edd}}$. 

In Fig. \ref{fig:BSk Impurity Background Parameters}  we vary the impurity parameter over 3 orders of magnitude, noting that the temperature profile is largely insensitive to the impurity parameter when $Q_{\rm{imp}} \lesssim 1$, but see large differences when $Q_{\rm{imp}} = 100$.  We interpret this as follows.  As the impurity parameter is increases, the thermal conductivity in the crust decreases, with $\kappa_{\rm{e}} \propto 1 / Q_{\rm{imp}}$ in the regime where electron-impurity scattering dominates over electron-phonon scattering.  This leads to less conduction of heat from the crust into the core,  raising the crust temperature, with increased cooling from crustal neutrino emission.

In Fig. \ref{fig:BSk Shallow Background Parameters} we vary the SCH parameter Q$_{\rm{S}}$ from 0 - 10 MeV per accreted nucleon, with the impurity parameter fixed at $Q_{\rm{imp}} = 1.0$. The increased levels of heat deposited in the outer layers of the star naturally leads to a hotter crust, introducing steeper temperature gradients since the outer boundary condition at the base of the H/He layer is fixed by that of the accretion rate.

For completeness, we have also plotted in the right-hand panels of Figs. \ref{fig:BSk Impurity Background Parameters} and \ref{fig:BSk Shallow Background Parameters} the corresponding values of the Coulomb parameter to check whether increased amounts of electron-impurity scattering or shallow crustal heating can melt the crust. Increasing the value of $Q_{\rm{imp}}$ does not make an appreciable difference to the state of the ions in the crust, and even an additional 10 MeV per nucleon of SCH does not melt the inner crust of the NS, melting only the outer crust at densities $\rho \sim$ 10$^{10}$ g cm$^{-3}$.
 
\section{Temperature Perturbation due to an Internal Magnetic Field}
\label{section: Temperature Perturbation due to an Internal Magnetic Field}

We now perturb our thermal background through the addition of a magnetic field \textbf{B}. We follow closely the prescription originally laid out in OJ. However, in their analysis only the accreted crust of the NS was considered, neglecting the NS core. In the following, we extend the procedure of OJ and include the possibility of having temperature perturbations in the NS core as well as the crust, making appropriate additions of the possibility of heat conduction via electrons, muons and neutrons in the core.

\noindent
\\
Fourier’s law (including contributions from each of the heat carriers) in tensorial form is written as

\begin{equation}\label{eq:PerturbedFourier}
    F^{a} = - \sum_{X} \kappa^{a b}_X \nabla_{b} T,
    \vspace*{3mm}
\end{equation}

\noindent
where $\kappa^{a b}_X$ is the thermal conductivity tensor, and X denotes the heat carrier involved (X = e, $\mu$, n). In the presence of a magnetic field, classical Larmor rotation of the electrons (and muons, where available) make the thermal conductivity anisotropic. The heat flux including contributions from the magnetic field were derived by \citet{1980SvA....24..303Y}. For a magnetic field $\textbf{B} = \text{B}\textbf{b}$, the heat flux carried by species $X$ can be written in vectorial form as

\begin{equation} \label{eq:Perturbed Flux 1 lepton}
    \textbf{F}_{X}  = - \, \kappa^{\perp}_{X} \bigl[ \nabla T + (\omega_B^X \tau_{X})^2 (\textbf{b} \cdot \nabla T) \cdot \textbf{b} + \omega_B^X \tau_{X} (\textbf{b} \times \nabla T) \bigr] \, ,
    \vspace*{3mm}
\end{equation}

\noindent
where $\kappa^{\perp}$ is the component of the thermal conductivity tensor perpendicular to the magnetic field, $\textbf{b}$ is the unit vector of the magnetic field (pointing in the direction of the magnetic field) and $\omega_B^X \tau$ is known as the `magnetisation parameter'. The magnetisation parameter is the product of the already familiar effective relaxation time $\tau$, and the electron (muon) gyrofrequency $\omega_B^X$, which is defined as 

\begin{equation}\label{eq:Gyrofrequency}
    \omega_B^{X} = \frac{eB}{m^{\ast}_{X} c} \, .
    \vspace*{2mm}
\end{equation}

\noindent
For the neutrons in the core, their contribution to the heat flux is unaffected by the addition of the magnetic field (since they are electrically neutral; $\omega_B^{X} = 0$ when $X = n$) and Eq. (\ref{eq:Perturbed Flux 1 lepton}) is simply

\begin{equation} \label{eq:Perturbed Flux 1 neutron}
    \textbf{F}_n  = -\kappa^0_n \, \nabla T \, ,
\end{equation}

\noindent
where $\kappa^0_n$ is the neutron part of the \textit{scalar} conductivity (given by Eq. (\ref{eq:ThermalConductivity}) when $X = n$).

\noindent
\\
In the core of the NS, nuclear equilibrium considerations imply the equality of the electron and muon chemical potentials, and therefore that the gyrofrequency of the two charge carriers are the same (i.e. $m^{\ast}_{e} = \mu_e / c^2 = \mu_{\mu} / c^2 \equiv m^{\ast}_{\mu}$). In the calculations that follow, we therefore assume the same $\omega_B$ for both muons and electrons. 

The addition of the magnetic field deflects a piece of the heat flux in the direction orthogonal to both the temperature gradients and the local magnetic field (a thermal analogue of the Hall effect), as can be seen from the final term in Eq. (\ref{eq:Perturbed Flux 1 lepton}). Purely to gain insight, we can compare equations (\ref{eq:PerturbedFourier}) and (\ref{eq:Perturbed Flux 1 lepton}) to write the conductivity tensor $\kappa_{a b}$ in a Cartesian basis, with the magnetic field orientated along the z-axis:

\begin{equation}\label{KTensor}
    \kappa = \left( \begin{matrix} \kappa^{\perp} & \kappa^{\wedge} & 0 \\ -\kappa^{\wedge} & \kappa^{\perp} & 0 \\ 0 & 0 & \kappa^{\parallel} \end{matrix} \right) \, ,
    \vspace*{3mm}
\end{equation}

\noindent
where $\kappa^{\parallel}$ is the component of the thermal conductivity parallel to the magnetic field and $\kappa^{\wedge}$ is the so-called Hall component. 

\noindent
\\
The components of the thermal conductivity tensor are related to the scalar conductivity ($\kappa^0$ in Eq. (\ref{eq:Perturbed Flux 1 neutron})) as

\begin{align} \label{eq:Concuctivity Components}
    \centering
        \kappa^{\parallel}_X = \kappa^0_X, \qquad \kappa^{\perp}_X = \frac{\kappa^0_X}{(1 + (\omega_B^X \tau_X)^2)} , \qquad \kappa^{\wedge}_X = \omega_B^X \tau_X \, \kappa^{\perp}_X  \, .
\end{align}
\vspace{1mm}

\noindent
In the direction perpendicular to the magnetic field, the thermal conductivity is suppressed, corresponding to a diminishing of the heat flow orthogonal to the magnetic field \citep{1980SvA....24..425U}. The ratio of the conductivities parallel and perpendicular to the magnetic field is given in terms of the magnetisation parameter as

\begin{equation}\label{eq:MagParameter}
    \frac{\kappa^{\parallel}_{X}}{\kappa^{\perp}_{X}} = 1 + (\omega^{X}_B \tau_{X} (T))^2 \, ,
    \vspace*{3mm}
\end{equation}

\noindent
where X in this instance denotes the \textit{charged} carrier involved (i.e. $X = e,\, \mu$). As noted above, the neutrons, whilst contributing to the thermal conductivity in the core, have no interaction with the magnetic field and therefore their contribution does not produce any anisotropy in the heat flow (Eq. (\ref{eq:MagParameter}) reduces to 1 when $X = n$). In regions where the electrons, muons, and neutrons coexist, the individual conductivity tensors $\kappa_{X}$ simply add linearly (via Eq (\ref{eq:PerturbedFourier})).

\subsection{The Thermal Perturbation Equations}

In cooling simulations of highly magnetised neutron stars, the regime $\omega_B \tau (T) \gg 1 $ is often considered as the large magnetic fields associated with these systems (in excess of $10^{14}$ G) would give rise to strong anisotropy \citep{2004A&A...426..267G, 2007PonsGeppert}. However, estimates for the strength of magnetic fields that could feasibly exist within NSs in LMXBs are much more conservative. The external field strengths in these systems are inferred to be of the order of just $10^8 - 10^9$ G (e.g. \citet{1999A&AT...18..447P}). Following OJ, we therefore assume to be in a regime whereby $\omega_B \tau (T) \ll 1 $ , such that we can treat the presence of a magnetic field as a perturbation on the heat flow of our spherically symmetric background. This limit inherently implies there is a maximum magnetic field strength that we may consider in order to still remain in the perturbative regime. This will be discussed in detail in Section \ref{sec: Perturbed Thermal Structure Results}. In general, we find that we may safely consider \textit{crustal} magnetic fields with strengths $B \lesssim 10^{13}$ G, and magnetic fields $B \lesssim 10^{8} $ G for \textit{core} magnetic fields (see Fig. \ref{fig:Magnetisation Parameter}). 

For sufficiently small magnetic fields, linearising Eqs. (\ref{eq:Perturbed Flux 1 lepton}) \& (\ref{eq:Perturbed Flux 1 neutron}) in both the temperature perturbation $\delta T$ and the magnetisation parameter $\omega_B \tau$ leads to

\begin{equation} \label{eq:Delta F}
\centering
\begin{split}
    \delta \textbf{F} = & - \kappa^0_{e} \bigl[ \boldsymbol{\nabla} \delta T + \Tilde{\omega} \tau_{e} [\textbf{B} \times \boldsymbol{\nabla} T_0] \bigr]  - \delta \kappa^{0}_{e} \boldsymbol{\nabla} T_0 \\
    & - \kappa^0_{\mu} \bigl[ \boldsymbol{\nabla} \delta T + \Tilde{\omega} \tau_{\mu} [\textbf{B} \times \boldsymbol{\nabla} T_0] \bigr]  - \delta \kappa^{0}_{\mu} \boldsymbol{\nabla} T_0 \\
    & - \kappa^0_n \boldsymbol{\nabla} \delta T - \delta \kappa^0_n \boldsymbol{\nabla} T_0,
\end{split}
\end{equation}
\vspace{2mm}
\noindent
where we have introduced (for convenience) the quantity 

\begin{equation}\label{eq:GFperFS}
    \Tilde{\omega} = \frac{e}{m^*_{X} \, c},
    \vspace*{3mm}
\end{equation}

\noindent
which OJ term the `gyromagnetic frequency per unit magnetic field strength' (note again that we have assumed $m^*_{\mu} \equiv m^*_{e}$). Like OJ, we shall use spherical harmonics to describe the dependence of all perturbed quantities on the spherical polar angles $(\theta, \phi)$, i.e.

\begin{equation}\label{eq:deltaTSH}
    \delta T = \sum_{l \,=\, 0}^{\infty} \sum_{m \,=\, -l}^{l} \delta T_{lm}(r) Y_{lm} (\theta, \phi) \, .
    \vspace*{3mm}
\end{equation}

\noindent
The perturbed heat flux $\delta \textbf{F}$ is a vector field and is written in terms of the scalar spherical harmonic $Y_{lm}(\theta,\phi)$ as

\begin{equation} \label{eq:VSH}
    \delta \textbf{F} = \mathlarger{\sum_{l \,=\, 0}^{\infty} \sum_{m \,=\, -l}^{l} \,} \, \biggl [ U_{lm}(r) \, Y_{lm} \hat{\textbf{r}} + V_{lm}(r) \, r\boldsymbol{\nabla} Y_{lm} + W_{lm}(r) \, (\textbf{r} \times \boldsymbol{\nabla} Y_{lm}) \biggr ] \, ,
    \vspace*{3mm}
\end{equation}

\noindent
where $\hat{\textbf{r}}$ is the radial unit vector. We follow \citet{PhysRevD.64.083008} and decompose the magnetic field into its constituent poloidal and toroidal parts  

\begin{equation} \label{eq:BDecomposed}
    \textbf{B} = \textbf{B}_{\text{pol}} + \textbf{B}_{\text{tor}} \, , 
    \vspace*{3mm}
\end{equation}

\noindent
and express them in terms of two scalar functions $\Phi \, (r, \theta, \phi )$ and $\Psi \, (r, \theta, \phi )$ such that:

\begin{equation} \label{eq:BDecomposedScalerPol}
    \textbf{B}_{\text{pol}} = - \boldsymbol{\nabla} \times (\textbf{r} \times \boldsymbol{\nabla} \Phi) \, ,
    \vspace*{-4mm}
\end{equation}

\begin{equation} \label{eq:BDecomposedScalerTor}
    \textbf{B}_{\text{tor}} = - \, \textbf{r} \times \boldsymbol{\nabla} \Psi \, .
    \vspace*{3mm}
\end{equation}

\noindent
Inserting Eqs. (\ref{eq:deltaTSH}), (\ref{eq:BDecomposedScalerPol}) and (\ref{eq:BDecomposedScalerTor}) into Eq. (\ref{eq:Delta F}) and comparing coefficients with Eq. (\ref{eq:VSH}), we obtain the first ODE describing the perturbed thermal structure as

\begin{equation} \label{eq:Perturbed Temp ODE}
    \frac{d \delta T_{lm}}{dr}  = - \frac{1}{\kappa} \Biggl[ \Biggl( \frac{d \kappa^0_e}{dT} + \frac{d \kappa^0_{\mu}}{dT} + \frac{d \kappa^0_n}{dT} \biggr) \frac{dT}{dr} \delta T_{lm}  + U_{lm} \Biggr] \, ,
    \vspace*{3mm}
\end{equation}

\noindent
where $\kappa = \kappa^0_e + \kappa^0_{\mu} + \kappa^0_n$ (See Eq. \eqref{eq:ThermalConductivitySum}). The result of this procedure also yields two additional algebraic expressions for the quantities $V_{lm}$ and $W_{lm}$, given by

\begin{equation} \label{eq:V_lm}
 V_{lm} = \frac{1}{r} \Biggl\{ \Biggr[ \biggl(\kappa^0_{e}\Tilde{\omega} \, \tau_{e} + \kappa^0_{\mu} \Tilde{\omega} \, \tau_{\mu} \biggr) \, \Psi_{lm} r \frac{dT}{dr} \Biggr] - \kappa \delta T_{lm} \Biggr\} \, ,
\end{equation}

\begin{equation} \label{eq:W_lm}
\centering
    W_{lm} = - \, \bigl(\kappa^0_{e}\Tilde{\omega} \, \tau_{e} + \kappa^0_{\mu} \Tilde{\omega} \, \tau_{\mu} \bigr) \, \Bigg[ \frac{1}{r} \Phi_{lm} \frac{d T_0}{dr} + \frac{d \Phi_{lm}}{dr}  \frac{d T_0}{dr} \Biggr] \,  ,
\end{equation}

\noindent
where $\Psi_{lm}$ and $\Phi_{lm}$ are the spherical decompositions of the scalar functions  $\Psi\, (r, \theta, \phi )$ and $\Phi\, (r, \theta, \phi )$ (in the same manner as Eq. \eqref{eq:deltaTSH}).

Next we write down the perturbed form of the energy conservation (Eq. (\ref{eq:Luminoisty})), which is

\begin{equation} \label{eq:PerturbedConserve}
    \nabla \cdot \delta \textbf{F} = \delta Q \, .
    \vspace*{3mm}
\end{equation}

\noindent
Recall from Eq. (\ref{eq:Luminoisty}) that Q is the net rate of heat energy generation per unit volume (Q$_h$ – Q$_{\nu}$). For simplicity, we will assume that the nuclear heating term $Q_h$ is independent of the temperature (and magnetic field strength). This is justified since the heat release calculated by \citet{Fantina} neglects thermal contributions to thermodynamic potentials and considers only ground-state transitions to compute the accreted crust. For highly magnetised stars ($B \gtrsim 10^{13}$ G) however, significant amounts of so-called \textit{Joule heating} can arise due to dissipation of the magnetic field in the solid crust (e.g. \citet{Miralles_1998}). However, as we shall see, our perturbation equations are only valid for magnetic field strengths $B \lesssim 10^{12}$, in which case the heating rate due to dissipation of the magnetic field is negligible compared to that of the heating rate due to DCH. The presence of the magnetic field therefore influences the net heat generated $Q$ within the star only via the dependence of the neutrino cooling $Q_{\nu}$ on the temperature (via the equations listed in Table \ref{NeutrinoCoreTable}). We can therefore rewrite Eq. (\ref{eq:PerturbedConserve}) as

\begin{equation}\label{eq:DivFQ}
    \boldsymbol{\nabla} \cdot \delta \textbf{F} = - \, \frac{d Q_{\nu}}{dT}  \delta T_{lm} Y_{lm} \, .
    \vspace*{3mm}
\end{equation}

\noindent
Taking the divergence of Eq. (\ref{eq:VSH}) is the standard result

\begin{equation}\label{eq:VSH Div}
    \boldsymbol{\nabla} \cdot \delta \textbf{F} = \mathlarger{\sum_{l \,=\, 0}^{\infty} \sum_{m \,=\, -l}^{l} \,} \biggl[ \frac{d U_{lm}}{dr} + \frac{2}{r} U_{lm} - \frac{1}{r} l (l+1) V_{lm} \biggr] Y_{lm} \, ,
    \vspace*{3mm}
\end{equation}

\noindent
where the final term in Eq. (\ref{eq:VSH}) has now vanished since 

\begin{equation} \label{eq: vanishing div}
     \boldsymbol{\nabla} \cdot \bigl( W_{lm} \textbf{r} \times \boldsymbol{\nabla} Y_{lm} \bigr)  = 0.
     \vspace*{3mm}
\end{equation}

\noindent
By comparing Eq. (\ref{eq:VSH Div}) with the RHS of Eq. (\ref{eq:DivFQ}) we obtain the second ODE describing the perturbed thermal structure as

\begin{equation}\label{eq:Perturbed U ODE}
    \frac{d U_{lm}}{dr}  = \frac{d Q_{\nu}}{dT} \delta T_{lm}  - \frac{2}{r} U_{lm} + \frac{1}{r} l(l+1) V_{lm}. 
    \vspace*{3mm}
\end{equation}

\noindent

We have a set of four equations ((\ref{eq:Perturbed Temp ODE}), (\ref{eq:V_lm}), (\ref{eq:W_lm}) and (\ref{eq:Perturbed U ODE})) in four unknowns ($U_{lm}, V_{lm}, W_{lm}$ and $\delta T_{lm}$), with the magnetic stream functions ($\Phi_{lm}$ and $\Psi_{lm}$, assumed known) playing the role of source terms.  Note, however, that equation (\ref{eq:W_lm}), giving $W_{lm}$ in terms of $\Phi_{lm}$, decouples from the other three equations.  Physically, we can say that the poloidal part of the magnetic field $\Phi_{lm}$ induces a purely toroidal perturbation in the heat flux $W_{lm}$, and therefore produces no perturbation in the temperature, i.e. does not couple to $\delta T_{lm}$.  Given that we are interested specifically in temperature perturbations, we will  not consider poloidal magnetic fields in our analysis, only toroidal ones.  This leaves us with equations (\ref{eq:Perturbed Temp ODE}), (\ref{eq:V_lm}) and (\ref{eq:Perturbed U ODE}) in the three unknowns $U_{lm}, V_{lm}$ and $\delta T_{lm}$. Such a result stands in contrast to the mass quadrupole more normally associated with magnetic neutron stars built directly from magnetic stresses (via Lorentz forces), whereby it is both the toroidal \textit{as well as} the poloidal component of the magnetic field that can generate the mountain (e.g. \citet{Haskell_2008}. 

A rigidly and steadily rotating triaxial star will emit mainly quadrupolar gravitational radiation.  If we set the spin axis to be along $Oz$, this corresponds to emission via only the $l=m=2$ mass quadrupole moment, giving gravitational waves at twice the spin frequency.  The corresponding temperature perturbation that sources this (via the `wavy capture layers' described in Section \ref{section:An Overview of the Problem}) will also be $l=m=2$, as described in \citet{2002}.

\subsection{Perturbed Boundary Conditions and Method of Solution}\label{Perturbed Boundary Conditions and Method of Solution}

The perturbation equations (\ref{eq:Perturbed Temp ODE}) \& (\ref{eq:Perturbed U ODE}) (together with the algebraic expression Eq. (\ref{eq:V_lm})) are a set of coupled ODEs that we solve using a set of (inner and outer) boundary conditions. In their analysis, OJ adopted the boundary conditions derived by UCB, who also modelled just the crust of the neutron star. The temperature perturbation at both the inner and outer boundaries (the crust-core interface and the base of the ocean respectively) were set to be zero, i.e. $\delta T_{\text{IB}} = \delta T_{\text{OB}} = 0$. UCB argued that this is a good approximation if the thermal conductivity is significantly higher in both the ocean and the core of the NS.

Indeed, the core thermal conductivity \textit{is} a few orders of magnitude greater than in the crust. However, consider our source term (the first term in Eq. (\ref{eq:V_lm})):

\begin{equation}\label{eq:SourceTerm}
     S \equiv \Tilde{\omega} \, \frac{dT}{dr} \, \Psi_{lm} \, \sum_{X} \kappa_X \, \tau_X \, .
    \vspace{3mm}
\end{equation}

\noindent
The source term is a function of both the conductivity as well as the strength of the toroidal magnetic field $\Psi_{lm}$. The magnetic field can therefore provide a non-zero temperature perturbation in the core even if the thermal conductivity there is much larger. This possibility was not explored in OJ, and so we implement a fully self-consistent calculation of the neutron star core here as a means to explore the possibility of a core magnetic field's influence on the magnitude of the temperature perturbation, particularly in the deep crust. 

In extending the computational domain of the calculation, we obtain our inner boundary condition by means of a regularity condition at the centre of the star as to avoid a singularity at $r = 0$. We expand all variables in Eqs. (\ref{eq:Perturbed Temp ODE}), (\ref{eq:V_lm}) and (\ref{eq:Perturbed U ODE}) via Taylor series near the origin to obtain an approximate solution at small radii. To leading order in $r$, we find that the temperature perturbation $\delta T$ and the radial flux perturbation $U$ depend on the second radial derivative of $\delta T$ (evaluated at the centre of the star) as 

\begin{equation}\label{eq:Perturbed Taylor U}
    U_{IB} \approx  - \, \kappa(\rho_c, T_{\rm{cent}}) \, \delta T''_{\rm{cent}}  \, r_{IB} \, ,
\end{equation}
\vspace{-3mm}
\begin{equation}\label{eq:Perturbed Taylor T}
     \delta T_{IB} \approx \frac{1}{2} \, \delta T''_{\rm{cent}} \, r_{IB}^2 \,  .
\vspace{3mm}
\end{equation}

\noindent
The inner boundary condition is then found via root-solving methods to obtain a solution for the second radial derivative $\delta T''$ that satisfies the outer boundary condition, which we shall now discuss.

Contrary to the situation in the core, and the claim made by UCB, both Fig. 5 of \citet{1999Potekhin} and Fig. 3 of \citet{2007Chugunov} suggest that the thermal conductivity in the ocean is actually lower than that of the crust (and the core). This naturally leads one to question the legitimacy of the $\delta T_{\rm OB} = 0$ boundary condition at the top of the crust, as clearly the ocean cannot be assumed to be perfectly conducting if the conductivity is low there. 

However, consider again the outer boundary condition in our unperturbed thermal background (Eq. (\ref{eq:T_OB})). The temperature at the base of the ocean is determined solely by the accretion rate, which we assume to be spherically symmetric. Therefore, if we assume there is no perturbation in the local accretion rate at the base of the ocean when we introduce the magnetic field, then there can be no perturbation in the temperature at the base of the ocean as well. In this case, this allows us to recover the $\delta T_{\rm OB} = 0$ boundary condition even if the ocean is not acting as a perfect conductor. 
 
It is also possible that a more accurate outer boundary condition for $\delta T$ could be derived by matching the crustal thermal calculation to a flux-temperature relation in the ocean \citep{2002}. In this case, the thermal profile would be then determined by not just the fraction of heat that is conducted up through the crust from the DCH/SCH, but also by the amount heat that is released due to compression of accreted material as it arrives at the NS surface \citep{1995ApJ...449..800B, Brown1998TheOA}.  

Rather than attempting such a large expansion of our computations, we have instead examined two other possible choices, in addition to the $\delta T_{lm}=0$ condition motivated above. We wish to test to what extent  the value of the temperature perturbation in the deep crust, where most of the quadrupole is generated \citep{2002}, is sensitive to which outer boundary condition we use.

Specifically, we consider the following three outer boundary conditions: (i) keeping the same $\delta T_{\text{OB}} = 0$ as did OJ (and UCB); (ii) assume a perfectly insulating condition for the perturbed heat flux:  $U_{\text{OB}} = 0$; (iii) assume perfect blackbody emission from the surface,  making use of the Stefan-Boltzmann Law, such that the perturbed flux would be

\begin{equation}\label{eq:deltaFOB}
   U_{\text{OB}} = 4 \sigma T^3 \delta T_{\text{OB}} \, .
   \vspace{3mm}
\end{equation}

\noindent
Blackbody emission would be applicable to isolated neutron stars emitting into vacuum, but is less clearly relevant for an accreting star, whose surface is covered in accreting material.   Note that for a transiently accreting NS, equation (\ref{eq:deltaFOB}) would be well motivated during the periods of quiescence, so may be relevant in a time-averaged sense.

As we shall soon see, the temperature perturbations in the deep crust \textit{are} in fact insensitive to the choice of outer boundary condition (see Fig. \ref{fig:Delta T Different Boundaries}), at least for densities $\rho > 10^{12}$ g cm$^{-3}$.

\subsection{The internal Magnetic Field}\label{The internal Magnetic Field}

To solve the perturbation equations, we need to specify the form of the internal magnetic field. Despite compelling evidence for a dipolar configuration of the external magnetic field around neutron stars, the internal field structure remains largely unknown. As such, we have (within reason) relative freedom in prescribing the internal field. We have shown that for small magnetic fields (such that $\omega_B \tau (T) \ll 1$), the poloidal component of an internal magnetic field is inconsequential in building our thermal mountain, and therefore consider a purely toroidal magnetic field. 

We are interested only in the component of the toroidal magnetic field that gives rise to quadrupolar gravitational wave radiation, and therefore again specialise to the $l = m = 2$ spherical harmonic. From our definition of the toroidal component of the magnetic field (Eq. (\ref{eq:BDecomposedScalerTor})), we write down the functional form of the toroidal field as

\begin{equation}\label{eq:Field Magnitude}
\begin{split}
     \boldsymbol{B}_{\rm{tor}} & =  - \, \textbf{r} \times \boldsymbol{\nabla} \bigl[ \Psi(r) \, Y_{22}(\theta, \phi) \big] \\
    & = - \frac{1}{2} \sqrt{\frac{15}{2\pi}} \, \Psi(r) \,  \bigl[ \text{sin} \, \theta  \, \text{sin}  2 \phi \, \boldsymbol{e}_{\theta} + \text{sin} \, \theta \, \text{cos} \, \theta  \, \text{cos} 2 \phi \, \boldsymbol{e}_{\phi} \bigr]  \, .
\end{split}
\end{equation}

\noindent
We find that the magnitude of Eq. (\ref{eq:Field Magnitude}) is a factor two larger than given in Eq. (34) of OJ, as well as a sign discrepancy in the final term containing $\boldsymbol{e}_{\phi}$. The magnitude of $\boldsymbol{B}_{\rm{tor}}$ is a function of position.  We follow OJ and parameterise magnetic field configurations in terms of the \emph{maximum} value of $|{\bf B}_{\rm tor}|$ within the star.  This will occur at the point where $\Psi(r)$ attains its maximum value, along the line $\theta = \pi / 2 $, $\phi = \pi / 4$. 

For any value of the magnetic stream function $\Psi_{22}(r)$, we define the form of the toroidal magnetic field on the domain

\begin{equation}\label{eq:Field Domain}
\boldsymbol{B}_{\rm{tor}} = \begin{cases}
          0 \quad &\text{if} \, r \leq R_{\rm{B, min}} \\
          \rm{Eq}. \, (\ref{eq:Field Magnitude}) \quad &\text{if} \, R_{\rm{B, min}} < r < R_{\rm{B, max}} \\
          0 \quad &\text{if} \, r \geq R_{\rm{B, max}} \, ,\\
     \end{cases}
     \vspace{3mm}
\end{equation}

\noindent
such that the magnetic field vanishes outside of the region of the NS defined by the inner and outer radii $R_{\rm{B, min}}$ and $R_{\rm{B, max}}$ respectively. We make a distinction between this region of the NS that the magnetic field permeates and the computational domain of the background and perturbed calculation ($R_{\rm{IB}}$ and $R_{\rm{OB}}$), since the two are not necessarily one and the same (and is something that will be explored in Section \ref{sec: Perturbed Thermal Structure Results}). 

In modelling the thermal profile of the NS core, we seek to allow for the possibility of the magnetic field to permeate the core (unlike OJ who confined the magnetic field to the accreted crust), with the expectation that non-vanishing temperature perturbations at the crust-core transition lead to greater asymmetries in the deep crust. One potential caveat to this procedure, however, is the feasibility of having a magnetic field in the core of a NS that is almost certainly superconducting over some density range (recall Fig. \ref{fig:p and n transition Temps}). Minimum energy considerations of superconducting matter implies any magnetic flux within the medium should be expelled due to the Meissner–Ochsenfeld effect (e.g. \citet{KHAN2003235}). 


This however, is not the whole story. In the seminal description of the properties of a proton superconductor in NS matter by \citet{Baym1969SuperfluidityIN}, it was argued that the conductivity of regular conducting matter is sufficiently large such that the characteristic timescale for the expulsion of magnetic flux is comparable with that of the age of the universe. In a more recent analysis of this phenomenon by \citet{2017Ho}, the authors suggest that actually the magnetic field may persist in the bulk of the NS core for at least $\gtrsim$ 10$^7$ years after the star's birth, as a result of disparities in the cooling timescale and the associated diffusion timescale of magnetic field itself. One caveat to this, however, is that NSs in LMXBs are expected to be much older than this, at $\sim$ 10$^9$ yr. Despite this, \citet{2017Ho} also estimated that even after 10$^7$ years, the field is likely only expelled from the innermost $\leq$ 100 m of the NS core (assuming a 10$^{11}$ G field). It is therefore possible that the field may not have been completely expelled from NS core, even after 10$^9$ years.

To this end, we consider two functional forms for the magnetic stream function $\Psi(r)$. In the first instance, we take the form of the toroidal field to be  

\begin{equation}\label{eq:ToroidalField1}
   \Psi(r) = C [(r - R_{\rm{B, \, min}})(r - R_{\rm{B, \, max}})]^2 \, ,
   \vspace{3mm}
\end{equation}

\noindent
where $C$ is a constant, such that the magnetic field vanishes outside of the region defined in Eq. (\ref{eq:Field Domain}) and is a maximum at the midpoint of the domain where $r = (R_{\rm{B, \, max}} - R_{\rm{B, \, min}}) / 2$). This form of $\Psi(r)$ allows us to consider two possibilities, where (i) the magnetic field permeates the entire star, extending over the full computational domain where $R_{\rm{B, \, min}} = R_{\rm{IB}}$ and $R_{\rm{B, \, max}} = R_{\rm{OB}}$, or (ii), the magnetic field is confined to only the \textit{crust} of the NS, such that $R_{\rm{B, \, min}} = R_{\rm{crust-core}}$ and $R_{\rm{B, \, max}} = R_{\rm{OB}}$, where $R_{\rm{crust-core}}$ is the radial location of the crust-core interface obtained from our TOV solution. The second case is equivalent to the form of the toroidal field used by OJ, following the prescription of one particular field configuration originally considered by \citet{2007Pons} (cf. their Eq. 35).

The second functional form for the magnetic stream function we consider is derived from Eq. (12) of \citet{2008Aguilera}, which we have modified to be

\begin{equation}\label{eq:ToroidalField2}
   \Psi(r) = \Psi_0 \, x \, \biggl( 1 - x \biggr)^2 \, \biggl( x - \frac{R_{\rm{B, \, min}}}{R_{\rm{B, \, max}}} \biggr)^{65} \, ,
   \vspace{3mm}
\end{equation}

\noindent
where $x = r / R_{\rm{B, \, max}}$ with $R_{\rm{B, \, min}} = R_{\rm{IB}}$ and $R_{\rm{B, \, max}} = R_{\rm{OB}}$ and $\Psi_0$ is a constant chosen such that magnetic field has a maximum value $B = 10^9$ G, consistent with the inferences of the external magnetic field of LMXBs. The value of the exponent in Eq. (\ref{eq:ToroidalField2}) was chosen (somewhat) arbitrarily in order to capture the possibility of a magnetic field that extends beyond the crust-core transition, but drops rapidly before reaching the centre of the star. This choice will allow us to explore whether non-vanishing temperature perturbations at the crust-core transition lead to greater asymmetries in the deep crust, whilst still remaining consistent with the results of \citet{2017Ho} that the magnetic field is expelled from the innermost part of the NS due to proton superconductivity, but not from the core completely.

\begin{figure}
    \centering
	\includegraphics[width=0.7\columnwidth]{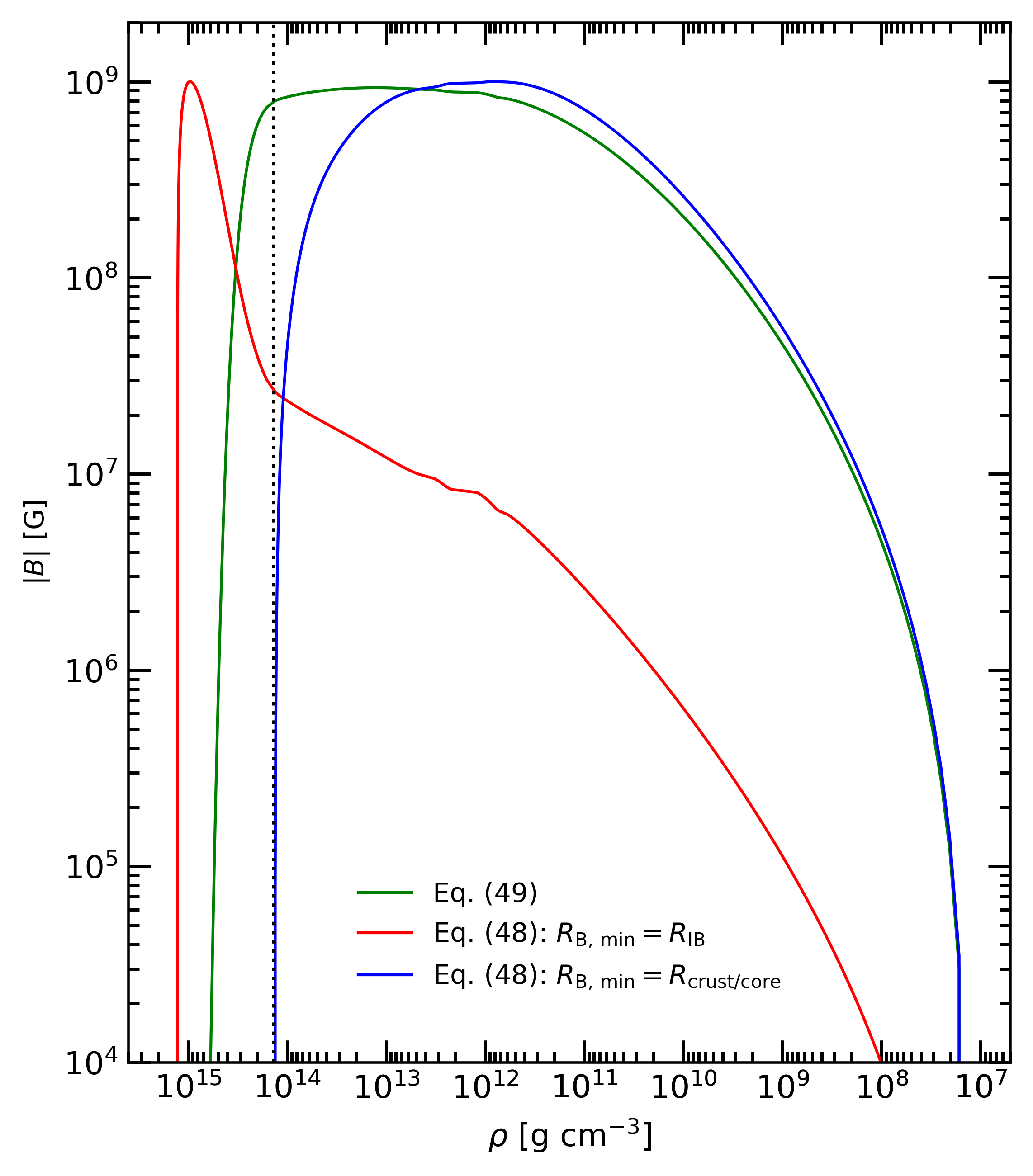}
    \caption{Magnitude of the internal toroidal magnetic field $B_{\rm{tor}}$ as a function of the density for three different field configurations. The blue curve shows a toroidal field that is confined to the crust only (Eq. (\ref{eq:ToroidalField1}) with $R_{\rm{B, \, min}} = R_{\rm{crust-core}}$). The red curve shows the magnitude of the toroidal field that permeates the entire star ((\ref{eq:ToroidalField1}) with $R_{\rm{B, \, min}} = R_{\rm{IB}}$). The green curve shows a toroidal magnetic field that penetrates only the outer core of the NS (Eq. (\ref{eq:ToroidalField2})). All are normalised to have the same maximum strength $B = 10^9$ G. The vertical dotted line indicates the average location of the crust-core transition for the BSk EoSs.}
    \label{fig:field shape}
\end{figure}

In Fig. \ref{fig:field shape} we plot the magnitude of the internal toroidal magnetic field as a function of the density for the low-mass BSk20 model in Table \ref{tab:Model properties}, for the three different configurations of $\Psi(r)$ discussed above.

The prescription of our magnetic field requires that the field vanishes outside of the region defined by $R_{\rm{B, min}}$ and $R_{\rm{B, max}}$ (Eq. (\ref{eq:Field Domain})). The red and blue lines denote the functional form of the magnetic field as described by Eq. (\ref{eq:ToroidalField1}) when $R_{\rm{B, \, min}}$ is set to $R_{\rm{IB}}$ and $R_{\rm{crust-core}}$ respectively, whilst the green line denotes the functional form of the magnetic field as described by Eq. (\ref{eq:ToroidalField2}). The green curve acts as an intermediary case between the two extremes whereby the field extends over the entire computational domain (the red line), and where it is confined to just the crust (the blue line).

\subsection{Results: Perturbed Thermal Structure}
\label{sec: Perturbed Thermal Structure Results}

In order to treat the presence of a magnetic field as a perturbation on the heat flow of a spherically symmetric background, one must be in the regime whereby $\omega_B \tau (T) \ll 1 $. For temperatures typical of NSs in LMXBs ($T \sim 10^8 - 10^9$ K), OJ found this procedure is safe for internal \textit{crustal} magnetic fields strengths $B \lesssim 10^{12}$ G. Given that we are now extending our computational domain, we need to take care to remain in the perturbative regime in the core as well as the crust.  

\begin{figure*} \centering
    \centering
	\includegraphics[width=\textwidth]{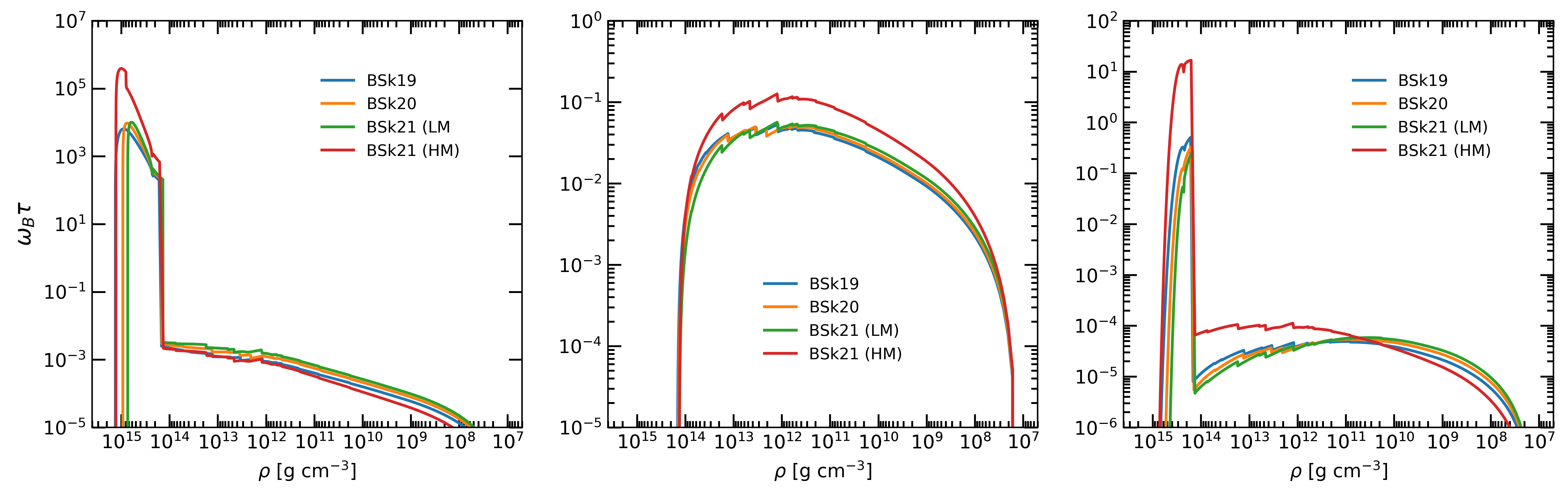}

    \caption{Magnetization parameter $\omega_B \tau (T)$ inside an accreting neutron star due to an internal toroidal magnetic field ($B = 10^{12}$ G) assuming each of  Eq. (\ref{eq:ToroidalField1}) with $R_{\rm{B, \, min}} = R_{\rm{IB}}$ (\textit{left}); Eq. (\ref{eq:ToroidalField1}) with $R_{\rm{B, \, min}} = R_{\rm{crust-core}}$ (\textit{centre}); and Eq. (\ref{eq:ToroidalField2}) (\textit{right}). Here $Q_{\text{imp}} = 1.0$, $Q_{\rm{S}} = 1.5$ MeV and $\dot{M} = 0.05 \dot{M}_{\text{Edd}}$.}
    \label{fig:Magnetisation Parameter}
\end{figure*}

As discussed in Section \ref{Perturbed Boundary Conditions and Method of Solution}, the thermal conductivity - which is proportional to $\tau$ (see Eq. (\ref{eq:ThermalConductivity})) - is a few orders of magnitude larger in the core than in the crust.  Before calculating the perturbed thermal structure, we plot in Fig. \ref{fig:Magnetisation Parameter} the magnetisation parameter $\omega_B \tau (T)$ for each of the three different magnetic field configurations discussed in Section \ref{The internal Magnetic Field}, for the four different BSk19-21 EoS models listed in Table \ref{tab:Model properties}.

We choose the maximum value of the magnetic field to be $B = 10^{12}$ G in all cases in order to compare with OJ. For the case where the magnetic field extends over the entire star (the left panel of Fig. \ref{fig:Magnetisation Parameter}, assuming the red curve in Fig. \ref{fig:field shape}) the magnitude of the magnetisation parameter in the \textit{core} is clearly in the regime $\omega_B \tau (T) \gg 1 $, far exceeding the condition in which our perturbation equations are valid. Notably, $\omega_B \tau (T)$ is greatest in the DUrca-cooled BSk21 high-mass model, an order of magnitude larger than any of the low-mass models. This is a result of the squared-temperature dependence of the scattering frequency $\nu$ in the NS core ($\tau \propto 1/ \nu \propto 1 / T^2$; see Table \ref{ThermCondTable} and Eqs. (\ref{eq:CoreRelaxationTime}) \& (\ref{eq:NeutronTau})). The right panel of Fig. \ref{fig:Magnetisation Parameter} (with $B$ described by the green curve in Fig. \ref{fig:field shape}) tells a similar story, where $\omega_B \tau (T) \gg 1 $ when DUrca is active. 

We shall therefore omit the high-mass BSk21 result from our subsequent calculations involving a \textit{core} magnetic field. In order to remain perturbative and recover the condition whereby $\omega_B \tau (T) \ll 1$ everywhere in the star, we restrict ourselves to magnetic field strengths $B \sim 10^{8}$ G when using Eq. (\ref{eq:ToroidalField1}) (when $R_{\rm{B, \, min}} = R_{\rm{IB}}$), and $B \sim 10^{9}$ G when using Eq. (\ref{eq:ToroidalField2}). These choices remain appropriate assumptions, given the inferences of the external magnetic field strength being $10^8 - 10^9$ G in LMXBs.

Note however that when the magnetic field is confined to the \textit{crust}, (the middle panel in Fig \ref{fig:Magnetisation Parameter}), we remain perturbative even when the magnetic field is assumed to be $10^{12}$ G, in agreement with \citet{Osborne_2020}. The largest magnetic field strengths we shall consider as `safe' for the crust-only configuration are $B \sim 10^{12}$ G when DUrca is active, and $B \sim 10^{13}$ G when DUrca is forbidden. 

These maximum values of the magnetic field to be considered `safe' in our calculations should be regarded as approximations, since the exact value of the magnetisation parameter within the NS depends not just on the assumed magnetic field strength, but also on the interior temperature profile (since $\tau \propto 1 / T^2$), which is a property of the background calculation. The results for $\omega_B \tau (T)$ shown in Fig. \ref{fig:Magnetisation Parameter} are specific to the choice $Q_{\text{imp}} = 1.0$, $Q_{\rm{S}} = 1.5$ MeV and $\dot{M} = 0.05 \dot{M}_{\text{Edd}}$ in the thermal background (Fig. \ref{fig:BSk Background Parameters}). The exact maximum allowed value of $B$ for any given thermal background will therefore vary slightly if one changes any of these quantities. This fact is accounted for in all subsequent calculations, and those few results we present that are out of the perturbative regime will be appropriately highlighted, such that any conclusions drawn from these results may be treated with correspondingly appropriate discretion. For example, in Section \ref{sec: The Resulting Deformations} we shall briefly consider results outside of the perturbative regime in order to discuss a potential proof-of-concept method to place upper limits on the strength of the magnetic field in the NS interior, which currently is poorly understood. 

Throughout this section, we shall present results for the temperature perturbations induced by the magnetic field in terms of the fractional temperature perturbation $\delta T / T$. We do this for convenience (since it is dimensionless) in order to make a more straightforward comparison between the asymmetry and the neutron star ellipticity $\varepsilon$ (i.e. the size of the mountain; Section \ref{sec: The Resulting Deformations}). Strictly speaking however, it is the magnitude of the perturbation $\delta T$ itself which is important. For example, a NS that is cold ($10^7$ K) or hot ($10^9$ K) would imply that $\delta T = 1\times 10^5$ K and $\delta T = 1\times 10^7$ K respectively for the same $1\%$ fractional asymmetry. Such a result could therefore be misleading, since only the latter would lead to significantly large thermal mountains (\citet{2002}). However, for the NS models that we consider here, the temperature in the crust is always $T \sim 10^8$ K, varying between models only by a factor of a few, even when the efficient DUrca process is permitted (see Fig. \ref{fig:BSk Background Parameters}). 

\begin{figure} \centering
    \centering
	\includegraphics[width=0.75\columnwidth]{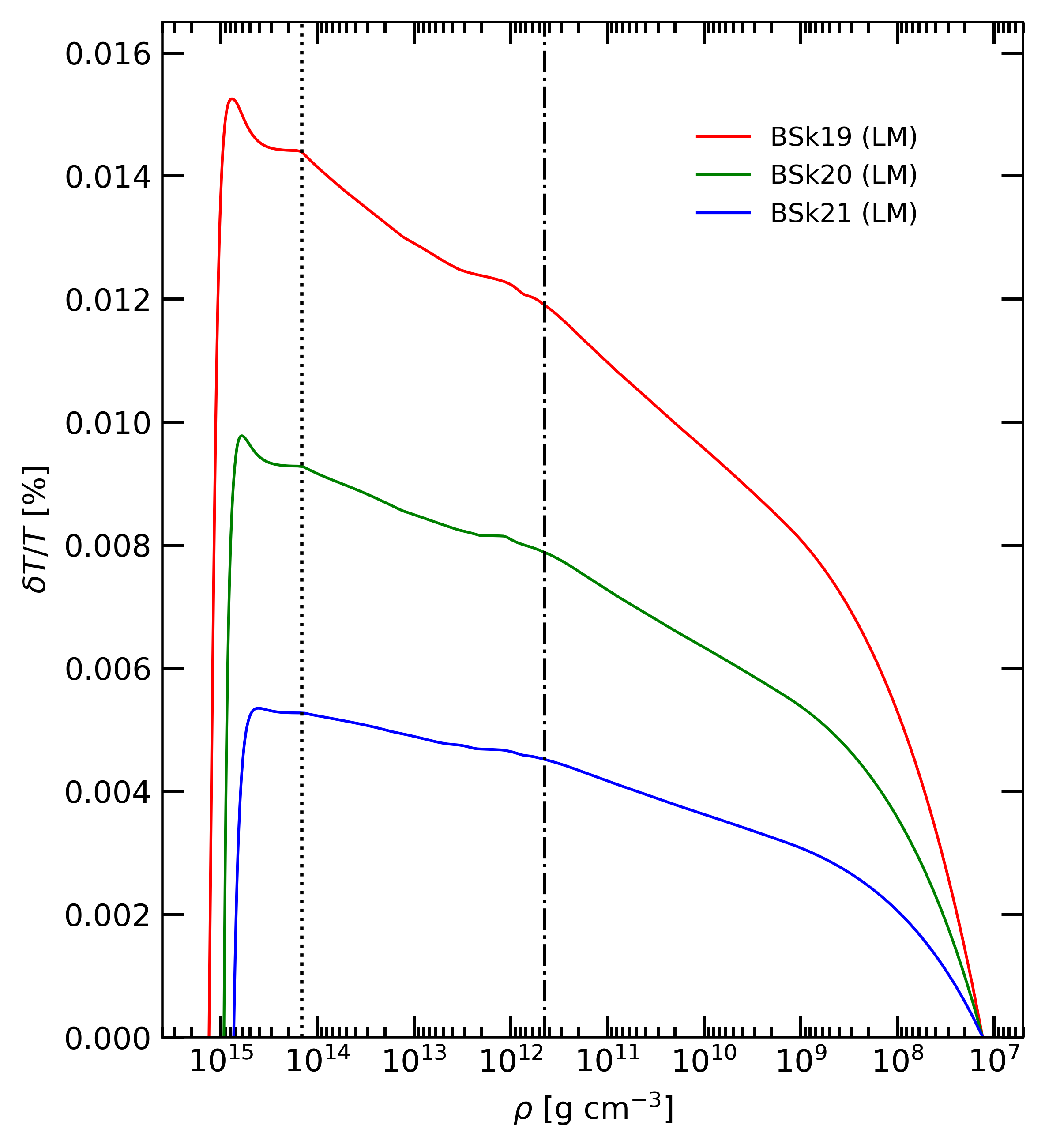}

    \caption{Magnitude of the fractional temperature perturbation $\delta T / T$ (as a percentage) inside an accreting neutron star due to a $B = 10^8$ G internal toroidal magnetic field (Eq. (\ref{eq:ToroidalField1}) with $R_{\rm{B, \, min}} = R_{\rm{IB}} \, ;$ Fig. \ref{fig:field shape}). The vertical dash-dotted line indicates the approximate location of the neutron drip point across the four models. The vertical dotted line shows the average location of the crust-core boundary for the three EoSs.  Here $Q_{\text{imp}} = 1.0$, $Q_{\rm{S}} = 1.5$ MeV and $\dot{M} = 0.05 \dot{M}_{\text{Edd}}$.}
    \label{fig:Perturbed temperature}
\end{figure}

In Fig. \ref{fig:Perturbed temperature} we plot $\delta T / T$ (as a percentage) of an accreting NS due to the presence of an internal toroidal magnetic field (assuming Eq. (\ref{eq:ToroidalField1}) with $R_{\rm{B, \, min}} = R_{\rm{IB}}$ - the red curve in Fig. \ref{fig:field shape}) for the three different BSk19-21 low-mass models listed in Table \ref{tab:Model properties}. The Python ODE solver \texttt{solve\_BVP} was again used for the integration, with the initial guess for the integration constructed by fixing the value of the temperature perturbation throughout the entire star to that of the results obtained by OJ ($\delta T \sim 10^{3}$ K).  We use the same model parameters as in Section \ref{Results: Background Thermal Structure}  (i.e. $Q_{\text{imp}} = 1$, $Q_S = 1.5$ MeV and $\dot{M} = 0.05 \dot{M}_{\text{Edd}}$). The internal toroidal magnetic field was set to have a magnitude $B = 10^8$ G at the midpoint of the star, extending over the full computational domain.  These calculations assume the outer boundary condition $\delta T_{\rm OB} = 0$ in order to compare with both UCB and OJ. When including the possibility of a core magnetic field, the fractional temperature perturbation in the inner crust ($10^{12} < \rho  < 10^{14}$ g cm$^{-3}$ ) of each model can be seen to lie in the region $\delta T / T \sim 0.005 - 0.013 \%$, and is largest at the point $r = (R_{\rm{B, \, max}} - R_{\rm{B, \, min}}) / 2$), corresponding to the location where the magnetic field is strongest.

In their analysis, OJ found typical values of the fractional temperature perturbation $\delta T / T$ to be a few times 10$^{-5}$ \% in the deep crust ($\rho \sim 10^{13}$ g cm$^{-3}$) when modelling only the crust of the NS with a $B = 10^9$ G magnetic field. Under this new analysis, we have modelled the crust and core of the NS, allowing for the possibility of the magnetic field to permeate the entirety of the star. In doing so, we have obtained values for the fractional temperature perturbation that are $\sim$ 2 orders of magnitude greater in the deep crust than the estimates obtained by OJ, even when the magnetic field is assumed to be an order of magnitude smaller.  

\begin{figure} 
    \centering
	\includegraphics[width=0.75\columnwidth]{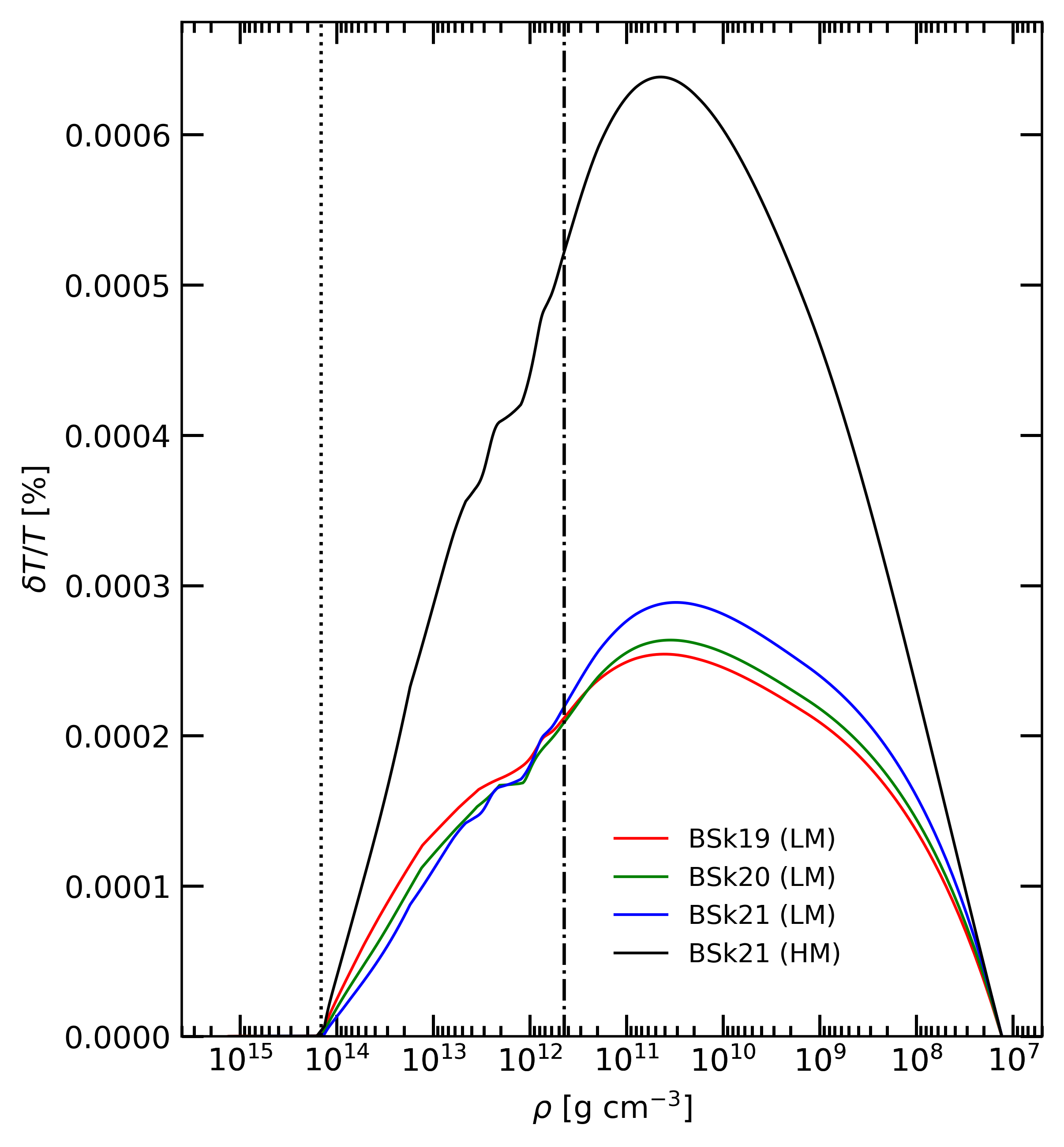}

    \caption{As for Fig. \ref{fig:Perturbed temperature}, but for a $B = 10^9$ G internal \textit{crustal} toroidal magnetic field (Eq. (\ref{eq:ToroidalField1}) with $R_{\rm{B, \, min}} = R_{\rm{crust-core}} \, ;$  Fig. \ref{fig:field shape}).  }
    \label{fig:Perturbed temperature Crust}
\end{figure}

Clearly, this is a substantial increase in the level of temperature asymmetry in the crust. To reconcile these findings, in Fig. \ref{fig:Perturbed temperature Crust} we modify our calculation. We use the same computational domain as the previous calculation (i.e. $R_{IB}$ and $R_{\rm OB}$ are unchanged) and continue to  use Eq. (\ref{eq:ToroidalField1}) for the functional form of $B$, but now confine the magnetic field to only the \textit{crust} of the NS. In this case, we have $R_{\rm{B, min}} = R_{\rm{crust-core}}$ and $R_{\rm{B, max}} = R_{\rm{OB}}$ (i.e. the blue curve in Fig. \ref{fig:field shape}). The internal toroidal magnetic field was chosen to have a maximum magnitude this time at $B = 10^9$ G, consistent with calculation performed by OJ. Additionally, since this calculation explores perturbations in only the crust of the NS, we also include the results for the high-mass BSk21 model, since $\omega_B \tau (T) = 0$ in the core when the field is removed there. 

Confining the magnetic field to the crust has a significant impact on both the magnitude of the temperature perturbation, as well as its distribution (as compared with Fig. \ref{fig:Perturbed temperature}). The shape of the curves in Fig. \ref{fig:Perturbed temperature Crust} match the results obtained by OJ markedly well (cf. their Fig. 4), also peaking at around $\rho \sim 10^{10}$ g cm$^{-3}$. Most interestingly, we see that the magnitude of the fractional temperature perturbation $\delta T / T$ is reduced to $\sim$ 10$^{-4} \%$ when the core magnetic field is removed, and therefore also similar in magnitude to the results obtained by OJ. 

To better understand this behavior, we refer back to Fig. \ref{fig:field shape}. When the inner boundary of the internal field is set to be at the crust-core interface, the field does not penetrate the core of the NS and the quartic nature of the field (see Eq. (\ref{eq:ToroidalField1})) forces it to drop off by many orders of magnitude over very narrow density ranges in both the deep crust ($R_{\rm{B, \, min}}$) and the top of the crust ($R_{\rm{B, \, max}}$) - see the blue curve in Fig. \ref{fig:field shape}. However, when the magnetic field \textit{is} allowed to penetrate the core of the NS, the gradient of the magnetic field strength varies much more slowly in the crust, particularly in the inner crust near the core/crust transition. Therefore, in the `crust-only' scenario, it is likely the case that the sharp decline of the magnetic field strength over a narrow density region suppresses the perturbations, since the source term Eq. (\ref{eq:SourceTerm}) (which is proportional to $\Psi_{lm}$) at each end of the integration is forced to become vanishingly small very quickly.

\begin{figure} 
    \centering
	\includegraphics[width=0.75\columnwidth]{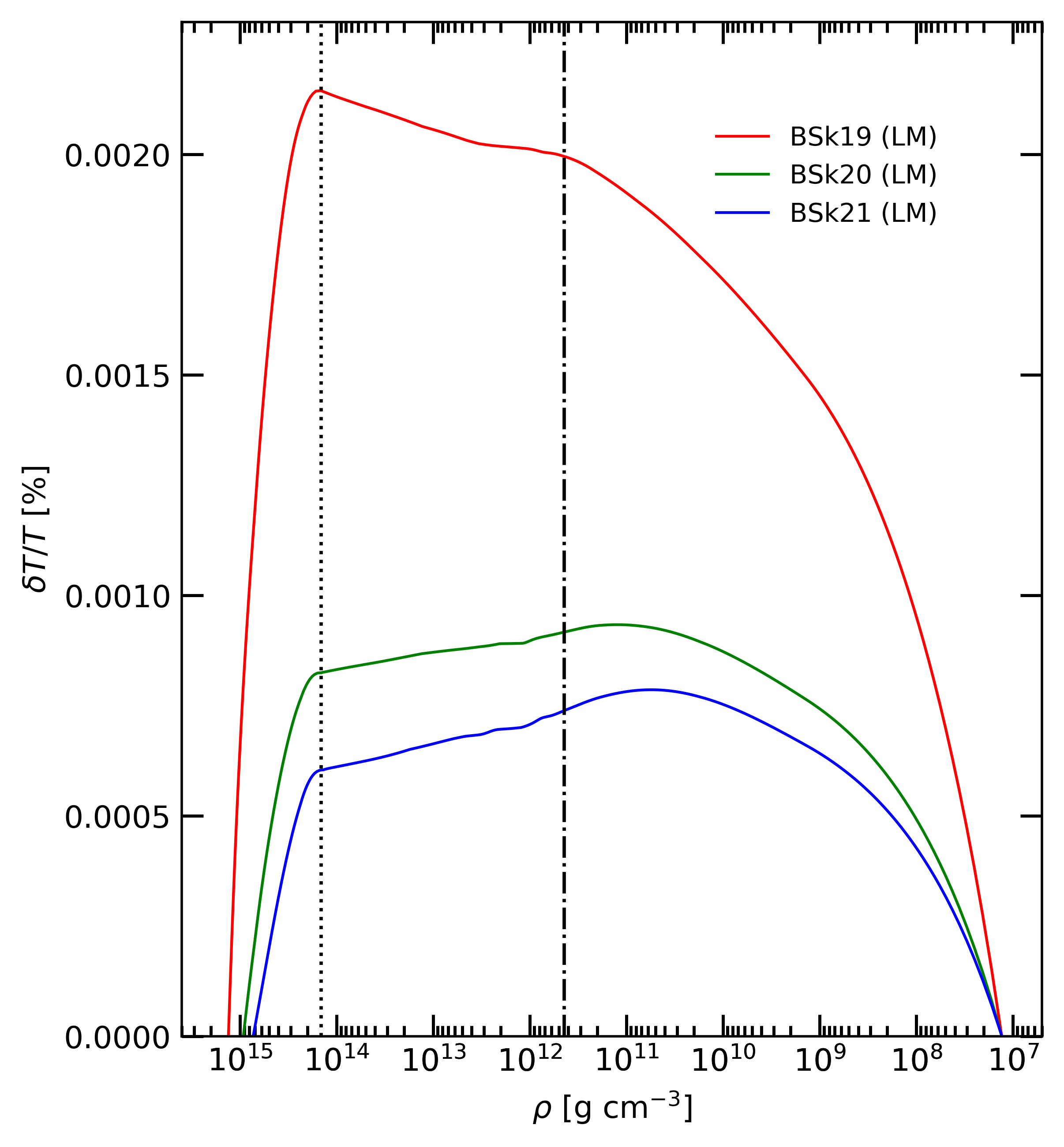}

    \caption{As for Fig. \ref{fig:Perturbed temperature}, but for a $B = 10^9$ G internal toroidal magnetic field that partially penetrates the core (Eq. (\ref{eq:ToroidalField2})).  }
    \label{fig:Perturbed temperature Crust and Core}
\end{figure}

The last case we need consider is the intermediate one, where the magnetic field only partially penetrates into the core, 
as described by Eq. (\ref{eq:ToroidalField2}) (the green curve in Fig. \ref{fig:field shape}).  We give our results in  Fig. \ref{fig:Perturbed temperature Crust and Core}, for the three low-mass models of Table \ref{tab:Model properties}.
As one might perhaps expect, being the intermediary case between Figs. \ref{fig:Perturbed temperature} and \ref{fig:Perturbed temperature Crust}, the fractional temperature asymmetry $\delta T / T$ in this case is larger than when the field is confined the crust, but smaller than when the field is allowed to extend over the entire star, lying in the region $\sim 10^{-4} - 10^{-3} \%$ depending on the EoS.

As was discussed in Section \ref{Perturbed Boundary Conditions and Method of Solution}, in addition to our $\delta T_{\rm OB} = 0$ outer boundary condition, we also consider two other conditions, corresponding to zero flux $U_{\rm OB} = 0$, and to the emission of blackbody radiation (Eq. (\ref{eq:deltaFOB})).  We do this in order to gauge the sensitivity of $\delta T$ in the inner crust to that of the choice of outer boundary condition.  

\begin{figure}
    \centering
	\includegraphics[width=0.75\columnwidth]{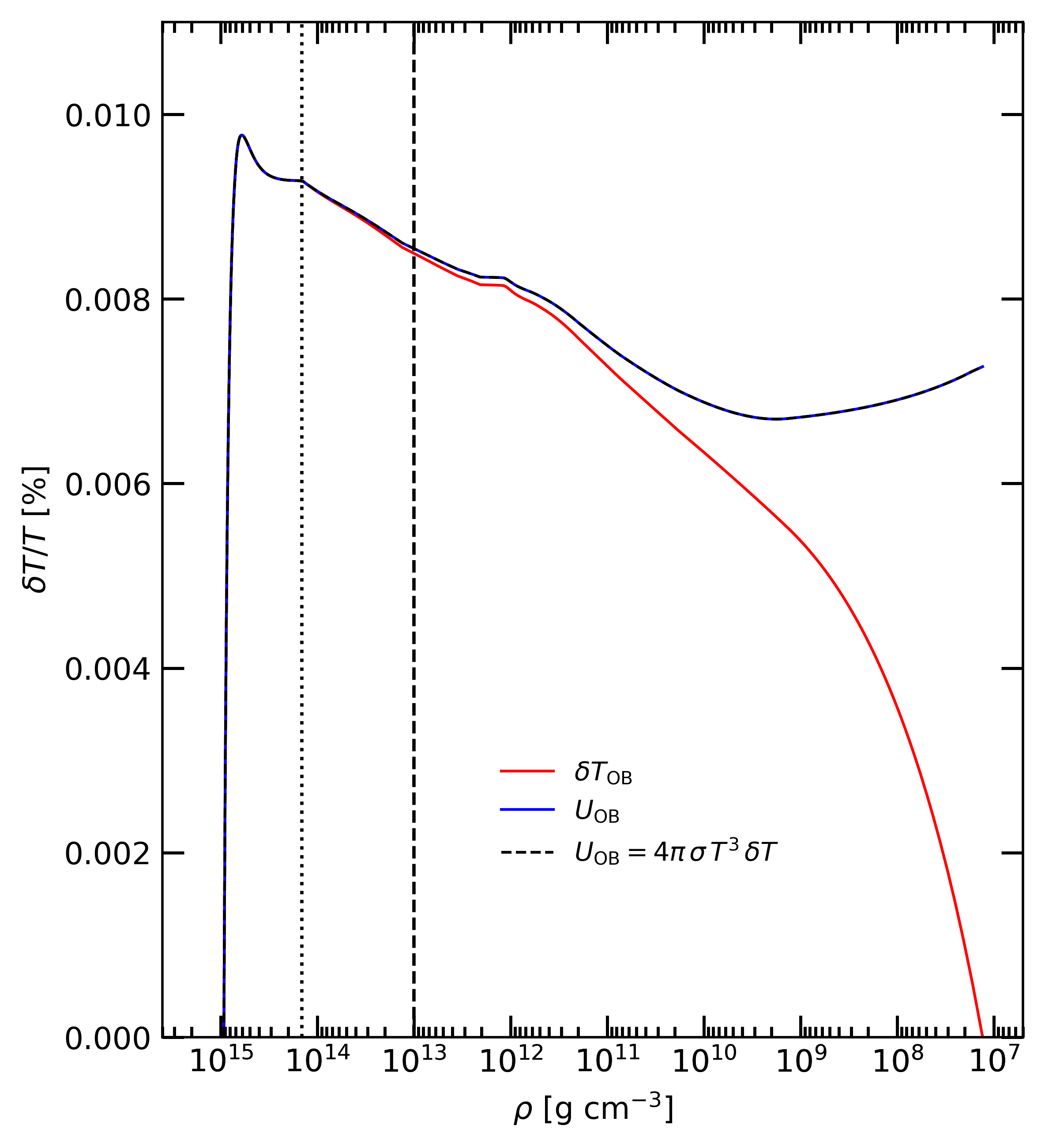}
    \caption{Magnitude of the fractional temperature perturbation $\delta T / T$ inside an accreting neutron star for the low-mass BSk20 EoS model in Table \ref{tab:Model properties} assuming the three different outer boundary conditions (indicated near the curves) due to a $B = 10^8$ G internal toroidal magnetic field (Eq. (\ref{eq:ToroidalField1}) with $R_{\rm{B, \, min}} = R_{\rm{IB}} \, ;$ Fig. \ref{fig:field shape}). The black-dashed and blue curves are close in magnitude, and therefore lie almost on top of one another. The vertical dotted line is the location of the crust-core transition. The vertical dashed line indicates the fiducial density $ \rho = 10^{13}$ g cm$^{-3}$, deep in the inner crust. Here $Q_{\text{imp}} = 1$, $Q_{\rm{S}} = 1.5$ MeV and $\dot{M} = 0.05 \dot{M}_{\text{Edd}}$. }
    \label{fig:Delta T Different Boundaries}
\end{figure}

The results are shown in Fig. \ref{fig:Delta T Different Boundaries}. For the cases of zero surface flux and blackbody radiation (the blue and black-dashed curves respectively), we find that the magnitude of the temperature perturbations are very similar to each other over the entire computational domain (never differing by more than $\sim 0.1 \%$). However, we see that allowing for a non-zero $\delta T_{\rm{OB}}$ causes $\delta T$ to diverge quickly from the $\delta T_{\rm{OB}} = 0$ result at densities $ \rho < 10^{12}$ g cm$^{-3}$ as one approaches the base of the ocean. At densities $ \rho > 10^{13}$ g cm$^{-3}$ however, the temperature perturbation can be seen to be largely insensitive to the outer boundary condition, with all three outer boundary prescriptions giving essentially the same results.  We will therefore continue to present results only for our preferred boundary condition of $\delta T_{\rm OB} = 0$, noting that other choices would have little impact on the temperature perturbation in the regions of the crust of importance for mountain building. We stress however that this applies strictly to temperature perturbations. Other quantities could be affected by one's choice of boundary conditions. The surface flux emanating from the crust (which is potentially observable; \citet{2002}), for example, would be intimately tied to the outer boundary condition. 

Additionally, an important aspect of the analysis done by OJ was to consider the effect of varying different properties of the background on the level of temperature asymmetry produced in the crust (cf. their Figs. 6 and 7). The thermal structure is a function of the many different components that enter the heat equation, some of which are properties of the equation of state (e.g. the critical temperatures), or constrained experimentally via observation (e.g. the shallow heating). 

\begin{figure} 
    \centering
	\includegraphics[width=0.75\columnwidth]{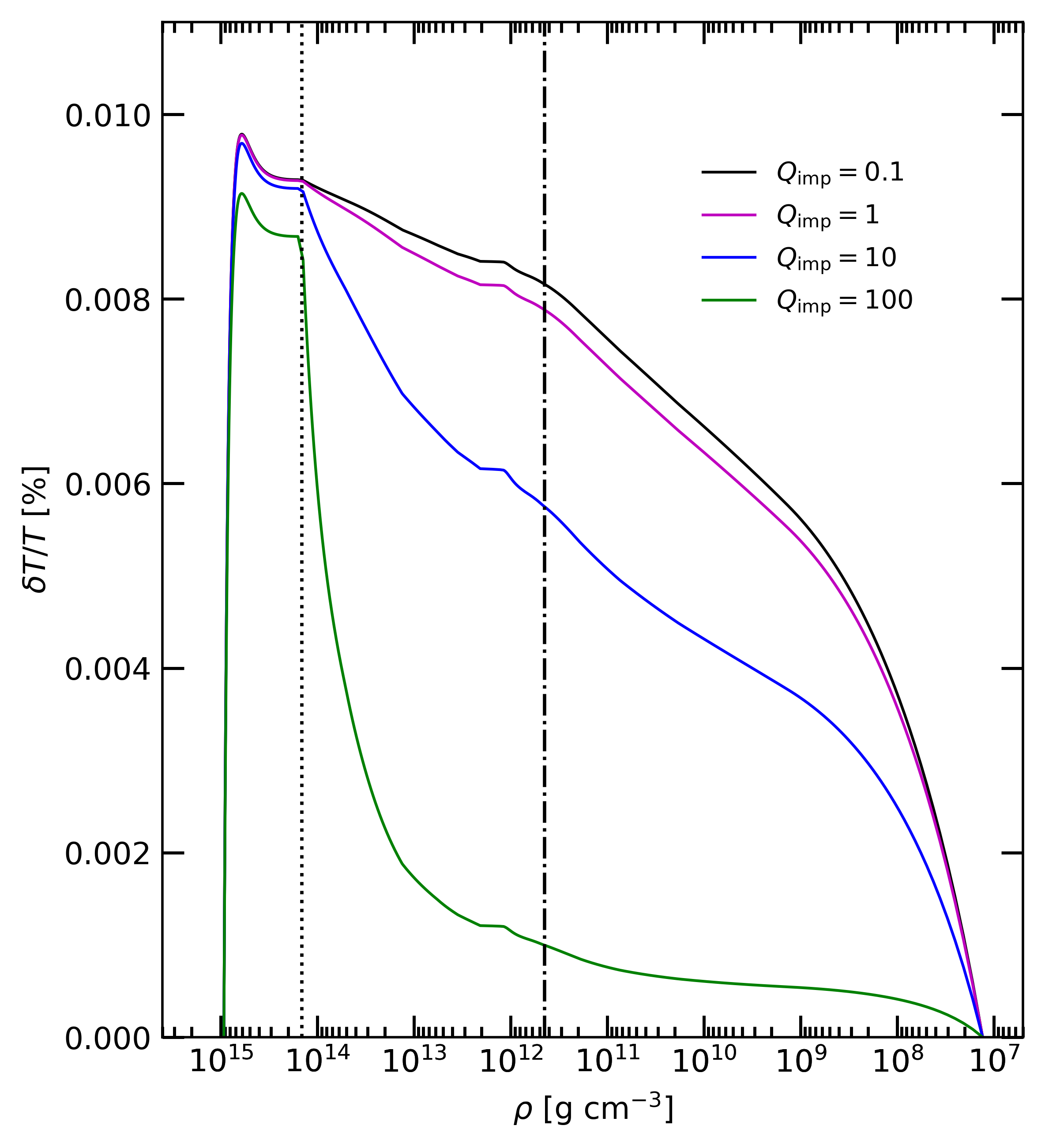}
    \caption{Magnitude of the fractional temperature perturbation $\delta T / T$ (as a percentage) inside an accreting neutron star for different values of the impurity parameter Q$_{\rm{imp}}$ (indicated near the curves) due to a $B = 10^8$ G internal toroidal magnetic field (Eq. (\ref{eq:ToroidalField1}) with $R_{\rm{B, \, min}} = R_{\rm{IB}} \, ;$ Fig. \ref{fig:field shape}). The vertical dotted line indicates the location of the crust-core transition, whilst the dash-dotted line is the location of neutron drip. Here $Q_{\rm{S}} = 1.5$ MeV and $\dot{M} = 0.05 \dot{M}_{\text{Edd}}$.}
    \label{fig:Vary Impurity}
\end{figure}

\begin{figure} 
    \centering
	\includegraphics[width=0.75\columnwidth]{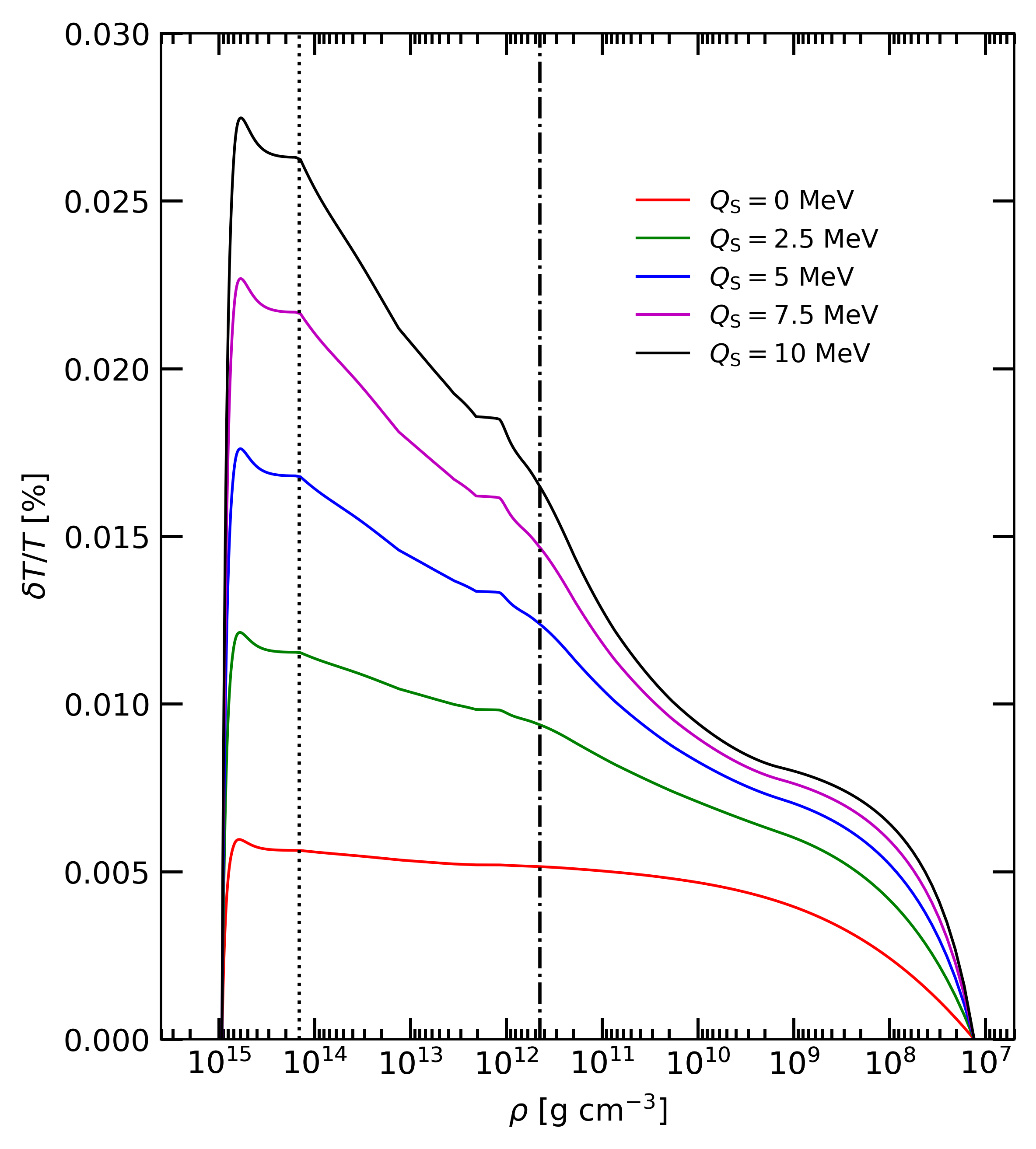}
    \caption{Magnitude of the fractional temperature perturbation $\delta T / T$ (as a percentage) inside an accreting neutron star for different values of the shallow heating parameter Q$_{\rm{S}}$ (indicated near the curves) due to a $B = 10^8$ G internal toroidal magnetic field (Eq. (\ref{eq:ToroidalField1}) with $R_{\rm{B, \, min}} = R_{\rm{IB}} \, ;$ Fig. \ref{fig:field shape}). The vertical dotted line indicates the location of the crust-core transition, whilst the dash-dotted line is the location of neutron drip. Here $Q_{\text{imp}} = 1.0$ and $\dot{M} = 0.05 \dot{M}_{\text{Edd}}$. }
    \label{fig:Vary Shallow}
\end{figure}

In Fig. \ref{fig:BSk Background Parameters} (and Figs. \ref{fig:Perturbed temperature} - \ref{fig:Perturbed temperature Crust and Core}), we assumed values of the shallow heating and impurity parameter to be 1.5 MeV and 1.0 respectively, in line with the respective average values of these parameters from observational estimates. However, to determine how sensitive our results are to these quantities, we show in Figs.  \ref{fig:Vary Impurity} and \ref{fig:Vary Shallow} how the temperature asymmetry $\delta T / T$ can vary with different values of the impurity parameter and shallow crustal heating term respectively (recall Figs. \ref{fig:BSk Impurity Background Parameters} \& \ref{fig:BSk Shallow Background Parameters} from Section \ref{Results: Background Thermal Structure}).

The effects of altering these quantities is quite modest, but still noteworthy. When we vary the impurity parameter, much like the background temperature profile we find that the temperature perturbation is relatively insensitive to $Q_{\rm{imp}}$ when $Q_{\rm{imp}} \lesssim 1$, but is reduced noticeably when $Q_{\rm{imp}} $ = 100; see Fig \ref{fig:Vary Impurity} . We interpret these findings in the following way, again considering the source term (Eq. (\ref{eq:SourceTerm})). Firstly, a larger $Q_{\rm{imp}}$ in the crust leads to steeper temperature gradients, as a larger thermal gradient is required to support the heat flux through the crust when the conductivity is reduced ($\kappa_{eQ} \propto 1 / Q_{\rm{imp}}$, see Eq. \ref{eq:ThermalConductivity} and Fig. \ref{fig:BSk Impurity Background Parameters}). But, when the conductivity is dominated by electron-impurity scattering ($Q_{\rm{imp}} >> 1$), our source term is

\begin{equation}\label{eq:SourceTerm2}
    S \propto  \kappa \, \tau \, \frac{dT}{dr} \sim \tau^2 \, \frac{dT}{dr} \sim \frac{1}{Q_{\rm{imp}}^2} \frac{dT}{dr},
    \vspace{3mm}
\end{equation}

\noindent
since $\kappa \propto \tau$ (see Eq. (\ref{eq:ThermalConductivity})). Therefore, although steeper temperature gradients act to increase the strength of the source term, the inverse-squared dependence of S on the value of $Q_{\rm{imp}}$ means the source term (and therefore the magnitude of the perturbations) is in fact made smaller when electron-impurity scattering is the dominant scattering mechanism in the accreted crust. 

For the shallow crustal heating term, the behavior of the temperature perturbations is more straightforward. As the amount of shallow heating is increased, so too is the magnitude of the perturbations in the inner crust; see Fig. \ref{fig:Vary Shallow}. This is because of the source term's dependence on the radial temperature gradient, that is steeper for increased amounts of shallow heating (see Fig. \ref{fig:BSk Shallow Background Parameters}).

Different LMXBs are observed to accrete at different (time-averaged) rates (see e.g. Table 2 of \citet{galloway_goodwin_keek_2017}). A larger accretion rate naturally leads to a greater amount of heat being deposited into the inner crust (see Eq. (\ref{eq:Heat})), resulting in a hotter crust with stepper temperature gradients, and therefore potentially larger temperature perturbations. The accretion rate is therefore also a natural parameter to explore to understand how the structure of the thermal background influences the strength of the induced temperature perturbations.

Additionally, whilst different LMXBs may accrete at different rates, the mass of the NS itself within the binary may also vary from system to system as well. As we have shown in Fig. \ref{fig:Perturbed temperature Crust}, the mass of the NS can be important when it comes to generating temperature perturbations, if the heavier NS is able to support the DUrca process.

\begin{figure*}
\centering
	\includegraphics[width=\textwidth]{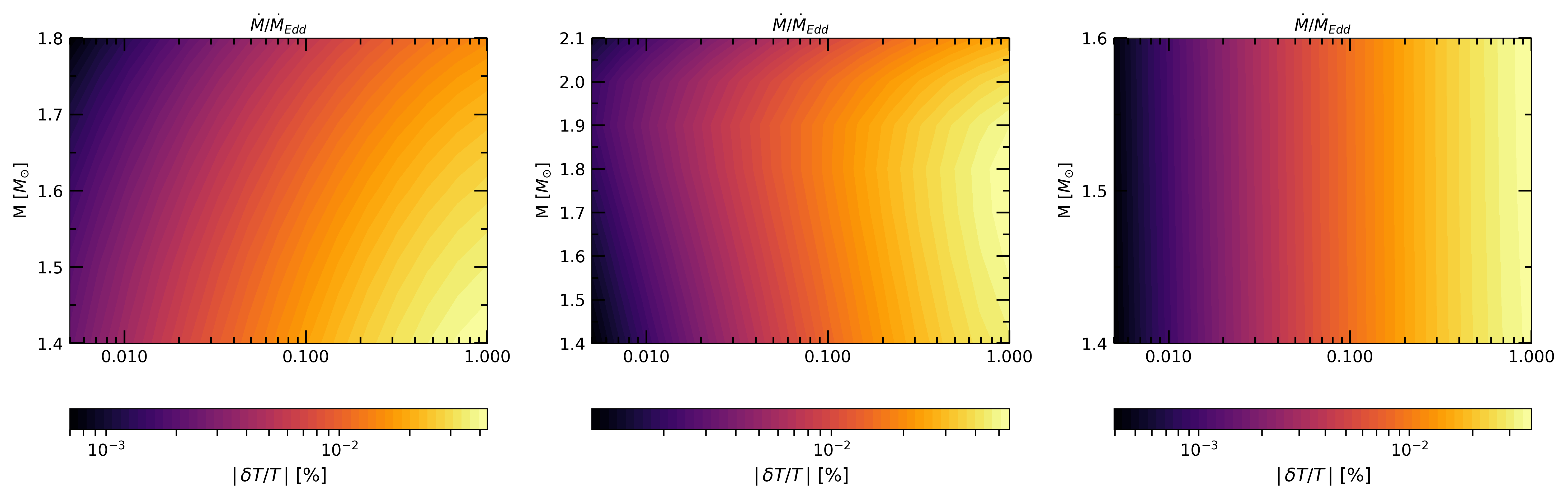}
    \caption{Magnitude of the fractional temperature perturbation $\delta T / T$ (as a percentage) as a function of both the accretion rate $\dot{M}$ and stellar mass $M$ for the BSk19 (\textit{left}), BSk20 (\textit{centre}), and BSk21 (\textit{right}) EoSs. All temperature perturbations are quoted at the fiducial density $10^{13}$ g cm$^{-3}$, assuming a $B = 10^8$ G internal toroidal magnetic field as described by Eq. (\ref{eq:ToroidalField1}) with $R_{\rm{B, \, min}} = R_{\rm{IB}}$. Here we set the impurity parameter $Q_{\rm{imp}} = 1$ and shallow heating term $Q_{\rm{S}} = 1.5$ MeV for all models.}
    \label{fig:Core Field Contours}
\end{figure*}

\begin{figure*}
\centering
	\includegraphics[width=\textwidth]{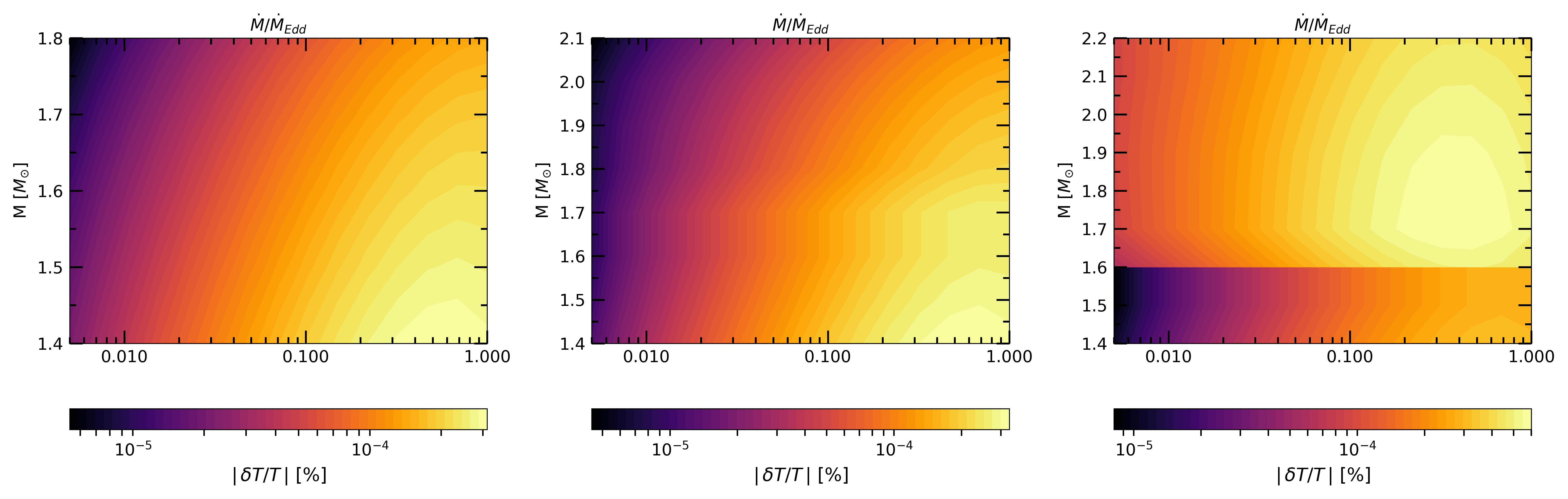}
    \caption{Same as Fig. \ref{fig:Core Field Contours}, but instead assuming a $B = 10^9$ G internal toroidal magnetic field this time described by Eq. (\ref{eq:ToroidalField1}) with $R_{\rm{B, \, min}} = R_{\rm{crust-core}}$. }
    \label{fig:Crust field Contours}
\end{figure*}

\begin{figure*}
\centering
	\includegraphics[width=\textwidth]{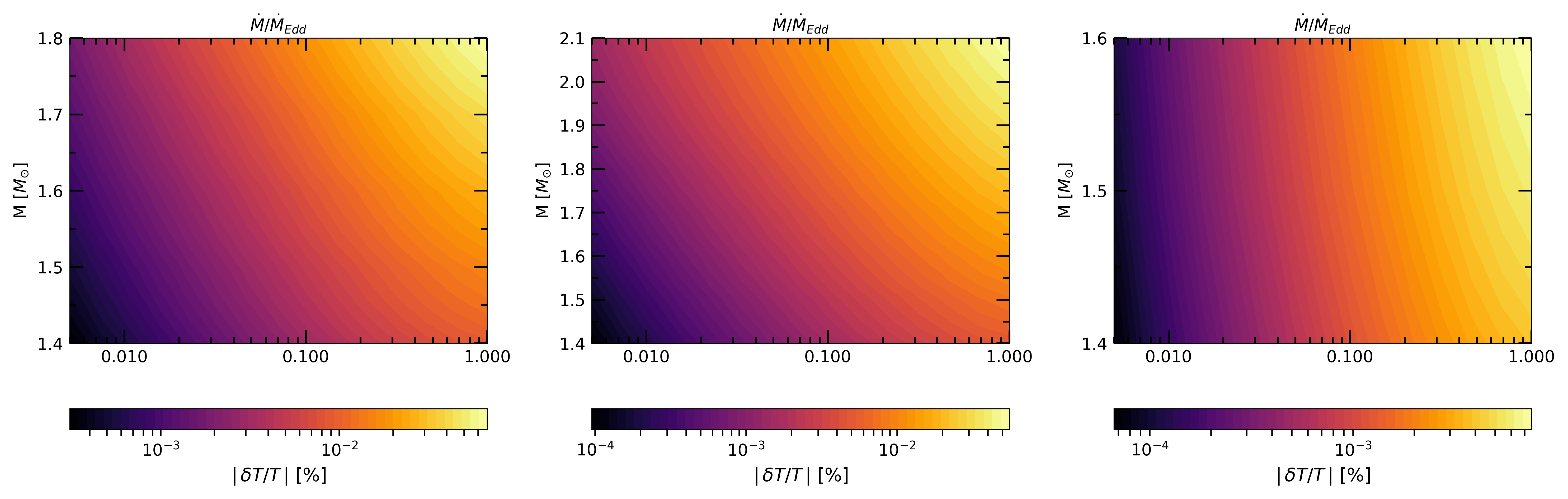}
    \caption{Same as Figs. \ref{fig:Core Field Contours} and \ref{fig:Crust field Contours}, this time assuming a $B = 10^9$ G internal toroidal magnetic field as described by Eq. (\ref{eq:ToroidalField2}). }
    \label{fig:Crust and core Field Contours}
\end{figure*}

To explore this issue, Figs. \ref{fig:Core Field Contours} - \ref{fig:Crust and core Field Contours} show the level of temperature asymmetry induced by the three different magnetic field configurations in Fig. \ref{fig:field shape} in the inner crust ($\delta T/ T$ quoted at $ \rho = 10^{13}$ g cm$^{-3}$) for a number of different NS models that accrete in the interval $0.005 - 1.0\dot{M}_{\rm{Edd}}$, with varying masses. We present the results of these calculations as contour plots Figs. \ref{fig:Core Field Contours} - \ref{fig:Crust and core Field Contours}, with the magnitude of the temperature perturbation in the inner crust for a particular EoS/accretion rate/stellar mass combination being represented by colourbars. For the magnetic field configurations that extend into the core of the NS (the red and green curves in Fig. \ref{fig:field shape}), we are limited to NS masses whereby DUrca is prohibited, so as to not enter the regime whereby $\omega_B \tau \gg 1$. Conversely, for the crust-only field, we consider masses varying between 1.4 M$_{\odot}$ and just below the TOV maximum of the particular BSk19-21 EoS model used (these being 1.86, 2.14, and 2.26 $M_{\odot}$ respectively; see Section \ref{section:Structure of an Accreted Crust}). As discussed above, in order to remain perturbative in all of our calculations, the maximum magnetic field strengths we consider in Fig. \ref{fig:Core Field Contours} is $B = 10^8$ G, and in Figs. \ref{fig:Crust field Contours} and \ref{fig:Crust and core Field Contours} we assume $B = 10^9$ G. 

\begin{figure*}
\centering
	\includegraphics[width=\textwidth]{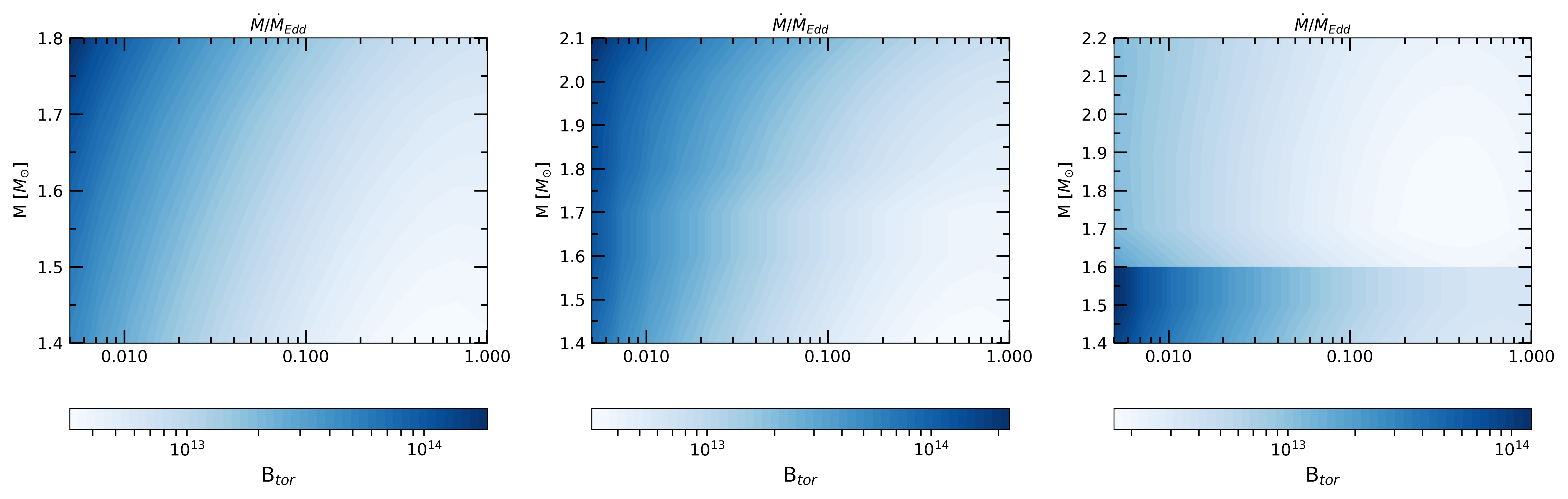}
    \caption{Magnitude of the \textit{crustal} magnetic field strength (as described by Eq. (\ref{eq:ToroidalField1}) with $R_{\rm{B, \, min}} = R_{\rm{crust-core}}$) required to reach $\delta T / T \sim$ 1\% for each of the temperature perturbations calculated in Fig. \ref{fig:Crust field Contours}. Due to the requirement $\omega_B \tau \ll 1$ however, we only consider trustworthy the results whereby the required magnetic field strength is $B \leq 10^{13}$ G (see Fig. \ref{fig:Magnetisation Parameter}), as magnetic fields stronger than this are outside of our perturbative regime.}
    \label{fig:Crust Magnetic field Contours}
\end{figure*}

The size of the temperature perturbations is determined largely by the accretion rate, getting larger as $\dot{M}$ increases, but does also have a slight dependence on the stellar mass as well. For the softer BSk19 EoS, a less-massive NS leads to a greater magnitude of $\delta T / T$ for the two magnetic field configurations described by Eq. (\ref{eq:ToroidalField1}), whilst a more massive NS leads to greater temperature perturbations for the magnetic field configuration described by Eq. (\ref{eq:ToroidalField2}). For the stiffer BSK20 model, there appears to be no clear trend for how the mass affects the size of the perturbations, with each field configuration having different masses that lead to their respective largest perturbations. 

Perhaps more interestingly, for the stiffest BSk21 model, a darker band for masses $ < 1.6 M_{\odot}$ is clearly visible for the case whereby the magnetic field is confined to the crust, with the largest temperature perturbations being produced even at low accretion rates when $M > 1.6 M_{\odot}$. The reason for this is the onset of DUrca at $1.6 M_{\odot}$ (see Section \ref{section:Neutrino Cooling}) leading to steeper background temperature gradients in the accreted crust. The outer boundary temperature is fixed by the accretion rate (see Eq. (\ref{eq:T_OB})) and therefore the same for a given accretion rate, irrespective of the mass of the NS. Consequently, a high-mass model whereby $M > 1.6 M_{\odot}$ must have a steeper temperature gradient in the crust in order to still satisfy the outer boundary condition set by the accretion rate. If one considers the source term in our problem, Eq. (\ref{eq:SourceTerm}), it can be seen that it is proportional to the radial background temperature gradient (dT/dr). Cooling of the NS from DUrca emission therefore increases the magnitude of the source term, in turn producing greater temperature asymmetries as compared to when DUrca is forbidden. 

For each of the calculations, the corresponding values of the Coulomb Parameter $\Gamma_{\text{Coul}}$ were also tracked to ascertain the physical state of the ions in the inner crust at each accretion rate. It was found that $\Gamma_{\text{Coul}}$ never fell below 300, much higher than the melting value of $\Gamma_{\rm m} \approx 175$, indicating the inner crust ($\rho > 10^{13}$ g cm$^{-3}$) remains solid even when approaching Eddington limit. This suggests that our temperature perturbations could, at least in principle, lead to some kind of elastic deformation of the NS crust. 

It is worth clarifying however that all of the calculations in Figs. \ref{fig:Core Field Contours} - \ref{fig:Crust and core Field Contours} were made assuming Q$_S$ = 1.5 MeV. For completeness, a calculation of the low-mass BSk20 EoS model was made assuming an accretion rate of $\dot{M} = 0.5\dot{M}_{\rm{Edd}}$ and accompanying shallow heating value of Q$_S$ = 10 MeV. This represents an extreme level of plausible crustal heating in the accreted crust, and serves to examine the state of the ions in the inner crust in this state of `maximal' heating. The Coulomb parameter at the fiducial density $ \rho = 10^{13}$ g cm$^{-3}$ in this scenario was found to be $\Gamma_{\text{Coul}} = 389$, and therefore we conclude that the crust remains solid even when the crust is - within observational constraints - maximally heated. 

In the analysis done by UCB, it was calculated that a fractional temperature asymmetry $\delta T / T \sim$ 1\% was required to produce a mass quadrupole (Q$_{22} \approx 10^{37 - 38}$ g cm$^{-2}$) large enough to balance accretion torques from the NS's companion. It is clear from Figs. \ref{fig:Core Field Contours} - \ref{fig:Crust and core Field Contours}, however, that even if the temperature asymmetry built from the magnetic field can translate into some kind of elastic deformation, none of our NS models produce perturbations large enough to create a mountain capable of generating GW emission at the required level. 

In Figs. \ref{fig:Core Field Contours} and \ref{fig:Crust and core Field Contours} we assumed magnetic fields strengths of $10^8$ and $10^9$ G respectively. These choices were made since they represent the largest magnetic fields strengths where our perturbation equations remain valid, and therefore the largest possible temperature asymmetries that may be achieved within the perturbative regime. However, magnetic fields in the core of the NS are probably larger than this value, even when taking account the inferences of the external field of LMXBs. It is therefore the case that temperature asymmetries formed in the deep crust as a result of a core magnetic field are also almost certainly larger than we have calculated here, but we cannot faithfully comment on their validity since our equations demand that $\omega_B \tau \ll 1$. 
In principle, this problem could be addressed with a non-perturbative calculation of the influence of a magnetic field on the heat conduction in the NS, but this is beyond the scope of what we have derived here. 

In Figure \ref{fig:Crust field Contours} however, we assumed a \textit{crustal} magnetic field strength of just $10^9$ G. Recall again the results of Fig. \ref{fig:Magnetisation Parameter} (specifically the middle panel), where we showed that in principle the magnetic field could be as as strong as $\sim 10^{13}$ G until our perturbation equations begin to break down. Since the perturbation equations are linear in $B$ (so long as one remains in the regime $\omega_B \tau \ll 1$), it is possible to simply re-scale our results to find the strength of the magnetic field that is required to produce a temperature asymmetry of $\sim$ 1\% for a given EoS/stellar mass and accretion rate combination. The result of such a calculation is shown in Fig. \ref{fig:Crust Magnetic field Contours}, which indicates that for \textit{strongly} accreting NSs, the minimum field strengths required to generate $\sim$ 1\% temperature asymmetries are in the region a few $ \times 10^{12}$ G. Fig. \ref{fig:Crust Magnetic field Contours} therefore suggests we require magnetic fields approx. 3 orders of magnitude larger than external inferences to produce temperature perturbations at the level that UCB estimated are required for GW torque balance to occur.

Additionally, though the results of Fig. \ref{fig:Crust Magnetic field Contours} also indicate that 1\% asymmetry may be achieved in \textit{weakly} accreting NSs as well, the required magnetic field strengths are of the order $ B > 10^{13}$ G. Such results are outside of the perturbative regime however, and so must be taken with caution. Although, such strong \textit{crustal} magnetic fields that far exceed $10^{12}$ G in accreting NSs are very unlikely to begin with, and therefore do not impact the overall picture regardless.

For comparison, OJ found that internal field strengths $\sim$ 10$^{13}$ G (with slight variations depending on the choice of a normal or superfluid core and the presence of some additional shallow heating) were required to produce temperature asymmetries at the percent level for their `crust-only' calculation. However, due to their more simplistic calculation of the crust however, magnetic field strengths at this level were enough to push them out of the perturbative regime.

\section{The Resulting Deformations}
\label{sec: The Resulting Deformations}

To sum up our results from the previous section, we have shown that for the majority of our NS models, we are unable to reach the $1\%$ temperature asymmetry in the deep crust that was required by UCB to generate significant gravitational wave emission. In most cases, we are limited by perturbative nature of our approach, whereby the condition $\omega_B \tau (T) \ll 1 $ everywhere in the star must be satisfied in order for Eqs. (\ref{eq:Perturbed Temp ODE}) and (\ref{eq:Perturbed U ODE}) to be valid. We found that when the magnetic field is allowed to permeate the core, the level of asymmetry in the deep crust ($10^{13}$ g cm$^{-3}$) is $\sim 2$ orders of magnitude larger than when the field is forced to be confined to the crust (compare Figs. \ref{fig:Perturbed temperature} and \ref{fig:Perturbed temperature Crust}). However, since the thermal conductivity is much larger in the core than in the crust, the condition $\omega_B \tau (T) \ll 1 $ is broken for magnetic fields greater than $\sim 10^8$ G (Fig. \ref{fig:Magnetisation Parameter}).

In reality, if the magnetic field \textit{is} able to permeate the core of the NS (recall the discussion around the Meissner-Ochsenfeld effect in Section \ref{The internal Magnetic Field}), then it is highly likely it will be greater in magnitude than the $10^8$ G we are limited to here.

The picture is also further complicated by the fact that the Meissner effect is a property that extends to only Type I superconductors. Paired protons in the NS interior may also form a Type II superconductor \citep{Baym1969SuperfluidityIN}, whereby rather than be expelled, the magnetic field is expected to be confined into isolated vortices. In fact, \citet{2008Akgun} have even shown that quadrupolar distortions from axisymmetric toroidal magnetic fields in Type II superconducting matter are possible. It is however worth bearing in mind that the deformations proposed by \citet{2008Akgun} are not a result of the same mechanism described here, but rather adds credence to the idea that a core magnetic field may be exploited to produce mountains. To this end, a reformulation of our mechanism in a non-perturbative regime could prove worthwhile. To reiterate, our perturbative approach can achieve asymmetries of the order $10^{-2} \%$ when the magnitude of a core magnetic field is just $10^8$ G. 

Given the uncertainties surrounding the feasibility of a core magnetic field however, we also stress the results we obtain from our `\textit{crust-only}' field configuration (Figs. \ref{fig:Perturbed temperature Crust} and \ref{fig:Crust field Contours}). Although the expected asymmetry from a crust-confined $B = 10^9$ G magnetic field is of the order $10^{-4} \%$, the results of Fig. \ref{fig:Crust Magnetic field Contours} show that if the strength of the magnetic field can exceed slightly above $10^{12}$ G, then the required asymmetry of $1\%$ may still be achieved. This is an order of magnitude smaller than was required by OJ to achieve the same level of asymmetry, and, crucially, is still valid even in the perturbative regime (Fig. \ref{fig:Magnetisation Parameter}). 

Up to now, we have discussed presented our results for the magnitude of the temperature asymmetry. The size of NS mountains, however, are described in terms of the neutron star ellipticity, defined as (e.g. \citet{2005Owen})

\begin{equation}
    \varepsilon = \frac{(I_{xx} - I_{yy})}{I_{zz}} = \sqrt{\frac{8\pi}{15}}\frac{Q_{22}}{I},
\end{equation}

\noindent
where $I$ is the moment of inertia (usually taken to be $10^{45}$ g cm$^{2}$) and $Q_{22}$ is the NSs mass quadrupole moment. Based on the calculations by UCB for the size of the mass quadrupole generated from temperature asymmetries in a single capture layer, OJ calculated a simple fitting formula for the neutron star ellipticity as (cf. their Eq. 2),

\begin{equation} \label{eq: EllipticityScale}
    \varepsilon \sim 5\times 10^{-8} \, \Biggl[ \frac{\delta T / T}{1\%} \Biggr] \, .
\end{equation}

\noindent
Using this simple scaling (evaluated at our fiducial density $ \rho = 10^{13}$ g cm$^{-3}$), one expects that a fractional temperature asymmetry at the percent level in the inner crust produces an ellipticity of $ \sim 5 \times 10^{-8}$, irrespective of the assumed origin of the perturbations. 

For the specific case of a magnetic field that is confined to the accreted crust (Eq. (\ref{eq:ToroidalField1}) with $R_{\rm{B, \, min}} = R_{\rm{crust-core}}$), the results of Fig. \ref{fig:Crust Magnetic field Contours} represent the magnetic field strengths required to produce an ellipticity of this size, generating a mountain large enough to balance accretion torques with the emission of gravitational waves. 

To place these results in to context, reconsider the results from the recent continuous GW search by \citet{Abbott_2021}, that constrained the ellipticity on the AMXP IGR J00291+5934 to be no more than $\varepsilon^{95\%} = 3.1 \times 10^{-7}$. In lieu of continuous GW detection, the results obtained by \citet{Abbott_2021} are still useful. Namely, the upper limit on the ellipticity of IGR J00291+5934 can be combined with our results to tentatively place an upper limit on the strength of the internal magnetic field within the NS. 

Since the temperature asymmetry (and therefore the ellipticity) have some dependence on the accretion rate $\dot{M}$, we can find an upper limit on the strength of the internal magnetic field by recalculating our temperature perturbations assuming an $\dot{M}$ specific to IGR J00291+5934.  The star's mass, however, is not known, so we present results for both a 1.4 $M_{\odot}$ and 2.1 $M_{\odot}$ star. As we have computed only steady-state solutions, we interpret our parameter $\dot{M}$ as the time-averaged accretion rate.   This also ensures consistency in our use of Eq. (\ref{eq:T_OB}) that requires burning at the base of the H/He layer to be stable. We estimate the average mass accretion rate in IGR J00291+5934 using data taken from Table 2 of \citet{De_Falco_2017}, for which the time-averaged X-ray flux $\langle F_x \rangle$ is given for four independent outburst/quiescence cycles that occurred in the years 2004 - 2015. Using this information, we assume the mass accretion rate to then be of the form 

\begin{equation} \label{eq: Mass Accretion}
    \dot{M} = \langle F_x \, \rangle \frac{4 \pi R d^2}{GM} \, . 
\end{equation}

\noindent
where $d$ is the distance of IGR J00291+5934 from Earth. We averaged the results of \citet{De_Falco_2017} to produce one value of $\langle F_x \rangle$ over all four outburst/quiescence cycles. At a distance of 3 kpc, we estimate the accretion rates to be 

\begin{equation} \label{eq: low-mass accretion rate}
    \dot{M}_{1.4 M_{\odot}} = 1.8 \times 10^{-12} M_{\odot} \, {\rm yr}^{-1} \approx 9 \times 10^{-5} \,  \dot{M}_{\rm{Edd}} \, ,
\end{equation}
\vspace{-3mm}
\begin{equation}\label{eq: high-mass accretion rate}
    \dot{M}_{2.1 M_{\odot}} = 1.2 \times 10^{-12} M_{\odot} \, {\rm yr}^{-1} \approx 5 \times 10^{-5} \,  \dot{M}_{\rm{Edd}} \, .
\end{equation}

\noindent
for a 1.4 and 2.1 $M_{\odot}$ star respectively. These accretion rates are appreciably smaller than those we considered in Section \ref{sec: Perturbed Thermal Structure Results}, and therefore we re-calculate the size of the temperature perturbation $\delta T / T$, again quoting the result at the density $ \rho = 10^{13}$ g cm$^{-3}$ in the inner crust specific to these values of $\dot{M}$. In doing so, we find that

\begin{equation}
    (\delta T / T) _{1.4 M_{\odot}} \sim 3.3 \times 10^{-8} \, B_8 \, \Rightarrow \, \varepsilon _{1.4 M_{\odot}} \approx 1.7 \times 10^{-13} \, B_8 \, 
\end{equation}

\vspace{-3mm}

\begin{equation} \label{eq: AMXP Durca DeltaT}
    (\delta T / T) _{2.1 M_{\odot}} \sim 4.3 \times 10^{-5} \, B_8 \, \Rightarrow \, \varepsilon _{2.1 M_{\odot}} \approx 2.2 \times 10^{-10} \, B_8 \, ,
\end{equation}
\vspace{1mm}

\noindent
assuming the BSk21 EoS and the field configuration given by Eq. (\ref{eq:ToroidalField1}) with $R_{\rm{B, \, min}} = R_{\rm{IB}}$. Using these results, it is possible to calculate the upper limit on the strength of the internal magnetic field required to produce the ellipticity constrained by \citet{Abbott_2021} as

\begin{equation}
    B^{\varepsilon^{95\%}}_{1.4 M_{\odot}} = \Biggl[ \frac{3.1\times10^{-7}}{1.7\times 10^{-13}} \Biggr] B_8 \approx 1.9 \times 10^{14} \, \text{G} \, ,
\end{equation}
\vspace{-3mm}
\begin{equation} \label{eq:IGR J00291+5934 high mass}
    B^{\varepsilon^{95\%}}_{2.1 M_{\odot}} = \Biggl[ \frac{3.1\times10^{-7}}{2.2\times 10^{-10}} \Biggr] B_8 \approx 1.4 \times 10^{11} \, \text{G} \, .
\end{equation}
\vspace{1mm}

\noindent
We do stress however that these upper limits must be taken with caution. In reference to Eq. \eqref{eq:Delta F}, it is the case that such large magnetic fields may predict the hall component $\kappa^{\wedge}$ of the field to be many orders of magnitude greater than the parallel component $\kappa^{\parallel}$. Such a result is clearly nonphysical (as can be seen by comparing  $\kappa^{\wedge}$ and $\kappa^{\parallel}$ in Eq. \eqref{eq:Concuctivity Components}), and is a consequence of the fact that the condition $\omega_B \tau (T) \ll 1 $ is broken for core magnetic fields greater than $\sim 10^8$ G (Fig. \ref{fig:Magnetisation Parameter}).

Though, given that methods to probe the structure of internal magnetic fields are so scarce, we present this method at least as a proof of concept to constrain the strength of the internal field. A calculation such as this is further motivation to return to a reformulation of our mechanism in the non-perturbative regime in the future, which would be able to explore magnetic fields in excess of the current limit of $\sim 10^8$ G. 

Regardless, in the low-mass case where DUrca is forbidden, the constraint on the upper limit of the magnetic field is not particularly informative: the upper limit is 5-6 orders of magnitude larger than externally inferred field strength. Our results for the high-mass star however, are more interesting: we obtain an upper limit on the internal toroidal magnetic field $\sim 10^{11}$ G. We must re-iterate that such magnetic fields push us far out of the perturbative regime however ($\sim 4$ order of magnitude; Fig. \ref{fig:Magnetisation Parameter}), and therefore must be viewed cautiously. 

Additionally, these results are also very much conditional on the accretion rate of the LMXB (and the mass of the NS; Figs. \ref{fig:Core Field Contours} - \ref{fig:Crust and core Field Contours}), and therefore the upper limits we present here are themselves limited by the low-level accretion of IGR J00291+5934. Analysis by \citet{Heinke_2009} also indicates that quiescent spectrum of IGR J00291+5934 is in fact consistent with the \textit{standard cooling model}, suggesting that it is a low-mass star ($M \lesssim 1.6 M_{\odot}$), therefore disfavouring the presence of DUrca cooling. We would therefore benefit greatly from further targeted searches towards millisecond accreting pulsars, with the hope the NS ellipticity can be accurately constrained for LMXBs that are accreting at much greater rates than IGR J00291+5934 (and those that exhibit evidence of enhanced cooling), as these systems would be more favourable for sustaining a thermal mountain. 

Lastly, we can also make one more comparison of our results of the ellipticities produced via the magnetic field for a \textit{thermal} mountains, to those produced explicitly by a ``conventional''  magnetic mountain, where Lorentz forces directly produce the distortion (e.g. \citet{1996A&A...312..675B, 2009MNRAS.395.2162L}). In particular, for a NS with a superconducting core, OJ estimated (cf. their Eq. (52), and also Eq. (2.6) of \citet{PhysRevD.66.084025}) that the ellipticity of a magnetic mountain is

\begin{equation}
    \varepsilon \sim \frac{B H_{c1} R^4}{GM^2} \approx 2 \times 10^{-13} \, B_8 \, ,
\end{equation}

\noindent
where $H_{c1}$ is the known as the `first critical field' ($\sim 10^{15}$ G). Since our results depend on the mass accretion rate, it is difficult to make a meaningful comparison between this ellipticity, and those obtained from our thermal mountains.
However, if the accretion rate is quite high, then Fig. \ref{fig:Crust and core Field Contours} suggests that our mechanism can produce much larger mountains ($\sim 2$ orders of magnitude) than those obtained via Lorentz forces for an equivalent-strength field when it is not confined to the crust and can penetrate (at least some of) the core.

\section{Summary and Conclusions}
\label{sec: Conclusions}

We have calculated the level of temperature asymmetry induced in the solid crust of an accreting neutron star due to the presence of an internal magnetic field. Our approach has expanded upon the work initially laid out in OJ, by extending the computational domain of the calculation to model the entire star (not just the crust), and allowing for the possibility of the magnetic field to penetrate the NS core. By exploiting the weak nature of magnetic fields of NSs in LMXBs, we have been able to treat the addition of a magnetic field onto a non-magnetic spherically symmetric background perturbatively. This has been shown to be safe for magnetic fields $\lesssim$ 10$^{8}$ G in the core and $\lesssim$ 10$^{13}$ G in the crust (Fig. \ref{fig:Magnetisation Parameter}), with the large disparity arising from the much higher thermal conductivity of the core compared to the crust.

In OJ, \textit{crustal} magnetic fields strengths in excess of 10$^{13}$ G were required to produce temperature perturbations at the percent level (the minimum amount of asymmetry required by UCB to produce significant GW emission), pushing outside the regime where their method was valid. We contrast those results of OJ with the ones presented here (Fig. \ref{fig:Core Field Contours} - \ref{fig:Crust and core Field Contours}). Under this new analysis, the presence of a magnetic field in at least some region of the NS core raises the expected level of temperature asymmetry in the deep crust by approx. 2-3 orders of magnitude (depending on the mass of the NS and the accretion rate). To provide confidence in these new results, we have also shown that when we again restrict the magnetic field to just the crust of the NS (whilst keeping the computational domain the same), the magnitude of the temperature perturbation reduces to the same order of magnitude as those found in OJ (compare our Fig. \ref{fig:Perturbed temperature Crust} with their Fig. 4). 

In extending the computational domain of the calculation, we have been able to introduce a more realistic description of the un-magnetised thermal background of the star. We have included all relevant neutrino emission and thermal transport mechanisms (Tables \ref{NeutrinoCoreTable} \& \ref{ThermCondTable}) applicable to a NS composed of standard \text{npe$\mu$} matter, in line with the current understanding of cooling theory of neutron stars in LMXBs (e.g. \citet{10.1093/mnras/sty825}). For completeness, we have also implemented a self-consistent calculation on the level of superfluidity to account for the suppression of both thermal transport and neutrino emission mechanisms. 

We have presented our results for three different modern accreting equations of state, provided by \citet{Fantina, Fantina_2022} (and \citet{2013A&A...560A..48P} for calculations associated with the NS core). One possibility we have not been able to properly explore, given the perturbative nature of our approach, is that the presence of the extremely efficient DUrca process (in NSs assuming the BSk21 EoS heavier than 1.6 $M_{\odot}$) may lead to much larger temperature asymmetries due to enhanced thermal conductivity which raises the magnetisation parameter - which is proportional to the source term (Eq. (\ref{eq:SourceTerm})) - by many orders of magnitude (Fig. \ref{fig:Magnetisation Parameter}). This means that the same level of asymmetry may be achieved when DUrca is active for a comparatively weaker magnetic field than when DUrca is forbidden, since the magnetisation parameter (Eq. \ref{eq:MagParameter}) is proportional to both the scattering relaxation time $\tau$ (which is itself proportional to the thermal conductivity $\kappa$; Eq. (\ref{eq:ThermalConductivity})) as well as the strength of the magnetic field $B$. 

Using our perturbative calculations, we found that the presence of DUrca  limits the strength of the magnetic field required to produce the upper-limit on the ellipticity of the AMXP IGR J00291+5934 to be $\sim 10^{11}$ G, but such a strong magnetic field pushes us out of the perturbative regime, and thus must be taken cautiously. In order to asses the importance of DUrca properly, a reformulation of our mechanism non-perturbatively could be used to constrain the (largely unknown) internal magnetic field strength of neutron stars, where the the condition $\omega_B \tau \ll 1$ need not be required. 

For all our background models, we have also computed the Coulomb parameter $\Gamma_{\rm{Coul}}$, which tracks the state of the ions in the crust at a given density. It is clear that the value of the temperature depends strongly on the accretion rate (and shallow heating to a lesser extent). The heating associated with these two quantities naturally increases the temperature of the crust, which will melt if $\Gamma_{\rm{coul}} < 175$. If the inner crust ($\rho \sim 10^{13}$ g cm$^{-3}$) is molten, a mountain cannot be created since the resultant fluid would not be capable of supporting shear stresses. However, we find that $\Gamma_{\rm{coul}} \geq 175$ is obtained for all our NS models at densities greater than $\sim 10^{10}$ g cm$^{-3}$ (Figs. \ref{fig:BSk Background Parameters} - \ref{fig:BSk Shallow Background Parameters}), indicating the inner crust is always solid and therefore (in principle) capable of supporting elastic strains.

We choose to quote most of our results at this fiducial density of $ \rho = 10^{13}$ g cm$^{-3}$ since it lies near three pycnonuclear reactions in the inner crust of the NS (see Tables A.3 - A.1 from \citet{Fantina}). This was a natural choice, for two reasons: (i) it is in this density region in which UCB found the largest mass quadrupole ($Q_{22}$) was generated, and (ii) these two reactions account for $\sim80\%$ of the total heat released from DCH in the entire crust.  Recall also that this location was found to be deep enough into the crust such that the detailed implementation of the outer boundary condition was not important (Fig. \ref{fig:Delta T Different Boundaries}). 

In comparison to the results obtained by OJ, we find these new results to be encouraging. Though we have been unable to produce the $1\%$ temperature asymmetry required by \citet{2002} for the majority of our NS models in order to reach the torque-balance limit, we have shown that $1\%$ temperature asymmetry within the accreted crust may be achieved for strongly accreting NSs, if the strength of the \textit{crustal} magnetic field can exceed slightly above $10^{12}$ G. We have also shown that asymmetries of the order $\sim 10^{-2}\%$ can be produced from \textit{core} magnetic fields of just $10^8$ G. Given core magnetic fields are actually likely many orders of magnitude larger than this value, it is not inconceivable that a non-perturbative calculation with stronger magnetic fields could yield temperature asymmetries at the percent level or above. 

There are also a number of caveats in our current model that should not be overlooked. Namely, (i) The use of a very simple outer boundary condition in the perturbed problem, (ii) a lack of an understanding of the magnetic field evolution itself in old accreting neutron stars, and the feasibility of a core magnetic field with the restriction of its strength to $< 10^8$ G, (iii) a want for a fully self-consistent calculation of the elastic response of the crust to our temperature perturbation (and hence a reliable magnitude of the NS ellipticity), (iv) improved microphysics, and (v) the omission of the possibly of heat conduction via superfluid phonons in the inner crust. 

In the first case, one solution could be to extend the model to include the low density ocean and the associated heat release from nuclear burning (see the discussion in Sec. \ref{Perturbed Boundary Conditions and Method of Solution}). Such a calculation was not performed here, however, since we are interested primarily in temperature perturbations generated in the deep crust, as this is where UCB calculated most of the mass quadrupole is generated. Indeed, at densities $\rho > 10^{12}$ g cm$^{-3}$ the temperature perturbations $\delta T$ have been shown to be insensitive to the choice of outer boundary condition (Fig \ref{fig:Delta T Different Boundaries}). 

In the second case, the lack of knowledge surrounding the nature of the evolution of the internal magnetic field means we cannot faithfully comment on the true nature of its structure inside old ($\sim 10^9$ yr) accreting NSs. Ambipolar diffusion is thought to be the main driver of the magnetic field evolution in the core of the NS, whilst Ohmic decay (resulting due to finite electrical conductivity) and Hall drift are responsible for the evolution of the crustal magnetic field (see \citet{Igoshev_2021} for a review). Calculating the magnetic field evolution is far beyond the scope of this paper, and as such we have calculated the level of temperature asymmetry in the presence of a magnetic field that extends over the entire star, when it is confined the crust, and when it is able to permeate the outermost region of the core. This therefore represents three regimes of possibility, with the actual level of temperature asymmetry in the NS resulting from a the magnetic field most likely existing somewhere between the two extremes. 

In the third case, the ellipticity scaling (Eq. (\ref{eq: EllipticityScale})) used in this work was derived from the results of UCB. They suggested that perturbations in the star’s density profile enter through a temperature dependence on the electron mean molecular weight, $\mu_e$, which was assumed to be equivalent to the threshold energy for electron capture and pycnonuclear events. Indeed, (Eq. (\ref{eq: EllipticityScale})) was fitted to a single capture layer in the deep crust with threshold energy $Q = 95$ MeV. However, this capture layer was \textit{artificially} added ad-hoc near the bottom of the crust by UCB, and only capture layers with threshold energies exceeding $\sim 90$ MeV were found to be able to produce large enough mass quadrupoles to balance accretion torques.

The modern accreting BSk19-21 EoSs which have been used in this work do not predict the existence of such capture layers. The deepest capture layer across the three BSk19-21 models is predicted by BSk21, with a maximum threshold energy $Q_{\rm{max}} = 69.1$ MeV taking place at a density $7.26 \times 10^{13}$ g cm$^{-3}$. The existence of additional capture layers deeper in the crust by UCB (i.e. those with $Q = \mu_e > 69.1$ MeV) is in this sense speculative. This suggests Eq. (\ref{eq: EllipticityScale}) likely overestimates the size of the ellipticity induced by $1\%$ temperature asymmetry, and therefore a self-consistent calculation of the crust's elastic response to a temperature perturbation is something that is to be addressed in future work. \citep{Hutchins_2023}.  

The fourth caveat we identify is related to a number of possible improvements to our thermal model. Throughout our analysis we have adopted the set of Brussels–Montreal functionals BSk19, BSk20, and BSk21. The composition tables provided by \citet{Fantina} and unified pressure-density relations presented in \citet{Fantina_2022} allowed us to implement a realistic and self-consistent calculation of the crust and core of an accreting neutron star that has been absent in previous works.

These models, however, are to some degree restrictive, and do not capture a number of pieces of physics that are expected to affect the thermal structure of accreting neutron stars. For one, the BSk models each assume the one-component approximation at each crustal layer. More realistic \textit{multicomponent} models have recently recently been considered by a number of authors \citep{Lau_2018, Shchechilin_2019,Schatz_2022}. Incorporating such models into our work would not only allow for a better justification of the method to smear heat deposited via DCH reactions (Sec. \ref{subsection:Accretion_Heating}), but also provide a better description of the effects on thermal conductivity due to the presence of impurities in the crustal layers (see Eq. (\ref{eq:Impurity}), Fig. \ref{fig:BSk Impurity Background Parameters}, Fig. \ref{fig:Vary Impurity}, etc.). 

Additionally, the BSk models also neglect the redistribution of unbound neutrons in the deep crust. Such an effect, particularly in the presence of neutron superfludity, has been shown to affect a range of different crustal parameters, including from the EoS and composition \citep{Gusakov_2020}, as well as the net heat generation within the accreted crust \citep{Gusakov_2021}. Such implications could also impact the scaling relation Eq. (\ref{eq: EllipticityScale}), since it is fitted to the results of UCB who also assume the OCP approximation.

It is also possible that our treatment of transport properties in the core of the NS may be improved in a couple of ways. Firstly, as part of our thermal model we implemented the results of \citet{1995NuPhA.582..697G} to describe lepton conduction in the NS core. More recent analysis by \citet{Shternin_2007}, however, has studied the inclusion of Landau damping on interactions involving relativistic electrons, which has been shown to modify the temperature dependence of the scattering frequencies. Whilst we have also sought to include the effects of baryon superfluidity in our thermal calculation, effects of proton superconductivity on transport properties was neglected since electrons are the primary carriers of heat inside the NS core. As shown by \citet{Schmitt_2018} however, proton superconductivity can also affect the temperature dependence of scattering mechanisms involving protons due to screening effects, which may in turn have consequences for the temperature profile of the star.

Lastly, the possibly of heat conduction via superfluid phonons is potentially an important piece of NS crust physics that is missing from our current model. The asymmetry-inducing mechanism considered here relies on the fact that the heat flow orthogonal to the magnetic field is suppressed since the heat-carrying electrons are charged. Conduction by electrically-neutral superfluid phonons may therefore enhance the heat flow orthogonal to the magnetic field line that would otherwise be suppressed. A significant 'short circuiting' of our asymmetry-inducing mechanism however is still unlikely, since \citet{PhysRevLett.102.091101} have shown magnetic field strengths in excess of 10$^{13}$ G - much larger than those associated with LMXBs - would be required to significantly enhance conduction by these superfluid phonons to a level where it might compete with the electron conductivity.

In short, whilst there is still much work to be done in the search for evidence of a continuous GW signal, the results here should be seen as encouraging as we enter deeper into the era of gravitational wave astronomy. With the proposed third generation ground-based interferometers, Cosmic Explorer and Einstein Telescope, the increased sensitivity of such instruments should lead to a wealth of new data, perhaps including the as-yet elusive continuous GW wave signal.

\section*{Acknowledgements}

We wish to thank Anthea Fantina and Nicolas Chamel for sharing data on the pressure-density relations for the BSk19-21 EoSs, as well as Ian Hawke for providing assistance with the construction of our numerical code. We also thank Nils Andersson and Andreas Schmitt for helpful conversations relating to magnetic fields in the presence of proton superconductivity. Additionally we acknowledge useful discussion with Ed Brown and Wynn Ho regarding various aspects of our thermal model. TH acknowledges support from the Science and Technology Facilities Council (STFC) through Grant No. ST/T5064121/1. DIJ also acknowledges support from the STFC via grant No. ST/R00045X/1.

\section*{Data Availability}

No new data were generated or analysed in support of this research.



\bibliographystyle{mnras}
\bibliography{Paper} 



\bsp	
\label{lastpage}
\end{document}